\documentclass{cimento}
\usepackage{graphicx} 
\usepackage{bbold}
\usepackage{amsmath}
\usepackage{amssymb}
\usepackage{lineno}

\title{Majorana quasiparticles in condensed matter}
\author{Ram\'on~Aguado\from{ins:x}\thanks{email: raguado@icmm.csic.es}}
\instlist{\inst{ins:x} Instituto de Ciencia de Materiales de Madrid (ICMM), Consejo Superior de Investigaciones Cient'ficas (CSIC), Cantoblanco, 28049 Madrid, Spain}
\PACSes{\PACSit{--.--}{\dots}
\PACSit{--.--}{\ldots}}

\begin{document}

\maketitle

\begin{abstract}
In the space of less than one decade, the search for Majorana quasiparticles in condensed matter has become one of the hottest topics in physics. The aim of this review is to provide a brief perspective of where we are with strong focus on artificial implementations of one-dimensional topological superconductivity. After a self-contained introduction and some technical parts, an overview of the current experimental status is given and some of the most successful experiments of the last few years are discussed in detail. These include the novel generation of ballistic InSb nanowire devices, epitaxial Al-InAs nanowires and Majorana boxes, high frequency experiments with proximitized quantum spin Hall insulators realised in HgTe quantum wells and recent experiments on ferromagnetic atomic chains on top of superconducting surfaces.
\end{abstract}
\tableofcontents

\section{Introduction}

In just five years, three giants of physics gave formal body to various, somewhat disjointed, ideas and constructed what we know today as quantum mechanics. Werner Heisenberg was first, producing his matrix formulation in the summer of 1925. Half a year later, Erwin Schr\"odinger put forward his non-relativistic wave equation. Soon thereafter, Paul Dirac completed the intellectual tour de force with, arguably, two of the most important building blocks of modern physics. First, he fully established in 1930 the most general framework of quantum mechanics, that of operators acting on the Hilbert space, in his highly influential book {\it Principles of Quantum Mechanics}. Two years before, he had presented the famous equation that bears his name, a relativistic extension of Schr\"odinger's wave equation \cite{Dirac1928}.  This fundamental work not only reconciled special relativity and quantum mechanics, but also led Dirac to predict the existence of antimatter, to understand the origin of spin and to lay the foundations of quantum field theory.

This quantum physics revolution had only just started when the young italian physicist Ettore Majorana \footnote{Ettore Majorana was very young when he joined {\it I ragazzi di Via Panisperna}, a research group in Rome led by Enrico Fermi. Majorana was known for being extremely humble about his work, and even considered it to be banal. In fact, he only wrote nine papers, the one about Majorana fermions being the last. Just a few months after he published this paper, he took the night boat to Palermo on March 23, 1938 to never be seen again (Fig. 1).  
Whether or not he could witness the enormous influence of his equation is a controversial subject since, despite several years of investigations, contradicting speculations about his fate are still debated: The most common theory is suicide but others defend that he retreated into a monastery or even travelled to America. His life and mysterious dissapearence have inspired many books, including various novels. It is also nicely portrayed in the italian movie {\it I ragazzi di Via Panisperna} (1988).} proposed in 1937 an alternative representation of the Dirac equation in terms of real wave functions. Majorana's representation had profound consequences: a real wave function describes a particle that, unlike electrons and positrons, is its own antiparticle. In his article, Majorana postulated that neutrinos could be one of these exotic relativistic fermions but, to date, their detection (through a rare nuclear weak process known as neutrinoless double beta decay) remains an experimental challenge \cite{RMP-MajoranaNeutrinos,Review-MajoranaNeutrinos2,Review-MajoranaNeutrinos3,Review-MajoranaNeutrinos4}. 
\begin{figure}[!tt]
\centering
\includegraphics[width=0.6\columnwidth]{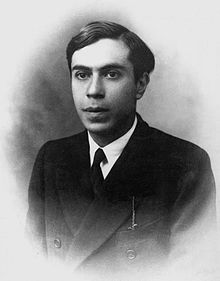}\\
\includegraphics[width=0.6\columnwidth]{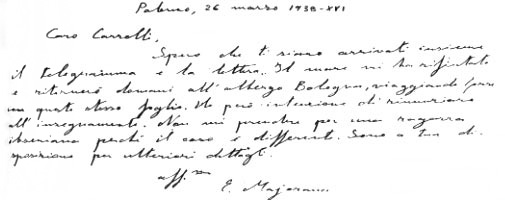}
\caption{Top: Portrait of Ettore Majorana (Copyright E. Recami and Maria Majorana, courtesy AIP Emilio Segre Visual Archives, E. Recami and Fabio Majorana Collection). Bottom: On the day of his disappearance, March 25 1938, Majorana sent a letter to Antonio Carrelli, Director of the Naples Physics Institute:
\emph{Dear Carrelli, I made a decision that has become unavoidable. There isn't a bit of selfishness in it, but I realize what trouble my sudden disappearance will cause you and the students. For this as well, I beg your forgiveness, but especially for betraying the trust, the sincere friendship and the sympathy you gave me over the past months. I ask you to remember me to all those I learned to know and appreciate in your Institute, especially Sciuti: I will keep a fond memory of them all at least until 11 pm tonight, possibly later too. E. Majorana}. The next day, Carelli received a telegram and a second letter (in the image) from Palermo:
\emph{Dear Carrelli, I hope you got my telegram and my letter at the same time. The sea rejected me and I'll be back tomorrow at the Hotel Bologna traveling perhaps with this letter. However, I have the intention of giving up teaching. Don't think I'm like an Ibsen heroine, because the case is different. I'm at your disposal for further details. E. Majorana}. He apparently bought a ticket from Palermo to Naples and was never seen again \cite{Holstein,Esposito}.}
\label{fig:1}
\end{figure}
The direct observation of Majorana fermions would profoundly impact various fields in physics. While for more than 60 years this impact was restricted to high energy physics --i. e elementary particle physics, nuclear physics, astrophysics, and cosmology--Majorana's idea is back in the spotlight \cite{Wilzeck}, but the stage has changed: the hunting for Majorana particles is happening now in new territory, that of condensed matter physics. Certainly, the search for Majoranas in condensed matter is becoming one of the hottest topics in physics nowadays, resulting in one of the most successful comebacks in physics of the last decades.

In condensed matter, the term Majorana fermion does not refer to an elementary particle, but rather to \emph{emergent quasiparticles}. Indeed, the Bogoliubov-de Gennes (BdG) equation that describes quasiparticle excitations in superconductors has the same mathematical structure as the Majorana equation \cite{Senthil-Fisher,Chamon,Beenakker-colliding}. The reason behind this similarity is the underlying particle-hole (charge conjugation) symmetry in superconductors: Unlike quasiparticles in a metal, with well defined charge, quasiparticles in a superconductor involve coherent superpositions of electrons and holes (with electron and hole excitations playing the role of particle and antiparticle, respectively). 

The race for detecting Majoranas in condensed matter \cite{Franz} has mostly concentrated on spinless superconductors with broken spin-rotation and time-reversal symmetry. This class of superconductors is intriguing since it realises nontrivial phases with topological edge states of Majorana character \footnote{During the last few years, the field of topological materials has exploded. Broadly speaking, these are materials with a bulk gap that can be in different phases characterised by the value of a topological invariant, not by a broken symmetry (for reviews see \cite{Hasan-Kane,Qi-Zhang}). Topological invariants count the number of protected edge states, such as the helical states in the Quantum Spin Hall effect or the chiral edge states in the Quantum Hall effect. In a topological superconductor, these nontrivial edge states have Majorana character.}. These can be Majorana edge modes propagating along the boundary of a two-dimensional $p_x+ip_y$ superconductor or Majorana zero-modes (MZMs) that bind to topological defects (vortex cores in two dimensions or the ends of a one-dimensional topological $p$-wave superconductor).  Remarkably, MZMs are not fermions but rather possess non-Abelian exchange statistics, a physical property that has no analog in high-energy physics. Although these ideas go back more than two decades \cite{Senthil-Fisher,Kopnin-Salomaa,Moore-Read,Volovik1,Read-Green1,Mackenzie,Kitaev1,Kitaev2,Volovik2,DasSarma1}, they remained relatively unknown outside their respective communities, since intrinsic materials with two-dimensional $p_x+ip_y$ pairing are almost nonexistent in nature (with only a few exceptions including the $\nu= 5/2$ fractional quantum Hall state \cite{Moore-Read}, the $p$-wave superconductor $Sr_2RuO_4$ \cite{Mackenzie,DasSarma1} and the A phase of superfluid ${}^3$He \cite{Kopnin-Salomaa,Volovik1,Volovik2}). 

This state of affairs changed with the seminal paper of Fu and Kane who demonstrated in 2008 the possibility of engineering $p$-wave superconducivity out of standard $s$-wave superconductors \cite{Fu-Kane1}. The conceptual breakthrough was to realise that $s$-wave pairing induced by the proximity effect in the surface of a three dimensional topological insulator (namely onto helical bands with spin-momentum locking), behaves effectively as a two-dimensional $p_x+ip_y$-wave superconductor. This work fuelled other theoretical proposals of similar spirit \cite{Fujimoto2008,Zhang2008,Sato2009,Lee2009,Sau2010,Alicea2010,Lutchyn2010,Oreg2010} and a great deal of experimental activity with the long term goal of generating, detecting and braiding non-Abelian MZMs in a controlled manner. To date it remains a great challenge to find direct experimental evidence of topological superconductivity, not to mention of non-Abelian statistics of MZMs. The successful achievement of such ambitious goal
would open the possibility of practical topologically protected, fault-tolerant, quantum computation \cite{Nayak:RMP08,DasSarma-review}. Thus,  if the detection of Majorana neutrinos were a major breakthrough in physics, then the unambiguous proof of non-Abelian braiding of Majorana quasiparticles would have similar, or even wider, relevance.

The search for Majorana particles in condensed matter is developing very fast and the aim of this review is to provide a brief perspective of where we are. For completeness, we include some introductory material and some technical sections but this is by no means a comprehensive view of the field of topological superconductivity and Majoranas in condensed matter (for this, we refer the interested reader to the fantastic review by Jason Alicea \cite{Alicea-review}). Several other reviews also exist, ranging from popular to very specialised \cite{Wilzeck,Franz,DasSarma-review,Beenakker-review,Flensberg-review,Stanescu-review,Franz-review,Beenakker-review2,Sato-Fujimoto-review}, with partial overlap with the introductory material here (the impatient/specialist reader can safely jump to sections \ref{detection} and \ref{progress}). The review strongly focuses on artificial implementations of topological superconductivity, with special emphasis on one-dimensional topological superconductivity in semiconducting nanowires and edges of two-dimensional topological insulators. For readers interested in intrinsic topological superconductors, we refer to the recent review by Masatoshi Sato and Yoichi Ando \cite{Sato-Andoreview}, where a very good overview of candidate materials for topological superconductivity $Sr_2RuO_4$, $Cu_xBi_2Se_3$, etc, as well as different routes for non-trivial paring (odd-parity superconductors, noncentrosymmetric superconductors with dominant triplet pairing, etc), are given. The relation between topological superconductivity and odd-frequency pairings has been reviewed by Tanaka, Sato and Nagaosa in Ref. \cite{Tanakareview}. Also worth mentioning is the recent  review by Lutchyn et al \cite{Lutchyn-review} which contains very useful sections with focus on materials science aspects of semiconductor-superconductor platforms and on next-generation experiments.

The review is organised as follows. In section \ref{Dirac-Majorana} a short overview of Majoranas in high energy physics is given, with dedicated subsections to the concept of antiparticle, charge conjugation symmetry, etc. In section \ref{MajoranasCM} the concept of Majorana quasiparticles in condensed matter is explained, with especial emphasis on the Bogoliubov-De Gennes description of superconductivity, the relation between the inherent particle-hole symmetry in this description and charge conjugation symmetry in high energy, the difference between Majorana fermions and Majorana zero modes in condensed matter, etc. This section contains two subsections with the canonical examples of topological superconductivity in two and one dimensions: the chiral $p_x+ip_y$ superconductor (subsection \ref{pwave}) and the Kitaev model (subsection \ref{Kitaev}). Section \ref{MajoranasCM} ends with a short discussion about non-Abelian braiding. The two most famous implementations of artificial topological superconductivity are described in section \ref{implementations}, where we explain in some detail the Fu and Kane model (subsection \ref{Fu-Kane}) based on topological insulators (both three-dimensional and two-dimensional) and the Lutchyn-Oreg models based on Rashba semiconductors (subsection \ref{Rashba}). Specific detection protocols are described in section \ref{detection}, while the state-of-the-art  (or at least my own subjective view on the subject) concerning detection of Majoranas in one-dimensional implementations is explained in detail in section \ref{progress}

\section{The Dirac and Majorana equations \label{Dirac-Majorana}}
\subsection{Dirac equation \label{1.1}}
One of the most iconic symbols of the pop culture of the twentieth century is Einstein's famous equation formulating the mass-energy equivalence of a relativistic particle at rest 
\begin{equation}
\label{relativistic energy-momentum1}
E=m c^2,
\end{equation}
with $c=3\times10^8 m/s$ being the speed of light in vacuum and $m$ the rest mass. The most general formulation of the energy of a relativistic particle, which takes into account a finite momentum ${\bf p}\equiv(p_x,p_y,p_z)$, is somewhat less known for non-experts:
\begin{equation}
 \label{relativistic energy-momentum2}
E^2=c^2{\bf p}^2+ m^2c^4.
\end{equation}

The energy and momentum enter on equal footing in Eq. (\ref{relativistic energy-momentum2}), which is a direct consequence of relativistic invariance \footnote{Eq. (\ref{relativistic energy-momentum2}) reflects the basic invariant of a Lorentz transformation $c^2t^2-{\bf x}^2$, where time and space are symmetric.}. Eq. (\ref{relativistic energy-momentum2}) took Paul Dirac busy for some months "trying to take the square-root of a matrix \footnote{One of the most famous anecdotes about Paul Dirac is the one quoted by Niels Bohr \cite{Gottfried}: "Dirac was the strangest man who ever visited my institute. During one of Dirac's visits I asked him what he was doing. He replied that he was trying to take the square-root of a matrix, and I thought to myself what a strange thing for such a brilliant man to be doing. Not long afterwards the proof sheets of his article on the equation arrived, and I saw he had not even told me that he had been trying to take the square root of the unit matrix!"}" when figuring out how to write a relativistic version of Schrodinger's wave equation. Dirac's struggle can be easily understood if one makes the standard operator substitution $E\rightarrow i\hbar\frac{\partial}{\partial t}$ and ${\bf p}\rightarrow -i\hbar\nabla$ in  Eq. (\ref{relativistic energy-momentum2}) to obtain 
\begin{equation}
 \label{relativistic energy-momentum3}
[\hbar^2\frac{\partial^2}{\partial t^2}-c^2\hbar^2\nabla^2+ m^2c^4]\Psi=0,
\end{equation}
which is just Klein-Gordon's equation:
\begin{equation}
 \label{Klein-Gordon}
[\frac{1}{c^2}\frac{\partial^2}{\partial t^2}-\nabla^2+ \frac{m^2 c^2}{\hbar^2}]\Psi=0.
\end{equation}
Eq. (\ref{Klein-Gordon}), unlike Schrodinger's equation,
\begin{equation}
i\hbar\frac{\partial \Psi}{\partial t}=-\frac{\hbar^2}{2m}\nabla^2\Psi,
\end{equation}
is quadratic with respect to both time and space differentiation. The double differentiation with respect to time is problematic since it does not give the correct wave function dynamics (given and initial condition $\Psi(t_0)$, the time evolution of the probability of a given variable cannot be predicted). 
Dirac's goal, when "the trying to take square-root of a matrix", was to write an equation which, unlike the Klein-Gordon equation, was linear in $\frac{\partial}{\partial t}$. In order to be covariant, the equation had also to be linear in $\nabla$, which lead Dirac to propose the following general form:
 \begin{equation}
 H_{Dirac}\Psi=( c\boldsymbol\alpha {\bf p}+\beta m c^2)\Psi.
\end{equation}
The four coefficients ${\boldsymbol \alpha\equiv(\alpha_x,\alpha_y,\alpha_z})$ and $\beta$ are determined by the requirement that a free particle must satisfy the relativistic energy-momentum relation, namely:
\begin{equation}
c^2{\bf p}^2+ m^2c^4=(c  \boldsymbol\alpha {\bf p}+\beta mc^2)^2.
\end{equation}
This gives Dirac's celebrated equation \cite{Dirac1928}:
\begin{equation}
 \label{Dirac}
i\hbar\frac{\partial}{\partial t}\Psi=H_{Dirac}\Psi=(c  \boldsymbol\alpha {\bf p}+\beta mc^2)\Psi.
\end{equation}
Importantly, $ \boldsymbol\alpha$ and $\beta$ cannot commute. In fact they are 4$\times$4 matrices whose form is such that when operating twice over $H_{Dirac}$ give Eq. (\ref{relativistic energy-momentum2}). While they are not unique, Dirac proposed the following form:
\begin{equation}
 \label{gamma-Dirac}
\alpha_i \equiv \sigma_x\otimes\sigma_i= \left( \begin{array}{ccc}
0 & \sigma_i \\
\sigma_i & 0  \end{array} \right) ,  \beta\equiv\sigma_z\otimes\mathbb{1} = \left( \begin{array}{ccc}
\mathbb{1} & 0  \\
0 & -\mathbb{1} \end{array} \right)
\end{equation}
 with $\sigma_i$ being the Pauli matrices.  
 Thus, the elimination of the awkward square root, while reconciling the principles of quantum theory with those of special relativity, leads quite naturally to a matrix structure describing the spin degree of freedom. By defining the so-called gamma matrices
\begin{equation}
\gamma^\mu\equiv(\beta;\beta\boldsymbol\alpha),
\end{equation}
one can write (from now on $\hbar=c=1$):
\begin{equation}
 \label{Dirac2}
(i\gamma^\mu \partial_\mu-m)\Psi=0.
\end{equation}
with the standard notation $\partial_\mu\equiv\frac{\partial}{\partial x^\mu}$. The gamma matrices obey the Clifford algebra $\{\gamma^\mu,\gamma^\nu\}=2\eta_{\mu\nu}$, with $\eta_{\mu\nu}$ the Minkowski tensor.
This notation explicitly expresses Dirac's equation in Lorentz covariant form. 
\subsection{The concept of antiparticle} As we have seen, Dirac's equation provides with an elegant way of taking the square root of the Laplacian operator. The square root form, however, means that both positive and negative energies are allowed. This can be easily seen if we seek for energy eigenvalues in Dirac's original representation in Eq. (\ref{Dirac}).
\[ \left( \begin{array}{ccc}
m & \boldsymbol\sigma{\bf p} \\
\boldsymbol\sigma{\bf p}  & -m  \end{array} \right) \left( \begin{array}{ccc}
\phi_A  \\
\phi_B \end{array} \right)=E \left( \begin{array}{ccc}
\phi_A  \\
\phi_B \end{array} \right)\]
where $\phi_A$ and $\phi_B$ are two-component spinors.  There are four independent solutions of this equation, two with $E>0$ and two with $E<0$, that can be obtained by solving the equations:
\begin{eqnarray}
&&\boldsymbol\sigma{\bf p}\phi_B =(E-m)\phi_A\nonumber\\
&&\boldsymbol\sigma{\bf p}\phi_A =(E+m)\phi_B.\
\end{eqnarray}
Using $\phi_A^{(s)}=\chi^{(s)}$ with
\[ \chi^{(1)}= \left( \begin{array}{ccc}
1\\
0 \end{array} \right), \chi^{(2)}= \left( \begin{array}{ccc}
0\\
1 \end{array} \right),\]
one can readily get $\phi_B^{(s)}=\frac{\boldsymbol\sigma{\bf p}}{E+m}\chi^{(s)}$. Similarly, the negative energy solutions fulfil $\phi_A^{(s)}=-\frac{\boldsymbol\sigma{\bf p}}{|E|+m}\chi^{(s)}$, which gives four solutions of the form ($s=1,2$):
\[ \phi^{(s)}_{E>0}= \left( \begin{array}{ccc}
\chi^{(s)}\\
\frac{\boldsymbol\sigma{\bf p}}{E+m}\chi^{(s)} \end{array} \right),  \phi^{(s+2)}_{E<0}= \left( \begin{array}{ccc}
\frac{-\boldsymbol\sigma{\bf p}}{|E|+m}\chi^{(s)}\\
\chi\sigma
 \end{array} \right).
\]

Instead of neglecting the seemingly nonsensical negative energy solutions, Dirac interpreted them as describing the \emph{antiparticle } of the electron \cite{Dirac1930}. Such interpretation was rather problematic since an electron with positive-energy would be unstable and decay into such negative-energy states by e. g. emitting a photon. To avoid this problem, Dirac hypothesised that the vacuum is a many-body quantum state in which all the negative-energy electron eigenstates are occupied. Decay of positive-energy states into such filled 'Dirac sea' would be forbidden owing to the Pauli exclusion principle. Dirac further argued that unoccupied negative-energy eigenstates, holes in the Dirac sea, would behave like a positively charged particle. Dirac's prediction of anti-matter in 1930 \cite{Dirac1930} was confirmed two years later by the experimental discovery of the positron.
\subsection{Majorana equation}The  matrices $\gamma^\mu$ contain both real and imaginary numbers and thus $\Psi$ must then be a complex field. This is expected since
electrons are electrically charged and the description of the charged particles requires complex fields. In his 1937 paper Majorana posed the question of whether
Dirac's equation must necessarily involve complex fields. Interestingly, he discovered the following set of purely imaginary gamma matrices: $\tilde{\gamma}^0=\sigma_2\otimes\sigma_1$, $\tilde{\gamma}^1=i\sigma_1\otimes\mathbb{1} $,$\tilde{\gamma}^2=i\sigma_3\otimes\mathbb{1}$ and $\tilde{\gamma}^3=i\sigma_2\otimes\sigma_2$, which results in the following equation \cite{Majorana1937}:
\begin{equation}
 \label{Majorana1}
(i\tilde{\gamma}^\mu\partial_\mu-mc)\tilde{\Psi}=0.
\end{equation}
Since the matrices $i\tilde{\gamma}$ are purely real, the corresponding fields are also real, which leads to the so-called reality condition
\begin{equation}
 \label{reality}
\tilde{\Psi}=\tilde{\Psi}^*. 
\end{equation}
Since an electrically charged particle is different from its antiparticle, the solutions to Majorana equation must necessarily describe neutral particles, equal to their own antiparticle. Majorana speculated that his real solutions might apply to neutrinos.
\subsection{Charge conjugation}
Formally, the statement that a particle equals its own antiparticle must be obtained by imposing charge conjugation symmetry. To make the charge degree of freedom explicit, let us couple the Dirac  equation to the electromagnetic field $A_\mu$ through the minimal substitution ($i\partial_\mu\rightarrow i\partial_\mu+eA_\mu$) \cite{Zee2003Book}:
\begin{equation}
 \label{Charge-Conj1}
[\gamma^\mu (i\partial_\mu+eA_\mu)-m]\Psi=0.
\end{equation}
The charge conjugate solution $\Psi_c$ must satisfy the same equation but with positive charge, namely: 
\begin{equation}
 \label{Charge-Conj2}
[\gamma^\mu (i\partial_\mu-eA_\mu)-m]\Psi_c=0.
\end{equation}
Our goal is now to establish a one-to-one correspondence between $\Psi_c$ and $\Psi$. Let us first take the complex conjugate of Eq. (\ref{Charge-Conj1}),
\begin{equation}
 \label{Charge-Conj1b}
[-\gamma^{\mu*} (i\partial_\mu-eA_\mu)-m]\Psi^*=0.
\end{equation}
Now, we define a matrix $\mathcal{C}$ which satisfies $-(\mathcal{C}\gamma^0)\gamma^{\mu*}=\gamma^\mu(\mathcal{C}\gamma^0)$, then Eq. (\ref{Charge-Conj1b}) can be written in the form of Eq. (\ref{Charge-Conj2}), namely
\begin{equation}
 \label{Charge-Conj2b}
[\gamma^\mu (i\partial_\mu-eA_\mu)-m]\mathcal{C}\gamma^0\Psi^*=0.
\end{equation}
This defines charge conjugation as $\Psi_c=\mathcal{C}\gamma^0\Psi^*$. The charge conjugation operator $\mathcal{C}$ is not unique. A possible choice, corresponding to the gamma matrices in Eq. (\ref{gamma-Dirac}), is $\mathcal{C}\gamma^0=i\gamma^2$.

In this language, the statement that a particle is equal to its own antiparticle is expressed as a wave function equal to its charge conjugate solution, namely
\begin{equation}
\label{pseudo}
\Psi=\Psi_c= i\gamma^2\Psi^*.
\end{equation}
Eq. (\ref{pseudo}), which in some papers is called pseudo-reality condition \cite{Jackiw1}, generalizes the reality condition in Eq. (\ref{reality}). Let us briefly discuss its meaning.

In particular, it is instructive to have a look at the Dirac equation in Weyl representation \footnote{The Weyl, or chiral, representation corresponds to the following choice of matrices \[
 \label{alpha-beta-Weyl}
\alpha_i = \left( \begin{array}{ccc}
-\sigma_i  & 0\\
0 & \sigma_i  \end{array} \right) ,  \beta = \left( \begin{array}{ccc}
0&\mathbb{1}   \\
\mathbb{1}&0 \end{array} \right)
\] which gives this very convenient form of gamma matrices \[
 \label{gamma-Weyl}
\gamma0= \left( \begin{array}{ccc}
0 & \mathbb{1} \\
 \mathbb{1} &0  \end{array} \right) ,  \gamma^i= \left( \begin{array}{ccc}
0&\sigma^i   \\
-\sigma^i&0 \end{array} \right).
\]}

\begin{equation}
\left( \begin{array}{ccc}
-m&i\partial_t-\boldsymbol\sigma{\bf p}  \\
i\partial_t+\boldsymbol\sigma{\bf p}&-m   \end{array} \right) \left( \begin{array}{ccc}
\psi_L \\
\psi_R  \end{array} \right)=0,
\end{equation}
where we have divided the four-component Dirac spinor into two two-component fields $\psi_L$ and $\psi_R$, which are mixed by the mass term \footnote{We note in passing that  $\psi_L$ and $\psi_R$ are the so-called left-handed and right-handed Weyl fermions. For $m=0$ both decouple and one gets the Weyl equations \cite{Peskin} that, for long time, were thought to correctly describe massless neutrinos. The discovery of neutrino oscillations provide evidence for mass terms which prevent neutrinos to be Weyl fermions.  An open question remains whether or not the mass term results from a constraint like in Eq. (\ref{pseudo-b}). Namely, whether neutrinos are Majorana fermions or not.}. Interestingly, the pseudo-reality condition in Eq. (\ref{pseudo}) can be rewritten as
\begin{equation} 
\label{pseudo-b}
\left( \begin{array}{ccc}
\psi_L  \\
\psi_R \end{array} \right)=\left( \begin{array}{ccc}
0 & i\sigma^y \\
-i\sigma^y  & 0  \end{array} \right) \left( \begin{array}{ccc}
\psi_L ^* \\
\psi_R^* \end{array} \right)=\left( \begin{array}{ccc}
i\sigma^y\psi_R^*  \\
-i\sigma^y\psi_L^*  \end{array} \right),
\end{equation}
and thus \emph{decouples} into two separate equations:
\begin{eqnarray}
\label{Majorana-decoupled}
&&(i\partial_ t+\boldsymbol\sigma{\bf p} )\psi_L+im\sigma^y\psi_L^*=0\nonumber\\
&&(i\partial_ t-\boldsymbol\sigma{\bf p} )\psi_R-im\sigma^y\psi_R^*=0,
\end{eqnarray}
that describe two decoupled Majorana fields of the form:
\begin{eqnarray}
\label{Majorana-spinor}
\Psi_L=\left( \begin{array}{ccc}
 \psi_L \\
-i\sigma^y\psi_L^*  \end{array} \right), \Psi_R=\left( \begin{array}{ccc}
i\sigma^y\psi_R^*  \\
 \psi_R
 \end{array} \right).
\end{eqnarray}
Note that the mass term in Eq. (\ref{Majorana-decoupled}) couples particles and antiparticles so the equation does not entail charge conservation. Formally, this can be expressed as the lack of global gauge invariance $\Psi({\bf r})\rightarrow e^{i\theta}\Psi({\bf r})$ which implies that Majorana particles do not  couple to the electromagnetic field (the minimal substitution to the electromagnetic gauge potential is not possible) and are thus necessarily charge neutral. 

Already at this level, it is interesting to note the connection between Eqs. (\ref{Majorana-decoupled}) and (\ref{Majorana-spinor}) and the equations we shall find when discussing superconducting settings. Indeed, as argued by e. g. Wilczek in Ref. \cite{Wilzeck2}, the mass term in  Eq. (\ref{Majorana-decoupled}) has the same structure as the anomalous pairing term in the BCS theory of superconductivity. Moreover, the four-component Nambu spinor that is routinely used in the BdG description of excitations in a superconductor is just an operator version of Eqs. (\ref{Majorana-spinor}). These remarkable analogies will be the subject of the next section.

Finally, the charge conjugation operation has a very clear meaning if we analyse the stationary solutions $\Phi_E=e^{iEt}\Psi$ of Eq. (\ref{Majorana-decoupled}): for each solution 
$\Phi_E$ there exists a charge conjugate solution at negative energy -$E$, namely,
\begin{equation}
\label{charge-cong-eigen}
  \Phi_{-E}=\mathcal{C}\Phi^*_E.
  \end{equation}
   
A quantum field theory can now be developed by constructing a superposition of energy eigenmodes $ \Phi_{E} $ with the appropriate creation an annihilation operators:
 \begin{equation}
 \label{Dirac-field}
 \hat\Psi({\bf r},t) =\sum_{E>0}a_Ee^{-iEt}\Phi_E+\sum_{E<0}b^\dagger_{-E}e^{-iEt}\Phi_E,
 \end{equation}
 where $a_E$ ($b^\dagger_E$) is a standard annihilation (creation) operator in second-quantization for a particle (antiparticle) at energy $E$. The above equation can be rewritten as
   \begin{equation}
   \label{second-quantized-charge-conj}
 \hat\Psi({\bf r},t) =\sum_{E>0}[a_Ee^{-iEt}\Phi_E+b^\dagger_{E}e^{iEt}\mathcal{C}\Phi^*_E].
 \end{equation}
The field $ \hat\Psi({\bf r})$ is a Dirac field operator that satisfies standard anticommutation rules $\{ \hat\Psi_i({\bf r}), \hat\Psi^\dagger_j({\bf r'})\}=\delta_{ij}\delta({\bf r}-{\bf r'})$.

 By demanding $a_E=b_E$ in Eqs (\ref{second-quantized-charge-conj}), one gets the field operator of a Majorana fermion which reads
    \begin{equation}
   \label{second-quantized-Majorana}
 \hat\Psi({\bf r},t) =\sum_{E>0}[a_Ee^{-iEt}\Phi_E+a^\dagger_{E}e^{iEt}\mathcal{C}\Phi^*_E].
 \end{equation}
 Owing to the pseudo-reality constraint 
 \begin{equation}
  \label{Majorana-field-pseudoreality}
 \hat\Psi_i=\mathcal{C}_{ij}\hat\Psi^\dagger_j, 
 \end{equation}
 the anticommutators take a Majorana form
 \begin{eqnarray}
  \{ \hat\Psi_i({\bf r}), \hat\Psi_j({\bf r'})\}=\mathcal{C}_{ij}\delta({\bf r}-{\bf r'})\nonumber\\
 \{ \hat\Psi_i({\bf r}), \hat\Psi^\dagger_j({\bf r'})\}=\delta_{ij}\delta({\bf r}-{\bf r'}).
 \end{eqnarray}
 
The Majorana character of the above field becomes particularly clear in Majorana representation where $\mathcal{C}=1$ which leads to the explicit equivalence \cite{Chamon}:
 \begin{equation}
 \label{Majorana-field-reality}
 \hat\Psi_i^\dagger({\bf r},t) =\hat\Psi_i({\bf r},t) 
\end{equation}
An important comment is in order here. Note that either the pseudo-reality condition in Eq. (\ref{Majorana-field-pseudoreality}) or the reality condition in Eq. (\ref{Majorana-field-reality}) is satisfied by the entire (time-dependent) quantum field $\hat\Psi_i({\bf r},t)$. On the contrary, it \emph{cannot} be satisfied at the level of stationary solutions with well defined energy (since $\Phi_{-E}$ and $\Phi_E$  are orthogonal). Interestingly, the only exception is the particular case $E=0$ where the Majorana condition can be fulfilled at the stationary level too. Thus far, most attention in condensed matter has centered around this important $E=0$ case which leads to the concept of MZMs in superconductors.

\section{Majoranas in Condensed Matter \label{MajoranasCM}}
\subsection{Bogoliubov quasiparticles as Majorana fermions}
As we anticipated, the BdG equation describing quasiparticle excitations in a superconductor and the Majorana equation are equivalent \cite{Chamon,Franz-review}. Such remarkable analogy is due to particle-hole (charge conjugation) symmetry, which is inherent to superconductors. To understand why, let us consider a generic Hamiltonian with BCS mean-field pairing:
\begin{equation}
{\cal H}=\int d{\bf r}[\sum_{\sigma,\sigma'}H_0^{\sigma,\sigma'}({\bf r})c^\dagger_{\sigma \bf r}c_{\sigma '\bf r}+(\Delta({\bf r})c^\dagger_{\uparrow \bf r}c^\dagger_{\downarrow \bf r}+H.c.)].
\end{equation}
Here $c^\dagger_{\sigma \bf r}$ creates an electron with spin $\sigma=\uparrow,\downarrow$ at position ${\bf r}$, $H_0^{\sigma,\sigma'}({\bf r})$ is a generic spin-dependent quadratic Hamiltonian (kinetic + single electron potentials) and $\Delta({\bf r})$ is the superconducting pair potential. Progress is achieved by using the four-component Nambu spinor
\begin{eqnarray}
\label{Nambu-spinor}
\hat\Psi({\bf r})=\left( \begin{array}{ccc}
 c_{\uparrow,{\bf r}} \\
c_{\downarrow,{\bf r}}\\  -c^\dagger_{\downarrow,{\bf r}} \\
c^\dagger_{\uparrow,{\bf r}}\end{array} \right)\equiv \left( \begin{array}{ccc}
\hat\psi({\bf r})\\
i\sigma^y\hat\psi^*({\bf r})\end{array} \right).
\end{eqnarray}
The above Hamiltonian can be cast in the so-called BdG form \cite{Bogoliubov,DeGennes}:
\begin{equation}
\label{BdG Hamiltonian1}
{\cal H}=\frac{1}{2}\int d{\bf r}\hat\Psi^\dagger({\bf r})H_{BdG}({\bf r})\hat\Psi({\bf r}),
\end{equation}
with 
\begin{eqnarray}
\label{BdG Hamiltonian2}
H_{BdG}({\bf r})=\left( \begin{array}{ccc}
H_0({\bf r})&\Delta({\bf r}) \\
\Delta^*({\bf r})&-\sigma^y H_0^{*}({\bf r})\sigma^y \end{array} \right).
\end{eqnarray}
The term $-\sigma^y H_0^{*}({\bf r})\sigma^y$ is the time-reversal of $H_0({\bf r})$ and appears since holes are the time-reversed version of electrons. The problem defined by Eqs. (\ref{BdG Hamiltonian1}) and (\ref{BdG Hamiltonian2}) can be solved by seeking for stationary solutions of the form 
\begin{equation}
\label{BdG-eigen}
H_{BdG}({\bf r})\Phi_n({\bf r})=E_n\Phi_n({\bf r}),
\end{equation}
with $\Phi_n({\bf r})=[u_{n\uparrow}({\bf r}),u_{n\downarrow}({\bf r}),v_{n\uparrow}({\bf r}),v_{n\downarrow}({\bf r})]^T$, such that the diagonalised Hamiltonian becomes
\begin{equation}
{\cal H}=\frac{1}{2}\sum_nE_na^\dagger_n a_n,
\end{equation}
with BdG quasiparticle operators defined as
\begin{equation}
\label{BdG-quasiparticle}
a_n=\int d{\bf r}\Phi^\dagger_n({\bf r})\hat\Psi({\bf r})=\int d{\bf r} [u^*_{n,\uparrow}({\bf r})c_{{\bf r},\uparrow}+u^*_{n,\downarrow}({\bf r})c_{{\bf r},\downarrow}-v^*_{n,\uparrow}({\bf r})c^\dagger_{{\bf r},\downarrow}+v^*_{n,\downarrow}({\bf r})c^\dagger_{{\bf r},\uparrow}].
\end{equation}

Note that since we have explicitly included the hole states, thus doubling the dimension of the Hamiltonian, the BdG description is redundant. Therefore, there must necessarily be some symmetry constraint between eigenstates that fixes the number of independent solutions. This constraint reads:
\begin{eqnarray}
PH_{BdG}({\bf r})P^\dagger=-H_{BdG}({\bf r}),
\end{eqnarray}
where the operator $P\equiv \mathcal{C}K$, defined in terms of the charge conjugation operator $\mathcal{C}=\tau^y\otimes\sigma^y$ and the complex conjugation operator $K$, reflects electron-hole symmetry \footnote{This representation is not unique. The BdG Hamiltonian is often given in the alternative representation 
\[
\hat\Psi({\bf r})=\left( \begin{array}{ccc}
 c_{\uparrow,{\bf r}} \\
c_{\downarrow,{\bf r}}\\  c^\dagger_{\uparrow,{\bf r}} \\
c^\dagger_{\downarrow,{\bf r}}\end{array} \right)\] which leads to the BdG Hamiltonian
\[
H_{BdG}({\bf r})=\left( \begin{array}{ccc}
H_0({\bf r})&-i\sigma^y\Delta({\bf r}) \\
i\sigma^y\Delta^*({\bf r})&-H_0^{*}({\bf r}) \end{array} \right).
\] In this case the charge conjugation operator is $\mathcal{C}=\tau^x$. See e. g. \cite{Beenakker-review2,Sato-Fujimoto-review}.}. This means that if there is a solution $\Phi_n({\bf r})$ at positive energy $E_n$, then there is also a solution $ \Phi_{m}({\bf r})$
 at $E_{m}\equiv-E_n$ such that 
 \begin{equation}
 \label{eigenstate-ehsymmetry}
 \Phi_{m}({\bf r})=P\Phi_n({\bf r}).
 \end{equation}
 Equivalently, $a^\dagger_n=a_{m}$. In other words, creating a Bogoliubov quasiparticle with energy $E$ or removing one with energy $-E$ are identical operations.

All the above symmetry considerations strongly suggest an interesting  connection between the BdG description of superconductivity and Majoranas, as we discussed in the previous section. Most importantly, the four-component Nambu spinor in Eq. (\ref{Nambu-spinor}) is nothing but an operator version of the Majorana wave functions in Eq. (\ref{Majorana-spinor}). Indeed, by particle-hole symmetry, the Nambu spinor fulfils the pseudo-reality condition \cite{Chamon}:
\begin{equation}
\hat\Psi({\bf r})=P\hat\Psi({\bf r})=C\hat\Psi^*({\bf r}).
\end{equation}

Thus the BdG theory for quasiparticle excitations in a superconductor already possess all the key properties of Majorana fermions. This interesting analogy has been emphasised and fully worked out by Chamon {\it et al}  \cite{Chamon}. Arguably, the Majorana nature of Bogoliubov quasiparticles was overlooked in the past since it is somewhat hidden in standard representation of Nambu spinors (in the electron-hole basis the spinors fulfill the pseudo-reality condition not Majorana's reality condition). However, a simple unitary transformation renders the field operator self-conjugate  \cite{Chamon}. 

We finally emphasise again that this Majorana character is satisfied by the entire (time-dependent) quantum field but not by the eigenmodes of well defined energy. This can be, in principle, a serious problem towards the observability of the Majorana character of Bogoliubov quasiparticles since they are typically probed in the
energy domain, rather than in the time domain. Despite this difficulty, however, some ideas already exist. For example, Beenakker has argued that the Majorana character of Bogoliubov quasiparticles could be revealed in high-frequency shot noise correlators that can detect the mutual annihilation of two colliding Bogoliubov quasiparticles originating from two identical superconductors (but with different superconducting phase) \cite{Beenakker-colliding}. This pairwise annihilation of Majorana fermions is akin to Majorana pair annihilation in particle physics.

Another route, which is by far the most explored, is to avoid the full time dependence and focus on zero-energy Bogoliubov quasiparticles instead, as we discuss in what follows. 
\subsection{Majorana zero modes}
Let us assume that there is an eigenstate of Eq. (\ref{BdG-eigen}) such that the Bogoliubov operator fulfills $a_n=a^\dagger_n$. According to Eq. (\ref{eigenstate-ehsymmetry}), this can only happen at $E=0$, so that a stationary solution with Majorana character is necessarily a zero energy solution of the BdG equations,
\begin{equation}
\label{BdG-Majorana}
H_{BdG}({\bf r})\Phi_0({\bf r})=0.
\end{equation}
The corresponding real space spinor satisfies $\Phi_{0}({\bf r})=P\Phi_0({\bf r})$, explicitly:
\begin{eqnarray}
\left( \begin{array}{ccc}
 u_{0,\uparrow}({\bf r}) \\
u_{0,\downarrow}({\bf r})\\  v_{0,\uparrow}({\bf r}) \\
v_{0,\downarrow}({\bf r})\end{array} \right)=\left( \begin{array}{cccc}
0&0&0&-1 \\
0&0&1&0\\ 0&1&0&0 \\
-1&0&0&0\end{array} \right)\left( \begin{array}{ccc}
 u^*_{0,\uparrow}({\bf r}) \\
u^*_{0,\downarrow}({\bf r})\\  v^*_{0,\uparrow}({\bf r}) \\
v^*_{0,\downarrow}({\bf r})\end{array} \right)=\left( \begin{array}{ccc}
 -v^*_{0,\downarrow}({\bf r}) \\
v^*_{0,\uparrow}({\bf r})\\  u^*_{0,\downarrow}({\bf r}) \\
-u^*_{0,\uparrow}({\bf r})\end{array} \right),
\end{eqnarray}
which implies that the most general form of a Majorana spinor is:

\begin{eqnarray}
\left( \begin{array}{ccc}
 u_{0,\uparrow}({\bf r}) \\
v^*_{0,\uparrow}({\bf r})\\  v_{0,\uparrow}({\bf r}) \\
-u_{0,\uparrow}^*({\bf r})\end{array} \right)=\left( \begin{array}{ccc}
-v^*_{0,\downarrow}({\bf r}) \\
u_{0,\downarrow}({\bf r})\\  u^*_{0,\downarrow}({\bf r}) \\
v_{0,\downarrow}({\bf r})\end{array} \right).
\end{eqnarray}

Applying the above constraint to the zero mode solution of Eq. (\ref{BdG-quasiparticle}),
\begin{equation}
a_0=i\int d{\bf r} [u^*_{0,\uparrow}({\bf r})c_{{\bf r},\uparrow}+u^*_{0,\downarrow}({\bf r})c_{{\bf r},\downarrow}-v^*_{0,\uparrow}({\bf r})c^\dagger_{{\bf r},\downarrow}+v^*_{0,\downarrow}({\bf r})c^\dagger_{{\bf r},\uparrow}],
\end{equation}
we get 
\begin{equation}
a_0=i\int d{\bf r} [u^*_{0,\uparrow}({\bf r})c_{{\bf r},\uparrow}+u^*_{0,\downarrow}({\bf r})c_{{\bf r},\downarrow}-u_{0,\downarrow}({\bf r})c^\dagger_{{\bf r},\downarrow}-u_{0,\uparrow}({\bf r})c^\dagger_{{\bf r},\uparrow}],
\end{equation}
which is clearly self-conjugate, namely $a_0=a_0^\dagger$, and therefore defines a MZM.

 In the previous derivation we have just assumed that the zero mode exists but that this is a rather peculiar situation. Indeed, the general symmetry obeyed by BdG eigenstates is 
$\Phi_{m}({\bf r})=P\Phi_n({\bf r})$ with $E_m=-E_n$, see Eq. (\ref{eigenstate-ehsymmetry}), and not $\Phi_{n}({\bf r})=P\Phi_n({\bf r})$ with $E_n=-E_n=0$. More importantly, if one of these zero modes exists, it \emph{cannot} acquire a nonzero energy $E$ by any smooth deformation of the Hamiltonian since \emph{finite energy BdG excitations always occur in pairs}. Namely, symmetry would require \emph{another mode} to appear at energy -$E$, in violation of unitarity. There is, however, a way out to the previous conundrum: if the gap separating the zero mode from other quasiparticle excitations \emph{closes}, nothing prevents the system to host pairs of standard BdG excitations should the gap open again. Such closing and re-opening of the superconducting gap is an instance of a \emph{topological transition}: broadly speaking, a transition that separates two phases characterised by the value of a topological invariant (instead of a broken symmetry). In this particular case, the topological invariant counts the number of Majorana zero modes. The superconductors hosting such exotic zero modes with Majorana character are called topological superconductors and the fact that the zero mode cannot acquire a nonzero energy without closing of the gap is called topological protection.

\subsection{Chiral $p_x+ip_y$ superconductors in two dimensions \label{pwave}}
As we mentioned in the introduction, topological superconductivity is expected to appear in spinless superconductors. Namely, in systems where Cooper pairing is established with only one active spin degree of freedom. In such case, the Pauli exclusion principle forces the Cooper pairs to have odd orbital parity resulting in $p$-wave superconductivity \footnote{The Pauli exclusion principle forces the total wave function to be antisymmetric. Since Cooper pairs in standard superconductors are antisymmetric spin singlets, their orbital part is symmetric ($s$-wave pairing). On the contrary spinless fermions have triplet-like pairing and the antisymmetry must be in the orbital part ($p$-wave pairing).}. As we discuss in this section, these superconductors host topological phases with Majorana excitations at their boundaries and at defects.

In two dimensions, the simplest system that exhibits a topological phase with Majorana excitations is the so-called chiral p-wave superconductor, which is described by the BdG Hamiltonian 
\begin{equation}
\label{chiral p-wave}
H_{BdG}=\left( \begin{array}{ccc}
\frac{p_x^2+p_y^2}{2m}-\mu &- i\Delta(p_x-ip_y)\\
i\Delta^*(p_x+ip_y) & -\frac{p_x^2+p_y^2}{2m}+\mu  \end{array} \right),
\end{equation}
with a quasiparticle spectrum $E(p_x,p_y)=\pm\sqrt{(\frac{p_x^2+p_y^2}{2m}-\mu)^2+|\Delta|^2(p_x^2+p_y^2)}$, which is always fully gapped for any $\mu\neq 0$. As discussed by Read and Green in their seminal paper \cite{Read-Green1}, at $\mu=0$ the system undergoes a topological phase transition between a (topologically nontrivial) weak-pairing phase and a (topologically trivial) strong-pairing phase.  Indeed, as one depletes the band, the bulk gap decreases and closes when the chemical potential touches the band bottom at $\mu=0$ (since p-wave pairing becomes zero). Let us consider this phase transition at $\mu=0$ in more detail. When $\mu$ is small we can use the low momentum expansion 
\begin{equation}
\label{chiral p-wave2}
H_{BdG}=\left( \begin{array}{ccc}
-\mu & -i\Delta(p_x-ip_y)\\
i\Delta^*(p_x+ip_y) & \mu  \end{array} \right)
\end{equation}

Following the original argument by Read and Green \cite{Read-Green1}, further insight can be gained by analysing the properties of the system as we cross the special point $\mu=0$. This can be  done by considering a domain wall where $\mu$ changes sign. If there is a continuous gapped interpolation between the $\mu>0$ and $\mu<0$ phases, then they are topologically equivalent. Conversely, as we discussed, a gap closing signals a boundary between topologically distinct phases. Following this line of thought, we want to seek for solutions of the BdG Hamiltonian with definite $p_y$ with an inhomogeneous mass profile:
\begin{equation}
\label{chiral p-wave2}
H_{BdG}=\left( \begin{array}{ccc}
-\mu(x) &-i \Delta(-i\partial_x-ip_y)\\
i\Delta^*(i\partial_x+ip_y) & \mu(x)  \end{array} \right),
\end{equation}
with
\begin{eqnarray}
\mu(x)=\left\{
\begin{array}{rl}
-\mu_0 <0,&\mbox{for $x<0$} \\
\mu_0 >0,&\mbox{for $x>0$}
\end{array}
\right. .
\nonumber
\end{eqnarray}
For $p_y=0$ and assuming also that $\Delta$ is real we arrive at $H_{BdG}=\Delta p_x\tau^y-\mu\tau^z$ which is a one-dimensional massive Dirac Hamiltonian (with Pauli matrices defined in electron-hole space). As discussed by Jackiw and Rebbi in their seminal paper \cite{Jackiw-Rebbi}, this one-dimensional Dirac Hamiltonian contains zero energy bound states at interfaces where the mass term changes sign (which is our case as we cross the $\mu=0$ critical point). In particular, Jackiw and Rebbi showed that the Dirac equation contains zero energy solutions bounded to the mass defect of the form
\begin{eqnarray}
\Phi(x)=
 \exp\left[-\int_0^x dx'\frac{\mu(x')}{\Delta}\right] |\phi_0\rangle,
\end{eqnarray}
with $|\phi_0\rangle$ being a normalised spinor which, in our case, should be an eigenstate of $\tau^x$ namely $|\phi_0\rangle=\frac{1}{\sqrt{2}}(1 1)^T$. The Bogoliubov quasiparticle operator corresponding to this zero mode can be constructed from a Nambu spinor in real space $[c(x), c^\dagger(x)]^T$ which gives:
\begin{equation}
\label{zero-mode-chiralRead-Green}
\gamma=\frac{1}{N}\int dx \exp\left[
-\int_0^x dx'\frac{\mu(x')}{\Delta}\right] [c(x)+c^\dagger(x)]
\end{equation}
which is a Majorana mode with $\gamma=\gamma^\dagger$.
 At finite $p_y$ the solution is 
 \begin{eqnarray}
\Phi(x)=
 \exp\left[
ip_y y-\int_0^x dx'\frac{\mu(x')}{\Delta}\right] |\phi_0\rangle.
\end{eqnarray} 
Clearly, these solutions are bound to the
domain wall and propagate in $y$ direction
along the wall with energy  $E=-\Delta p_y$.  These chiral Majorana fermions can be thought as the superconducting analog of the chiral edge modes of the Quantum Hall effect.

Similar arguments apply to vortex cores in the two-dimensional chiral p-wave superconductor. Let us consider a disk of radius $R$ with $\mu<0$ surrounded by a region with $\mu>0$ for $r>R$. We know from our previous analysis that there will be chiral Majorana fermions around the disk at $r=R$. Importantly, there cannot be an exact zero mode at the interface since the finite size spectrum is of the form $E=\frac{n\Delta}{R}$, where the lowest angular momentum $n$ is shifted from $0$ to $1/2$. This shift comes from the antiperiodic boundary condition which is due to a Berry phase contribution along the closed curve \cite{Stone-Roy}. Thus, while the chiral edge modes are gapless in the thermodynamical limit, there is always a minimum edge excitation energy  $E=\frac{\Delta}{2R}$ in finite size systems. By inserting a flux quantum $\Phi_0=\frac{hc}{2e}$, the boundary conditions become periodic so integer angular momenta are allowed. This, in particular, implies that zero energy solutions can occur.
Physically, an interface between a trivial and a topological region is formed when a magnetic flux, i. e. a vortex, threads the superconductor.  The emergence of Majorana bound states located at the vortex core can be demonstrated by writing Eq. (\ref{chiral p-wave2}) in polar coordinates $\partial_x+i\partial_y\rightarrow e^{i\theta}(\partial_r+\frac{i}{r}\partial_\theta)$ in the presence of a vortex core located at $r=0$, and following the same steps that we used for the previous derivation of the Jackiw-Rebbi zero modes, which gives \cite{Read-Green1}:
\begin{eqnarray}
\Phi(r,\theta)\sim\frac{1}{\sqrt{r}} \exp\left[
-\int_0^r dr'\frac{\mu(r')}{|\Delta(r)|}\right] \left( \begin{array}{ccc}
-ie^{i\theta/2}\\
 ie^{-i\theta/2} \end{array} \right).
\end{eqnarray}
These Majorana bound states in the vortex core of an spinless p-wave superconductor are equivalent to half-quantum vortices in spinful p-wave fluids, first found by Volovik \cite{Volovik1}.

\subsection{One-dimensional $p$-wave superconductivity: the Kitaev model \label{Kitaev}}
\begin{figure}[!tt]
\centering
\includegraphics[width=1\columnwidth]{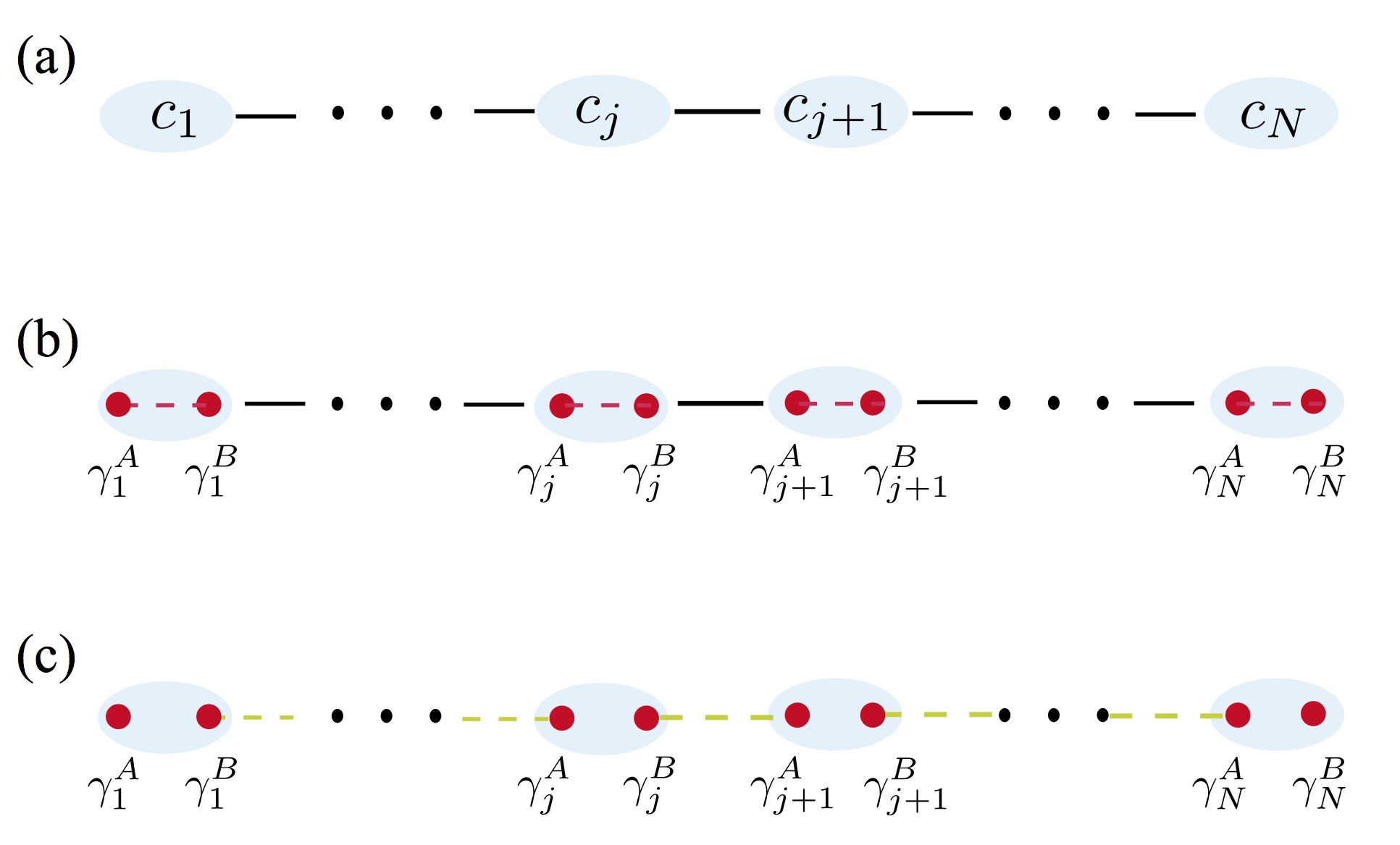}\\
\caption{Kitaev chain. a) One-dimensional chain of spinless fermions with tight-binding parameters as described by Eq. (\ref{kiatev0}). b) Trivial phase of the Kitaev chain: in the Majorana representation of Eq. (\ref{fermionMajoTrans}), the Majoranas on each site are coupled forming a standard fermion. c) Non-trivial phase: Majoranas on each site are decoupled while Majoranas on adjacent sites become coupled. This leaves two unpaired Majoranas at the ends of the chain.}
\label{fig:2}
\end{figure}
The simplest models that exhibit topological superconductivity are BdG Hamiltonians of spinless fermions in one dimension. 
A particularly enlighting model is Kitaev's model \cite{Kitaev1}, which is essentially a toy lattice model that describes the one-dimensional version of the spinless $p$-wave superconductor of the previous subsection. Owing to its simplicity, the model grasps the main property we are seeking, a topological phase with emergent MZMs,  in a rather intuitive fashion. 

The model describes a chain with N sites of spinless fermions with long-range $p$-wave superconductivity (Fig. \ref{fig:2}):
\begin{equation}
\label{kiatev0}
H=-\mu\sum_{j=1}^{N}\Big(c^{\dagger}_{j}c_{j}-\frac{1}{2}\Big)\,+\,\sum_{j=1}^{N-1}\Big[- t\,\big(c^{\dagger}_{j}c_{j+1}+c^{\dagger}_{j+1}c_{j}\big)\,+\,\Delta\,(c_{j}c_{j+1}\,+\,c^{\dagger}_{j+1}c^{\dagger}_{j})\Big]\,,
\end{equation}
where $\mu$ represents the onsite energy, $t$ is the nearest-neighbor hopping amplitude and $\Delta$ is the is the p-wave pairing amplitude (assumed real for the moment). This model is quite simple but already contains all the relevant ingredients for topological superconductivity. First note that time-reversal symmetry is broken (since electrons do not have spin degeneracy) Furthermore, the superconducting pairing is rather non-standard (it couples electrons with the same spin in contrast to standard s-wave pairing). Note also that electrons on adjacent sites are paired.

In order to reveal the nontrivial properties of the model, let us first consider a chain with open boundary conditions, and then write the fermionic operators in terms of two new operators $\gamma_{j}^{A}$ and $\gamma_{j}^{B}$ as:
\begin{equation}
\label{fermionMajoTrans}
c_{j}=\frac{1}{2}\Big(\gamma_{j}^{A}+i\gamma_{j}^{B}\Big)\,,\quad c_{j}^{\dagger}=\frac{1}{2}\Big(\gamma_{j}^{A}-i\gamma_{j}^{B}\Big)\,.
\end{equation}
Using standard fermionic anticommutation algebra for $c_{j}$ it is very easy to verify that the new operators satisfy the following algebra
\begin{equation}
\label{MajAlgebra}
\{\gamma_{i}^{A},\gamma_{j}^{B}\}=2\delta_{ij}\delta_{AB}\,,\quad \gamma_{j}=\gamma^{\dagger}_{j}\,,\quad \gamma_{j}^{2}=\gamma^{\dagger2}_{j}=1\,.
\end{equation}
Thus $\gamma_{j}^{A}$ and $\gamma_{j}^{B}$ are Majorana operators.

This decomposition can be understood as the decomposition of a complex Dirac fermion into real and imaginary parts that correspond to Majorana fermions. The inverse transformation gives us the Majorana operators,
\begin{equation}
\label{MajTrans}
\gamma_{j}^{A}=c_{j}+c_{j}^{\dagger}\,,\quad \gamma_{j}^{B}=i(c^{\dagger}-c_{j})\,.
\end{equation}
In terms of these new operators, the Hamiltonian in Eq.\,(\ref{kiatev0}) reads
\begin{equation}
\label{kitaev0a}
H=-\frac{i\mu}{2}\sum_{j=1}^{N}\gamma^{A}_{j}\gamma^{B}_{j}+\frac{i}{2}\sum_{j=1}^{N-1}\Big[\omega_{+}\gamma^{B}_{j}\gamma^{A}_{j+1}+ \omega_{-}\gamma^{A}_{j}\gamma^{B}_{j+1}\Big]\,,
\end{equation}
where $\omega_{-}=\Delta-t$ and $\omega_{+}=\Delta+t$ represent hopping amplitudes between Majorana fermions in neighbouring sites. Let us consider now different possibilities depending on the value of $\omega_{-}$ and $\omega_{+}$. For $t=\Delta=0$, the Hamiltonian is trivial and given by 
\begin{equation}
\label{kitaev0b}
H=-\frac{i\mu}{2}\sum_{j=1}^{N}\gamma^{A}_{j}\gamma^{B}_{j}\,.
\end{equation}
This Hamiltonian just expresses the fact that Majorana operators from the same physical site are paired together to form a standard fermion. A less obvious case occurs for $t=\Delta$, namely $\omega_{-}=0$ and $\mu=0$. In this case, Eq.\,(\ref{kitaev0a}) becomes,
\begin{equation}
\label{kitaev0c}
H=it\sum_{j=1}^{N-1}\gamma^{B}_{j}\gamma^{A}_{j+1}\,.
\end{equation}
Despite its innocent-looking form, Eq. (\ref{kitaev0c}) is rather nontrivial. First, notice that Majorana operators on the same site are now decoupled  (remember that they originally represented a single fermionic degree of freedom!). Furthermore, long range coupling is established since Majorana operators on neighbouring sites are now coupled. Finally,  the Majorana operators at the end of chain $\gamma_{1}^{A}$ and $\gamma_{N}^{B}$ seem to have disappeared from the problem. In order to reveal the deep meaning of all these features in full, let us rewrite the Hamiltonian by defining a new set of fermionic operators 
\begin{equation}
\begin{split}
d_{j}&=\frac{1}{2}\Big(\gamma_{j}^{B}+i\gamma_{j+1}^{A} \Big)\,,\quad d_{j}^{\dagger}=\frac{1}{2}\Big(\gamma_{j}^{B}-i\gamma_{j+1}^{A} \Big)\,.
\end{split}
\end{equation}

In terms of these new operators, Eq.\,(\ref{kitaev0a}) reads
\begin{equation}
\label{fermioMan}
H=2t\sum_{j=1}^{N-1}\Big(d_{j}^{\dagger}d_{j}-\frac{1}{2} \Big)\,.
\end{equation}
The fermionic operators $d_j$ diagonalise the superconducting problem and therefore describe Bogoliubov quasiparticles with energy $t=\Delta$. Importantly, the diagonalised Hamiltonian contains $N-1$ quasiparticle  operators while the original problem has $N$ sites. The missing fermionic degree of freedom is hiding in the highly delocalised combination:
\begin{equation}
\label{newopMaj}
f=\frac{1}{2}\Big(\gamma_{1}^{A}+i\gamma_{N}^{B}\Big)\,,\quad f^{\dagger}=\frac{1}{2}\Big(\gamma_{1}^{A}-i\gamma_{N}^{B}\Big)\,.
\end{equation} 
This fermionic operator does not appear in the Hamiltonian and thus has zero energy. This is an obvious consequence of the fact that the Majorana operators at the ends of the chain commute with the Hamiltonian
$[H,\gamma_{1}^{A}]=[H,\gamma_{N}^{B}]=0$. Furthermore, this is a standard fermion operator which, as usual, can be empty or occupied. However, this fermionic state is special since both the empty and occupied configurations are degenerate, owing to their zero energy. Note that this ground state degeneracy is rather peculiar since both states differ in fermion parity. This is very different from standard superconductors where, despite the breakdown of particle number conservation, fermion parity is conserved (with even being the parity of the ground state). In contrast, the ground state degeneracy found here, corresponding to different fermionic parities, is unique to topological superconductors and has profound consequences as we shall discuss in the next subsection. 

To make connection with the results of the previous subsection and to investigate the bulk properties for arbitrary parameters, we now consider periodic boundary conditions which, using translational invariance $c_j=\frac{1}{\sqrt{N}}\sum_p e^{ipj}c_p$, allows to write Eq.\,(\ref{kiatev0}) in momentum space $p$ as,
\begin{equation}
H=\sum_p\xi_p(c^\dagger_pc_p-\frac{1}{2})-\sum_p t\cos pa+\Delta\sum_p(c^\dagger_pc^\dagger_{-p}e^{ipa}+c_kc_{-p}e^{-ipa}),
\end{equation}
with $\xi_p=-(\mu+2t\cos pa)$ and $a$ the lattice spacing. Dropping the unimportant constant, the above Hamiltonian can be written in BdG form as
\begin{equation}
H=\frac{1}{2}\sum_{p}\psi_{p}^{\dagger}H_{BdG}\psi_{p}\,\quad 
\psi_{p}\,=\,\begin{pmatrix}
c_{p}\\
c_{-p}^{\dagger}
 \end{pmatrix}\,,
\end{equation}
with $H_{BdG}=\xi_{p}\tau_{z}+\Delta_{p}\tau_{y}=\mathbf{h}\cdot\mathbf{\tau}\,,$
where $\mathbf{h}=(0,\Delta_{p},\xi_{p})$, $\Delta_{p}=-2\Delta\sin pa$ and $\mathbf{\tau}=(\tau_{x},\tau_{y},\tau_{z})$ the Pauli matrices in electron-hole space. The excitation spectrum of $H_{BdG}$ is then given by
\begin{equation}
\label{quasipKiatev}
E_{p,\pm}=\pm\sqrt{(\mu+2t\cos pa)^{2}+4\Delta^{2}\sin^{2}pa}\,.
\end{equation}
This spectrum is mostly gapped except in special cases: for $\Delta\neq0$, the energy gap closes when both elements inside the square root  vanish simultaneously. The normal-state dispersion $\xi_{p}$ vanishes at $\pm p_F$, where the Fermi wavevector is determined by the condition $\mu+2t\cos p_Fa=0$. The pairing term, on the other hand, vanishes at $p=0$ and $p=\pm\pi/a$, which is a direct consequence of its $p$-wave nature. Thus the system is gapless only when the Fermi wavevector equals $0$ or $p=\pm\pi/a$. This happens when the chemical potential is at the edges of the normal state dispersion, namely $\mu=-2t$ (for $p_F=0$) or $\mu=2t$ (for $p_F=\pm\pi/a$. The lines $\mu=\pm 2t$ define phase boundaries corresponding to two distinct topological phases. These phases are distinguished by the presence or absence of unpaired
MZMs at the ends in the geometry with open boundary conditions (bulk-boundary correspondence). These phases are characterised by a $\mathbb{Z}_2$ topological invariant, the Majorana number $M=(-1)^{\nu}$, where $\nu$ represents the number of pairs of Fermi points. The topological superconducting phase occurs for an odd number of pairs of Fermi points, namely $M=-1$, while an even number corresponds to the trivial one. Owing to the bulk-boundary correspondence, one thus expects unpaired MZMs for the open chain when $M=-1$. Indeed, the special point $\mu=0$ and $\Delta=t$ discussed above for the open chain is well within the topological phase.

In general, for a small but non-zero $\mu$, the Majorana bound states are not really localised at the ends of the wire, but their wave-functions exhibit an exponential decay into the bulk of the wire. The non-zero spatial overlap of the two Majorana wave-functions results in a non-zero energy splitting between the two Majorana states. Of course that for long wire's lengths, the splitting can be so small that the two Majorana states can be considered to be degenerate. 
Moreover, the Majoranas can also split when the higher-energy states in the bulk
come very close to zero energy, hence the Majorana modes are protected as long as the bulk energy gap is finite. 
This follows from the particle-hole symmetry involved in the problem, where the spectrum has to be symmetric around zero energy. Therefore, trying to move the Majorana zero modes 
from zero energy individually is impossible, as it would violate particle-hole symmetry.
\subsection{Non-Abelian braiding of Majorana zero modes  \label{braiding}}
Arguably, the most fascinating property of Majorana zero modes is their non-Abelian quantum statistics which has no counterpart in standard particle physics. Quantum mechanics dictates that, in three dimensions, particles either obey Fermi-Dirac or Bosonics statistics. This simply means that the wave function of a system of $N$ indistinguishable particles is necessarily symmetric (bosons) or antisymmetric (fermions) upon particle exchange i.e. $|\Psi_1,\Psi_2,...,\Psi_N\rangle=\pm|\Psi_2,\Psi_1,...,\Psi_N\rangle$. From this point of view, one could say that fermions and bosons are boring since exchanging them leaves the ground state invariant, up to a sign. Two dimensions are more interesting. Now, there are more possibilities that go beyond the fermionic or bosonic cases since it is possible to have \emph{anyon} statistics where the wave functions can pick up arbitrary phases under exchange, 
$|\Psi_1,\Psi_2,...,\Psi_N\rangle=e^{i\theta}|\Psi_2,\Psi_1,...,\Psi_N\rangle$, which generalizes the boson/fermion cases ($\theta=0,\pi$). The real fun occurs in systems with a \emph{degenerate} many-body ground state. In such case, particle exchange leads also to an \emph{state exchange}, where the system goes from one ground state to another. This form of statistics is dubbed \emph{non-Abelian}  \cite{Moore-Read,Read-Green1} since the unitary transformation $U_{AB}$ that operates in the subspace of degenerate ground states, $|\Psi_1,\Psi_2,...,\Psi_N\rangle_B=U_{AB}|\Psi_2,\Psi_1,...,\Psi_N\rangle_A$, is generally a \emph{noncommuting} matrix. Thus, the final state of the system after exchanging particles depends on the order of the exchange operations.
\begin{figure}[!tt]
\centering
\includegraphics[width=1\columnwidth]{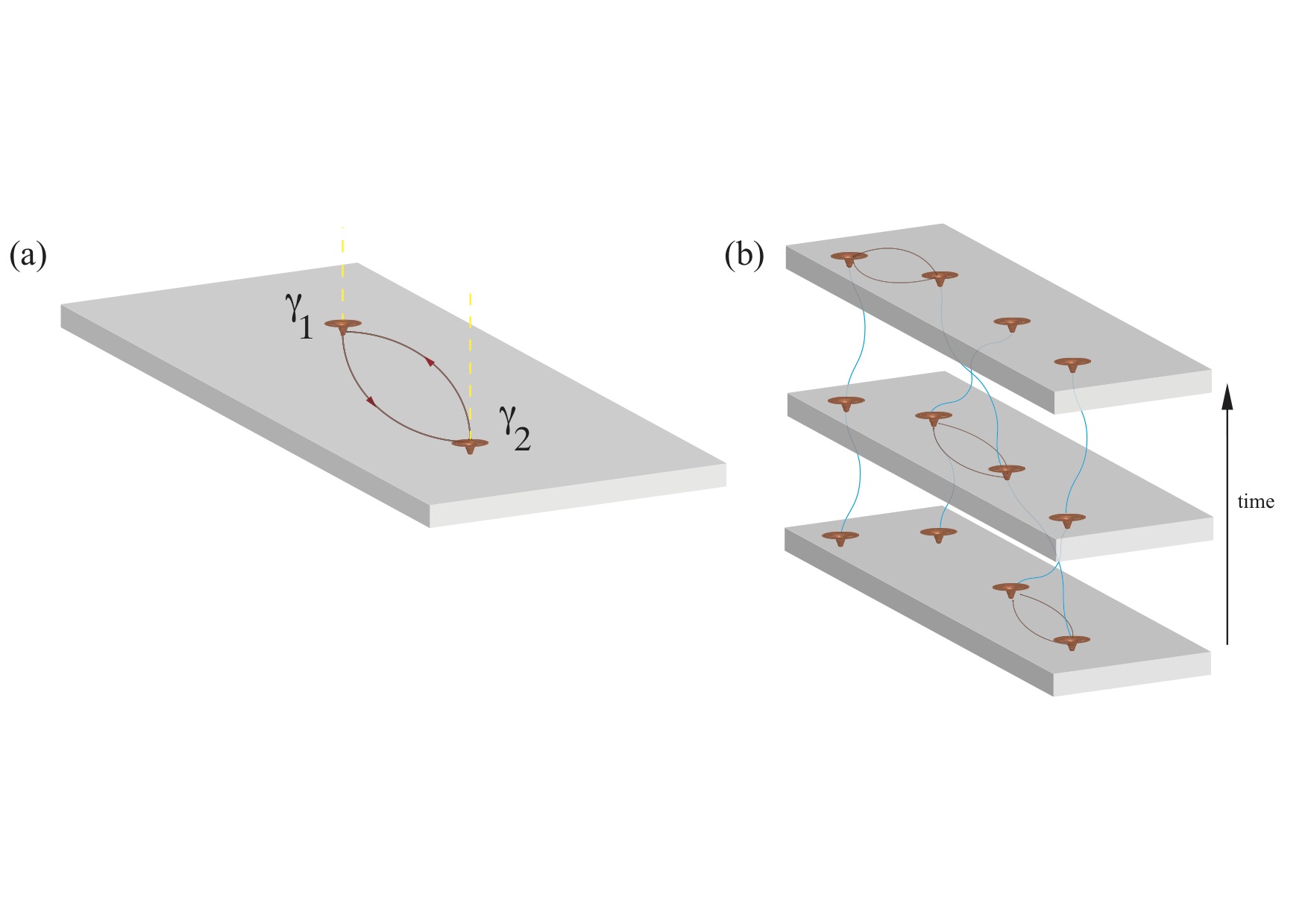}\\
\caption{Non-Abelian statistics. a) Cartoon of a two-dimensional $p$-wave superconductor with vortices hosting Majorana zero modes described by $\gamma_1$ and $\gamma_2$. Inside the vortex core the superconducting gap vanishes, and encircling the vortex entails to a jump phase by $2\pi$. This is represented by branch cuts emanating from the vortex cores (yellow dashed lines), such that the superconducting phase is single valued away from the branch cut and jumps by $2\pi$ across the branch cut. If we exchange the vortices, one of them necessarily crosses the branch cut of the other which, for the clockwise exchange shown in the example, implies $\gamma_1\rightarrow -\gamma_2$ and $\gamma_2\rightarrow \gamma_1$. b) The final state of the system after exchanging several pairs of Majorana zero modes (timeline in the figure) depends on the order of the exchange operations much like braiding cords in a necklace, hence the name "braiding".}
\label{fig:3}
\end{figure}

As we shall discuss now, this exotic form of statistics can be accomplished by exchanging Majorana zero modes. To understand how this works, it is crucial to recall the concept of fermion parity that we briefly mentioned in the previous subsection when discussing the Kitaev model. In particular, let us write a fermion operator in terms of two Majorana operators
\begin{equation}
\label{fermion-from-Majorana}
f=\frac{1}{2}\Big(\gamma_{1}+i\gamma_{2}\Big)\,,\quad f^{\dagger}=\frac{1}{2}\Big(\gamma_{1}-i\gamma_{2}\Big)\,,
\end{equation} 
and the corresponding parity operator
\begin{equation}
\label{parityoperator}
P\equiv 1-2\hat{n}=1-2f^\dagger f=-i\gamma_1\gamma_2,
\end{equation}
whose eigenvalues $\pm 1$ refer to even ($n=0$) and odd ($n=1$) fermionic parity, respectively. If the fermion operator in Eq. (\ref{fermion-from-Majorana}) is constructed from non-overlapping Majoranas (namely, true zero modes), the ground state is twofold degenerate since both fermion parities correspond to zero energy. In general, the number state corresponding to $N$ of such fermions $|n_1,n_2,...,n_N\rangle$ is $2^N$-fold degenerate, corresponding to each $n_i$ being zero or one. This ground state degeneracy is precisely the one that gives rise to non-Abelian statistics under exchange of Majoranas. 

In order to illustrate the above idea, let us follow Ivanov's reasoning \cite{Ivanov} which is particularly elegant yet simple. The starting point is to think about vortices in a superconductor and, in particular, to consider the topological case where each vortex traps a Majorana zero mode \footnote{Intuitively, this can be understood by considering the possible energies of the subgap states which are trapped by the vortex owing to the suppression of the superconducting gap inside the vortex core, $E_n\sim n\Delta^2/E_F$. For the chiral $p$-wave superconductor, the quantum number $n$ takes integer values, and thus zero modes can be trapped, as compared to standard Caroli de Gennes Matricon states \cite{Caroli} in s-wave superconductors where $n$ is half-integer. This difference can be traced back to the zero-point energy which is zero for massless Dirac fermions, owing to the cancellation of the zero-point energy by a Berry phase.  For a full discussion see \cite{Jackiw-Rossi}.}. Encircling a vortex necessarily implies a jump  in the superconducting phase by $2\pi$, which is taken into account by branch cuts associated to each vortex (Fig.\ref{fig:3}a). If we now exchange two vortices in a clockwise manner, one of them crosses a branch cut and acquires a  $2\pi$ phase shift, while the other does not. This is translated into the following transformation rule for the associated Majorana operators  ($\gamma_1$ and $\gamma_2$ in  Fig.\ref{fig:3}a) \footnote{The above transformation rule can be easily understood since a change of phase $\phi$ in the superconductor translates into an phase shift of $\phi/2$ in the corresponding fermion operators $f\rightarrow e^{i\phi/2}f$ and $f^\dagger\rightarrow e^{-i\phi/2}f^\dagger$, which yields $\gamma=(f+f^\dagger)\rightarrow -(f+f^\dagger)$ when $\phi=2\pi$.}:
\begin{equation}
\label{braiding1}
\gamma_1\rightarrow-\gamma_2\,,\quad \gamma_2\rightarrow\gamma_1.
\end{equation}
The unitary operator that implements this transformation reads 
\begin{equation}
\label{braiding2}
U_{12}=e^{-i\frac{\pi}{4}\gamma_1\gamma_2}=(1+\gamma_1\gamma_2)/\sqrt{2},
\end{equation}
which, up to a phase, is the braiding transformation of Ising anyons first discussed in the context of the $5/2$ quantum Hall state \cite{Nayak:RMP08}.

In order to understand the physical consequences of the process of exchanging Majoranas, let us consider its effects on fermionic number states. The simplest case is to consider a situation with only two Majoranas which define a single fermion $f=(\gamma_{1}+i\gamma_{2})/2$. As we discussed, there are two possible degenerate ground states $|0\rangle$ and $|1\rangle=f^\dagger|0\rangle$ depending on the occupation $\hat{n}=f^\dagger f$ of this single fermion. The effect of Majorana exchange on these states is:
\begin{equation}
\label{braiding3}
U_{12}|n\rangle=e^{\frac{\pi}{4}(1-2n)}|n\rangle.
\end{equation}
Thus, exchanging two Majorana zero modes only generates occupancy dependent phase factors. Physically, this can be understood since fermionic parity is conserved and cannot change upon Majorana exchanges. 
The simplest case where one can fully appreciate the profound consequences of Majorana exchange is the case of two pairs of Majorana modes. As before, we consider the effect that Majorana exchange has on number states which, in this case, are defined in terms of the two possible fermions $f_1=(\gamma_{1}+i\gamma_{2})/2$ and $f_2=(\gamma_{3}+i\gamma_{4})/2$. While exchanges between Majoranas belonging to the same fermion (namely $U_{12}$ and $U_{34}$) only leads to overall phase factors as before, 
\begin{eqnarray}
\label{braiding4}
U_{12}|n_1,n_2\rangle=e^{\frac{\pi}{4}(1-2n_1)}|n_1,n_2\rangle\nonumber\\
U_{34}|n_1,n_2\rangle=e^{\frac{\pi}{4}(1-2n_2)}|n_1,n_2\rangle,
\end{eqnarray}
the exchange of two Majoranas belonging to different fermions (like e.g. $U_{23}$) is more interesting and generates superposition of \emph{different} number states \cite{Ivanov},
\begin{eqnarray}
\label{braiding5}
U_{23}|n_1,n_2\rangle=\frac{1}{\sqrt{2}}\big[|n_1,n_2\rangle+i(-1)^{n_1}|1-n_1,1-n_2\rangle\big].
\end{eqnarray}
As expected, it is the total parity $n_1+n_2$ which is conserved in this case, since superpositions such as $|0,0\rangle+i|1,1\rangle$ (or $|1,0\rangle-i|0,1\rangle$) involve states of the same parity.

The non-Abelian nature of Majorana exchange is explicit since the above operators do not commute when the same Majorana operator is involved in the exchanges 
\begin{equation}
[U_{i-1,i}U_{i,i+1}]=i\gamma_{i-1,i}\gamma_{i,i+1}. 
\end{equation}
Thus, the final state of the system after exchanging several pairs of Majorana zero modes depends on the order of the exchange operations, much like braiding cords in a necklace, hence the name non-Abelian "braiding". This idea is pictorially represented in Fig. \ref{fig:3}b.

Since the fermion states above store quantum information non-locally (they are defined in terms of Majoranas that are far apart and, therefore, cannot be measured by local noise operators), non-Abelian braiding provides an attractive platform for fault-tolerant quantum computation. Unfortunately, braiding of Majoranas is not enough for universal quantum computing. This can be easily seen by representing a full spin with three Majorana operators as \cite{Tsvelik,Shnirman,Mao}
\begin{eqnarray}
&&\sigma_x=-i\gamma_2\gamma_3\nonumber\\
&&\sigma_y=-i\gamma_1\gamma_3\nonumber\\
&&\sigma_z=-i\gamma_1\gamma_2, 
\end{eqnarray}
such that the Hilbert space spanned by four Majoranas is that of two decoupled qubits (corresponding to even and odd parity, i.e either $|\uparrow\rangle\equiv|11\rangle$, $|\downarrow\rangle\equiv|00\rangle$ or $|\uparrow\rangle\equiv|10\rangle$, $|\downarrow\rangle\equiv|01\rangle$). In this qubit representation, the braid operations in Eqs. (\ref{braiding4}) and (\ref{braiding5}) can be written as 
\begin{eqnarray}
&&U_{12}=U_{34}=e^{-i\frac{\pi}{4}\sigma_z}\nonumber\\
&&U_{23}=e^{-i\frac{\pi}{4}\sigma_x},
\end{eqnarray}
which means that braiding Majoranas is equivalent to performing single-qubit rotations in the fermionic parity basis by an angle $\pi/2$. Thus, by braiding Majoranas it is not possible to generate arbitrary single qubit rotations. Furthermore, it is not possible to create two qubit gate operations and entanglement. In principle, one can supplement braiding with other quantum gates in order to obtain a universal set of operations but, unfortunately, these gates are not topologically protected.  Another option is to couple Majorana-based qubits to standard qubits such as superconducting qubits \cite{Hassler} or spin-qubits \cite{Flensberg-spinqubit}. A full discussion about these issues and possible ways to overcome these limitations are discussed in the excellent review by Das Sarma, Freedman and Nayak \cite{DasSarma-review} which focuses on topological quantum computation with Majorana zero modes. 

We finish this part by mentioning that another important obstacle towards Majorana-based quantum computation is the process of braiding itself. Indeed, it is not obvious how to implement Ivanov's scheme in a realistic device where physical quasiparticle excitations must be braided and manipulated (although some clever ideas already exist such as the one based on networks of quantum wires \cite{Alicea-braiding}). Further in the review, we shall briefly discuss a very recent idea which replaces the braiding operation by a parity measurement protocol \cite{Karzig2016,Plugge2017}. A simpler way to test the ground state degeneracy in the topological phase is to perform so-called "fusion rules" experiments which rely on fusing Majoranas in various sequences such that different intermediate fermionic charge configurations result in different final charge configurations  \cite{Alicea-braiding,Aasen2016}.

\section{Physical implementations \label{implementations}}
Despite their conceptual importance, the above early ideas for obtaining Majoranas out of chiral $p$-wave superconductors \cite{Senthil-Fisher,Kopnin-Salomaa,Moore-Read,Volovik1,Read-Green1,Mackenzie,Kitaev1,Kitaev2,Volovik2,DasSarma1} received little attention from the experimental community, at least as compared to the great deal of experimental activity that novel proposals have launched \cite{Wilzeck,Franz,Beenakker-Kouwenhoven}.  There may well be many reasons for this but, probably, that main one is that $p$-wave pairing is very rare in nature. Indeed, there are only a few cases where $p$-wave superconductivity can appear intrinsically in a material, with $Sr_2RuO_4$ and the $\nu=5/2$ quantum Hall state being the two paradigmatic examples. While there are various experiments that seem to be consistent with spin-triplet Cooper pairing in $Sr_2RuO_4$, no unambiguous proof of Majorana physics (to the best of our knowledge) has been reported. Part of the problem is probably due to the coexistence of both $p+ip$ and $p-ip$ pairings in typical crystals, which complicates the interpretation. Similarly, there are no unambiguous experimental evidence of $p$-wave pairing using the $\nu=5/2$ quantum Hall state. \footnote{For a recent review on chiral superconductors see Ref. \cite{Kallin-Berlinsky}.}. It took a decade until this state of affairs changed after various ingenious theoretical proposals demonstrated how to 'cook' $p$-wave superconducting correlations starting from conventional 
$s$-wave superconductors. As we shall see in this section, we only need three main ingredients for the clever recipe: the superconducting proximity effect, spin-orbit coupling and time-reversal symmetry breaking \footnote{In what follows, we just discuss the Fu-Kane and Lutchyn-Oreg models which are, arguably, the most influential theoretical proposals to engineer topological superconductivity in one dimension. The list of alternatives is, by no means, restricted to these two platforms and we can mention in passing other options, not discussed in this review, such as carbon nanotubes \cite{nanotube1,nanotube2,nanotube3,nanotube4},  Ge/Si hole nanowires \cite{GeSiwires} or topological insulator nanowires \cite{TIwires}.}.
\subsection{The Fu and Kane model \label{Fu-Kane}}
In 2008, Liang Fu and Charlie Kane published their highly influential paper \cite{Fu-Kane1} where they showed that one can engineer $p$-wave superconductivity out of $s$-wave superconductors by virtue of the superconducting proximity effect with the helical surface of a three dimensional topological insulator. 
\begin{figure}[!tt]
\centering
\includegraphics[width=1\columnwidth]{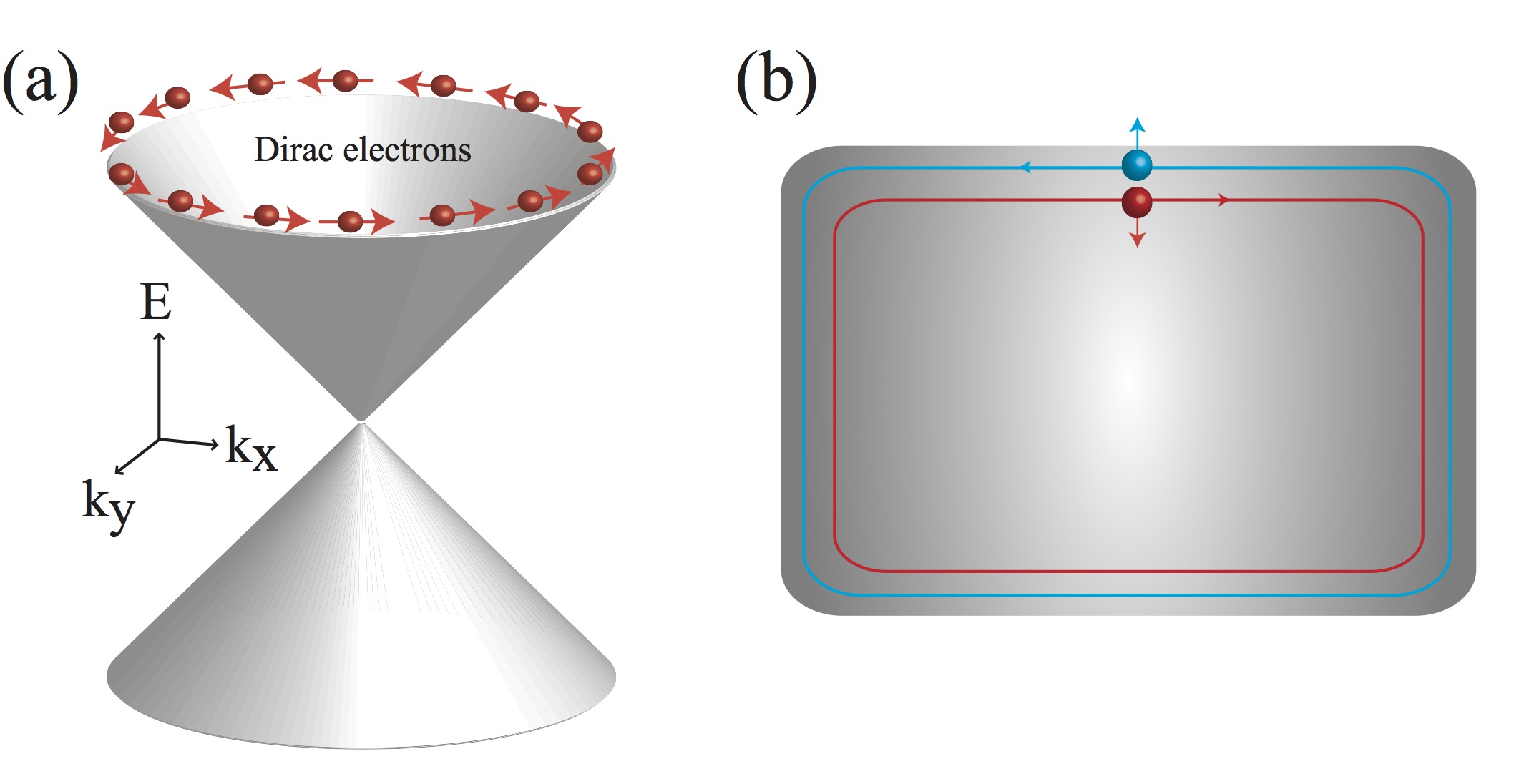}\\
\caption{a) The surface states of a 3D topological insulator can be described as a single Dirac cone. While encircling the Dirac cone, the electron spins wind by 2$\pi$ which converts conventional
$s$-wave pairing into $p$-wave, as described by Eq. (\ref{BCS-pwave}). b) The edge states of a two-dimensional topological insulator are counter-propagating helical channels. As long as time-reversal symmetry is preserved, backscattering between these Kramer's partners is supressed, thus giving rise to a symmetry-protected topological phase with quantized conductance (quantum spin Hall effect).}
\label{fig:4}
\end{figure}
In order to illustrate the idea, let us follow the original argument by Fu and Kane and consider the low-energy Dirac Hamiltonian describing the surface of a topological insulator:
\begin{equation}
H_0 = \int  d{\bf r}^2 r \Psi^\dagger [-iv_F(\partial_x\sigma^y-\partial_y\sigma^x)-\mu]\Psi,
\label{eq:Dirac-cone}
\end{equation}
where $v_F$ is the Dirac cone velocity, $\mu$ the chemical potential and $\sigma$ the Pauli matrices in spin space. Eq. (\ref{eq:Dirac-cone}) describes surface eigenstates with energies 
$\epsilon_\pm ({\bf k})=\pm v_F|{\bf k}|-\mu$ which are the upper and lower branches of a single Dirac cone with definite spin-helicity. (Fig.\ref{fig:4}a). Their key observation is that standard singlet $s$-wave pairing effectively behaves as $p$-wave when projected onto the basis of helical electrons.  This can be seen by analysing the consequences that a standard BCS pairing term of the form
\begin{equation}
\label{BCS}
H_S=\int d{\bf r}^2\Delta (\psi^\dagger_{\uparrow}\psi^\dagger_{\downarrow}+H.c)
\end{equation}
has on the above helical bands. As expected, the spectrum is now gapped, $E_\pm({\bf k})=\sqrt{\epsilon^2_\pm ({\bf k})+\Delta^2}$, but the nature of the gapped states is by no means trivial. Their physical meaning becomes clear when we rewrite Eq. (\ref{BCS}) in the helical basis. Indeed, when expressed in this new basis, Eq. (\ref{BCS}) contains $p$-wave pairing terms of the form:
\begin{equation}
\label{BCS-pwave}
H_S= \int\frac{d^2{\bf k}}{(2\pi)^2}\frac{\Delta}{2}\large(\frac{k_x+ik_y}{|{\bf k}|}\large) \left[\psi^\dagger_+({\bf k})\psi^\dagger_+({\bf -k})+\psi^\dagger_-({\bf k})\psi^\dagger_-({\bf -k})+H. c.\right]
\end{equation}
with $\psi_{\pm}({\bf k})= \frac{\psi_{{\bf k}, \uparrow}\pm e^{-i\phi_k}\psi_{{\bf k}, \downarrow}}{\sqrt{2}}$ and $ \phi_k = \tan^{-1}(k_x/k_y)$.
Physically, Eq. (\ref{BCS-pwave}) describes \emph{intraband} pairing which is possible owing to the helical nature of the electrons (Fig.\ref{fig:4}a). 

If one of the two helical sectors can be projected out (like, for instance doping with electrons the Dirac cone such that $\psi_{-}\rightarrow 0$) this system is a particular realization of the weak pairing phase of the Read and Green's chiral $p_x+ip_y$ superconductor discussed in Section \ref{pwave}. Therefore, as long as the bulk does not contribute, the surface of the proximitized 3D topological insulator forms a 2D topological superconductor. Similar to the Read and Green's model, this system hosts chiral Majorana edge states at boundaries between the topological superconducting surface and a trivial magnetic insulator, which breaks time-reversal symmetry \footnote{Note the subtle difference between the boundary of Read and Green's chiral $p_x+ip_y$ superconductor (which breaks time-reversal symmetry) and vacuum (which does not) and Fu and Kane's one, where time reversal symmetry is broken at the trivial side}. Moreover, a $h/2e$ vortex binds Majorana zero modes \cite{Fu-Kane1}.

Similar ideas can be applied to proximitized two dimensional topological insulators \cite{Fu-Kane2}. If the chemical potential is within the bulk gap of the topological insulator, the only relevant electronic degrees of freedom are a pair of one-dimensional spin-filtered counterpropagating edge  states that exhibit perfect spin-momentum locking. These low-energy modes can be described by a one-dimensional Dirac equation of the form:
\begin{equation}
H_0 = \int dx \psi^\dagger(x) [-iv_F(\partial_x\sigma^x-\mu)]\psi(x).
\label{eq:Dirac-cone-1D}
\end{equation}
Since at the Fermi energy there is just one single propagating spin channel in each direction\footnote{Due to Kramers' theorem, these helical edge states are protected from backscattering as long as time-reversal symmetry is preserved, which gives rise to the quantum spin Hall effect.}, one expects that these channels provide a good platform to engineer a one-dimensional topological superconducting phase by virtue of the proximity effect. However, these edges constitute the boundary of a two-dimensional system, with no end, such that the expected Majoranas in the topological phase cannot be localized. In order to localize them, one must include boundaries where time-reversal symmetry is broken \cite{Fu-Kane2}.
In order to gain intuition of how this happens, let us consider the two different ways of gapping out the spectrum in Eq. (\ref{eq:Dirac-cone-1D}): either through an $s$-wave pairing superconductor $H_S=\int dx \Delta (\psi^\dagger_{\uparrow}\psi^\dagger_{\downarrow}+H.c)$, or by applying a Zeeman field in a direction perpendicular to the spin quantization axis of the edge states (chosen along the x direction in Eq. (\ref{eq:Dirac-cone-1D})) $H_Z=-E_Z \int dx \psi^\dagger(x) \sigma^z\psi(x)$. Further insight about the nature of the proximitized edges can be gained by considering the Hamiltonian of the combined system:
\begin{eqnarray}
\label{BCS-pwave-1D}
H&=& \int\frac{d k}{(2\pi)}\{\epsilon_+ (k)\psi^\dagger_+(k)\psi_+(k)+\epsilon_- (k)\psi^\dagger_-(k)\psi_-(k)\}\nonumber\\
&+&\frac{\Delta_p}{2}\{ \psi^\dagger_+(k)\psi^\dagger_+(-k)+\psi^\dagger_-(k)\psi^\dagger_-(-k)+H. c.\}\nonumber\\
&+&\Delta_s\{ \psi^\dagger_+(k)\psi^\dagger_-(-k)+\psi_-(-k)\psi_+(k)\},
\end{eqnarray}
with $\epsilon_\pm (k)=-\mu\pm \sqrt{(v_Fk)^2+E^2_Z}$, that explicitly contains intraband $p$-wave pairing terms (second line)  coexisting with interband $s$-wave pairing (third line) of the form: 
\begin{eqnarray}
\label{p-wave-gaps}
\Delta_p&=&\frac{v_Fk\Delta}{\sqrt{(v_Fk)^2+E^2_Z}}\nonumber\\
\Delta_s&=&\frac{E_Z\Delta}{\sqrt{(v_Fk)^2+E^2_Z}}
\end{eqnarray}
\begin{figure}[!tt]
\centering
\includegraphics[width=1\columnwidth]{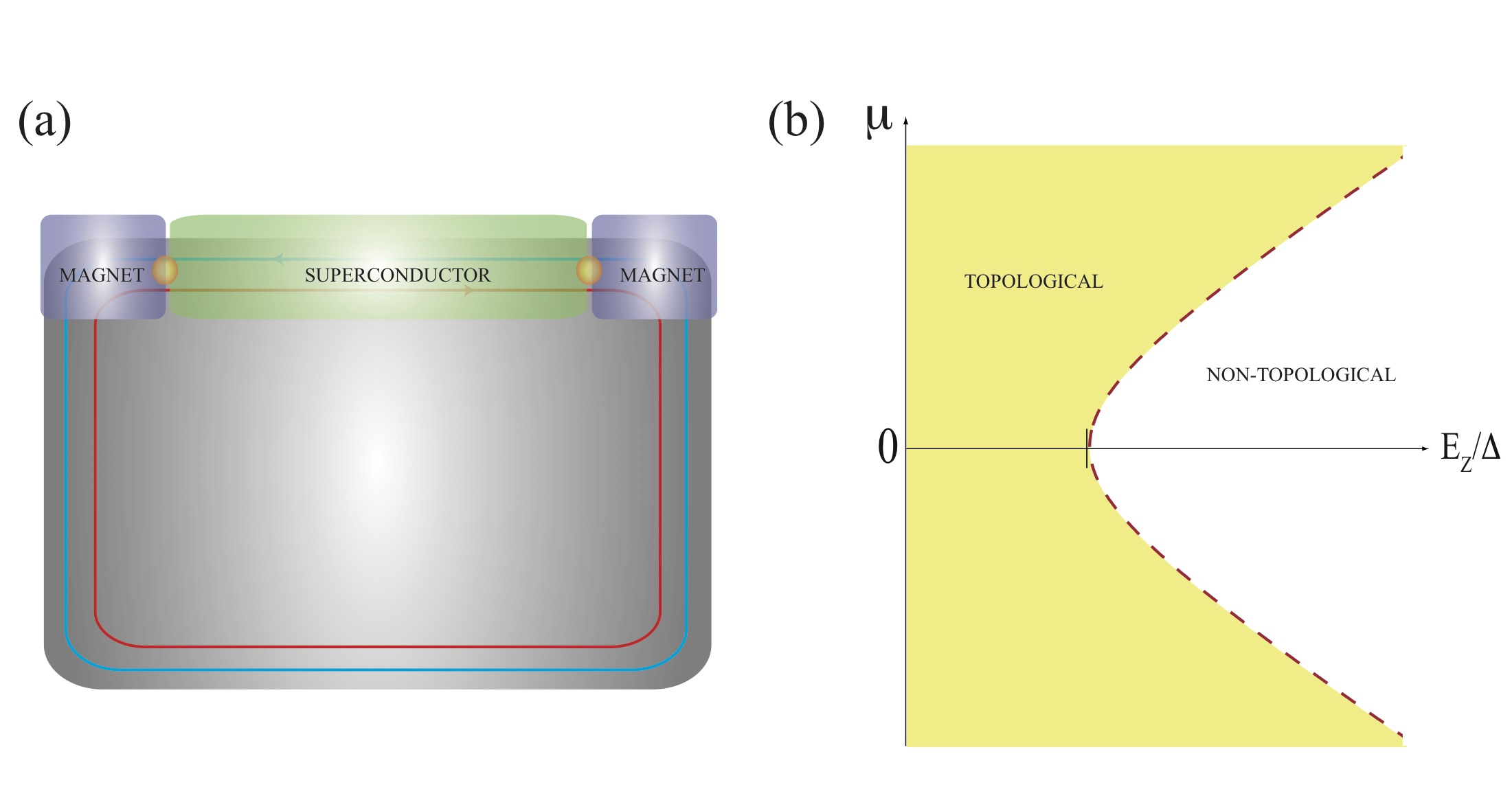}\\
\caption{a) The edge of a two-dimensional topological insulator becomes a one-dimensional topological superconductor when proximitized by a superconductor (green). This topological superconductor supports Majorana bound states (yellow circles) at boundaries which break time reversal symmetry, like, e.g., at the boundaries with a ferromagnetic insulator (purple). b) Phase diagram, the proximitized edge is a one-dimensional topological superconductor when $E_Z<\sqrt{\mu^2+\Delta^2}$.}
\label{fig:5}
\end{figure}
Once we have shown that this combined system effectively contains $p$-wave terms, the last step is to demonstrate that this system undergoes a topological transition. For simplicity, let us consider  $\mu=0$. In this case the spectrum of the combined system reads $E(k)=\pm \sqrt{(v_Fk)^2+(\Delta\pm E_Z)^2}$, which is just two copies of a massive one-dimensional Dirac spectrum with masses $\Delta\pm E_Z$. As we showed in section \ref{pwave}, one-dimensional Dirac Hamiltonians contain zero energy bound states at interfaces where the mass term changes sign \cite{Jackiw-Rebbi}, namely at points where the system undergoes a topological transition through a gap closing when $E_Z=\pm\Delta$. In order to see how zero modes emerge, let us consider spatially-varying mass profiles of the form $\Delta(x)$ and $E_Z(x)$ and seek for zero energy eigenstates the BdG hamiltonian that describes the above system:
\begin{equation}
\label{zero-mode-Fu-Kane1}
H_{BdG}(x)= -iv_F\partial_x\sigma^x\tau^z+E_Z(x)\sigma^z+\Delta(x)\tau^x.
\end{equation}
Solving $H_{BdG}(x)\Phi_0(x)=0$, we find 

\begin{eqnarray}
\Phi_0(x)=\frac{\phi(x)}{\sqrt{2}}\left( \begin{array}{ccc}
0\\
1\\  -1 \\
0\end{array} \right)
\end{eqnarray}
with an envelope function 
$\phi(x)\sim e^{-\frac{1}{v}\int_0^xdx'[E_Z(x')-\Delta(x')]}$ that describes the boundary between the mass terms. Using the Nambu representation in Eq. (\ref{Nambu-spinor}), we obtain an operator
\begin{equation}
\label{zero-mode-Fu-Kane2}
\gamma=\frac{1}{\sqrt{2}}\int dx\phi(x)[c_\downarrow(x)+c^\dagger_\downarrow(x)]
\end{equation}
which is indeed a Majorana operator with $\gamma=\gamma^\dagger$. As we anticipated, the interface between the two mass terms in Eq. (\ref{zero-mode-Fu-Kane1}) is a natural host for Majorana zero modes, thus providing a physical platform to generate the Jackiw-Rebbi's zero modes that we obtained in Eq. (\ref{zero-mode-chiralRead-Green}) for the chiral $p$-wave superconductor.

For the more general situation $\mu\neq0$, it is instructive to make connection with the Kitaev's model. In the $E_Z>>\Delta$ limit and with $\mu$ close to the bottom of the upper band, one may project out the $\psi_-$ band in Eq. (\ref{BCS-pwave-1D}) and perform a low momentum expansion which gives an effective Hamiltonian in real space:
\begin{eqnarray}
\label{BCS-pwave-1D-lowmomentum}
H= \int dx\psi^\dagger_+(x) [-\frac{\partial^2_x}{2m_{eff}}-\mu_{eff} ]\psi_+(x)-\frac{\Delta_{eff}}{2}[ \psi^\dagger_+(x)i\partial_x\psi^\dagger_+(x)+H. c.],
\end{eqnarray}
with $m_{eff}=v_F^2/E_Z$, $\mu_{eff}=\mu-E_Z$ and $\Delta_{eff}=v_F\Delta/E_Z$. Interestingly, the effective model described by Eq. (\ref{BCS-pwave-1D-lowmomentum}) is equivalent to the low-energy limit of Kitaev's model (i. e. the low-density limit, near $\mu=-t$, of Eq. (\ref{kiatev0})). A similar mapping holds for $\mu$ near the top of the lower band, for a full discussion see e. g. Ref. \cite{Alicea-review}. Having established the connection with the Kitaev model for $E_Z>>\Delta$, we can conclude that the proximitized edge of a 2D topological insulator will host a topologically trivial (strong pairing) phase when $|\mu|\lesssim E_Z$  and a topologically nontrivial (weak pairing) phase when $|\mu|\gtrsim E_Z$. The full phase diagram, valid at any $\mu$, $\Delta$ and $E_Z$, can be obtained by nullifying the quasiparticle gap extracted from Eq. (\ref{BCS-pwave-1D})
\begin{equation}
\label{closing1}
E_\pm(k)=\pm\sqrt{\Delta_p^2+\Delta_s^2+\frac{\epsilon^2_+ (k)+\epsilon^2_- (k)}{2}\pm (\epsilon_+ (k)-\epsilon_- (k))\sqrt{\Delta^2_s+\mu^2}},
\end{equation}
which vanishes for $E_Z^2=\Delta^2+\mu^2$. By considering the above analysis for the $E_Z>>\Delta$ case, we can conclude that the proximitized edge is a one-dimensional topological superconductor when $E_Z<\sqrt{\mu^2+\Delta^2}$ (Fig. \ref{fig:5}).
\subsection{The Rashba semiconductor model \label{Rashba}} The above ideas of engineering $p$-wave superconductivity by proximitizing helical electrons are by no means restricted to topological insulators. In 2010, four seminal papers demonstrated that a promising strategy can be envisioned by proximitizing semiconductors with strong spin-orbit coupling such as InAs or InSb. Sau et al \cite{Sau2010} and Alicea \cite{Alicea2010} focused on semiconductor heterostructures, while Lutchyn et al 
\cite{Lutchyn2010} and Oreg et al \cite{Oreg2010} demonstrated that the idea can be simplified even further by using one-dimensional semiconducting nanowires instead of heterostructures. Less than two years after the conceptual breakthrough by Fu and Kane \cite{Fu-Kane1}, these semiconductor proposals helped to convince the remaining skepticals: if certain amounts of disbelief still lingered in the community about experimentally engineering $p$-wave superconductivity, this last set of remarkable theory predictions surely eliminated them. 

In order to illustrate the main ideas behind the semiconductor platforms, let us start with the Hamiltonian describing a semiconductor 2DEG with Rashba spin-orbit coupling: 
\begin{equation}
\label{Rashba2D}
H_{0}=\int d{\bf r}^2\Psi^\dagger\left[-\frac{\nabla^2}{2m}-\mu-i\alpha(\partial_x\sigma^y-\partial_y\sigma^x)\right]\Psi
\end{equation} 
where $m$ is the electron's effective mass, $\mu$ the chemical potential and $\alpha$ is the strength of the spin-orbit interaction, which aligns the electron spins within the plane of the 2DEG and perpendicular to their momentum. 
Although this Hamiltonian superficially resembles the one describing the Dirac cone of a three-dimensional topological insulator, compare the spin-orbit-coupling term in Eq. (\ref{Rashba2D}) with Eq. (\ref{eq:Dirac-cone}), there is an important difference between both systems, since the model for a two-dimensional semiconductor contains a kinetic energy contribution which is absent in the topological insulator case. This kinetic energy contribution turns out to be very relevant since there are, in general, two spin-orbit-split Fermi surfaces which prevent the system to behave as "spinless" upon superconducting pairing. This difficulty can, however, be overcome by the inclusion of a Zeeman term that competes with the Rashba coupling and tries to align the spins perpendicular to the plane (i. e. along the $z$ direction):
\begin{equation}
H_Z=E_Z\int d{\bf r}^2\Psi^\dagger\sigma^z\Psi.
\end{equation} 
Importantly, this Zeeman term opens up a gap that separates the two spin-orbit-split bands such that for chemical potentials $\mu<E_Z$ the system has a single Fermi surface. In this regime, the system behaves as a topological superconductor when proximitized with an $s$-wave superconductor. In this topological superconducting phase, the edge of this two-dimensional hybrid semiconductor-superconductor system supports chiral Majorana modes, since time-reversal symmetry is explicitly broken by the Zeeman term, and with vortices that support Majorana zero modes \cite{Sau2010},  in close analogy with the topological insulator case.

In order to understand how topological superconductivity comes about in these semiconducting systems, let us focus from now on in the one dimensional nanowire case \cite{Lutchyn2010,Oreg2010}, which can be understood in a very intuitive fashion. Our starting point is the Hamiltonian of a one dimensional semiconducting nanowire with  Rashba spin-orbit interaction and in the presence of a Zeeman field perpendicular to the Rashba field (here, we assume that the magnetic field is applied parallel to the nanowire axis):
\begin{equation}
\label{Rashba1D}
H=\int dx \Psi^\dagger\left[-\frac{\hbar^2\partial^2_x}{2m}-\mu-i\alpha\partial_x\sigma^y+E_Z\sigma^x\right]\Psi,
\end{equation} 
with eigenvalues $\varepsilon_{k,\pm}=\frac{\hbar^{2}k^{2}}{2m}-\mu\pm\sqrt{E^2_Z+\alpha^{2}k^{2}}$.

\begin{figure}[!tt]
\centering
\includegraphics[width=1\columnwidth]{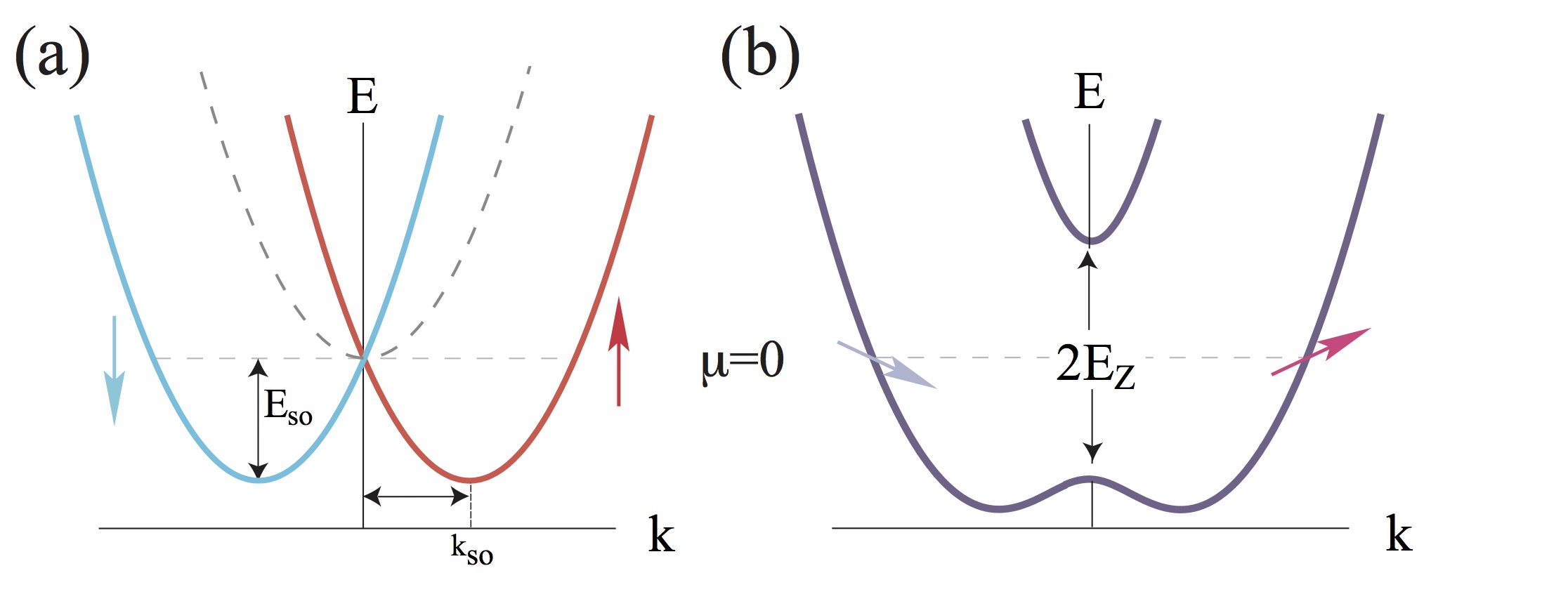}\\
\caption{Electronic bands of a one-dimensional nanowire with Rashba spin-orbit coupling. a) The Rashba coupling removes the spin degeneracy of the parabolic band with $\alpha=0$ (dashed parabola). As a result, each spin projection along the Rasbha axis resides in a different parabolic band. The minima of each parabola are shifted along $k$ by an amount $\pm k_{SO}$, while the energy is shifted by an amount $-E_{SO}$. This spectrum resembles a one-dimensional Dirac cone at low $k$ but deviates from it at large $k$ (band upturn for $k\gtrsim k_{SO}$) owing to the kinetic energy contribution. As a result, for any $\mu$ above the band bottom the system has two Fermi surfaces. b) A Zeeman term perpendicular to the Rashba axis opens a gap at $k=0$. When $\mu$ resides within this gap there is only a Fermi surface with spin projection locked to momentum (this regime is often defined as the helical regime). The degree of spin canting depends on $k$: spins at low momenta are almost aligned with the Zeeman field while canting towards the Rashba axis occurs for larger $k$.}
\label{fig:6}
\end{figure}

Since the Rashba and Zeeman fields are orthogonal, they compete in fixing the spin quantization axis: in the absence of Zeeman field, $E_Z=0$, the Rashba term removes the spin degeneracy of the one-dimensional parabolic band and gives rise to two parabolas shifted relative to each other along the momentum axis (each by an amount $k_{SO}=m\alpha/\hbar^2$) and displaced down in energy by an amount $E_{SO}=m\alpha^2/2\hbar^2$ (Fig. \ref{fig:6}a). These parabolas correspond to spin up and spin down projections along the spin quantization axis fixed by the Rashba coupling (here $\sigma^y$). On the other hand, a finite Zeeman $E_Z\neq0$ perpendicular to the Rashba axis mixes both spins and hence removes the spin degeneracy at $k=0$ by opening up a gap of size $2E_Z$. As a result, the degree of spin canting depends on $k$ (spins at low momenta are almost aligned with the Zeeman field while canting towards the Rashba axis occurs for larger $k$), see Fig. \ref{fig:6}b. This can be explicitly seen by writing the relation between the helical spinors, namely the ones that appear in the eigenstates that diagonalize the Hamiltonian density in Eq. (\ref{Rashba1D}), and the spinors along the Zeeman field:
\begin{equation}
\label{H0Vec}
 \phi_{\pm}(k)=\frac{1}{\sqrt{2}}\begin{pmatrix}
\pm \gamma_{k}\\
1
 \end{pmatrix}\,.
\end{equation}
with $\gamma_{k}=\frac{(i\alpha k+E_Z)}{\sqrt{E^{2}_Z+\alpha^{2}k^{2}}}$.
It turns out that this spin canting is crucial  for obtaining $p$-wave superconductivity: at $E_Z=0$ opposite momenta have opposite spins and thus can form singlets along $\sigma^y$, but at $E_Z\neq 0$ these singlet components get reduced in favour of finite triplet components along $\sigma^x$. As we did before for topological insulators, this physics is made explicit by writing an $s$-wave pairing term in the helical basis which gives:
\begin{eqnarray}
\label{BCS-pwave-Rashba}
H&=& \int\frac{d k}{(2\pi)}\frac{\Delta^p_+}{2}\{ \psi^\dagger_+(k)\psi^\dagger_+(-k)+H. c.\}+\frac{\Delta^p_-}{2}\{ \psi^\dagger_-(k)\psi^\dagger_-(-k)+H. c.\}\nonumber\\
&+&\Delta_s\{ \psi^\dagger_+(k)\psi^\dagger_-(-k)+\psi_-(-k)\psi_+(k)\},
\end{eqnarray}
with effective gaps
\begin{eqnarray}
\label{pairing-NW}
\Delta^{s}(k)&=&\frac{E_Z\Delta}{\sqrt{E_Z^2+\alpha^2k^2}}\nonumber\\
\Delta^{p}_{\mp}(k)&=&\frac{\pm i\alpha k\Delta}{2\sqrt{E_Z^2+\alpha^2k^2}}.
\end{eqnarray}
Again, as we have seen many times along the review, the projection of BCS pairing onto helical bands gives rise to both $s$-wave pairing (here the interband term $\Delta^{s}(k)$) and $p$-wave pairing  (here the intraband pairing terms 
$\Delta^{p}_{\mp}$, with $\mp$ corresponding to the lower/upper band).  Note that these pairings are essentially the same as the ones that we obtained for the two-dimensional topological insulator case, see Eq. (\ref{p-wave-gaps}), but with $\alpha$ playing the role of a Fermi velocity. Despite this similarity, both systems are not completely equivalent, as we already mentioned for the two-dimensional Rashba case \footnote{For a full discussion about the differences between topological insulators and Rashba semiconductors for engineering $p$-wave superconductivity see Ref. \cite{Potter-Lee}.} : while at low momenta the Rashba nanowire model has the same form as the two-dimensional topological insulator edge Hamiltonian, the standard kinetic energy contribution introduces important changes. In particular, note that the bands of the nanowire resemble a Dirac cone only at small $k$, but their lower half eventually bends upward at large $k$, owing to the parabolic contribution of the kinetic energy. As a result, for any chemical potential $\mu$ above the bottom of the conduction band there are always two Fermi surfaces (Fig. \ref{fig:6}a). This situation changes when $\mu$ resides within the Zeeman gap (Fig. \ref{fig:6}b) where there is only a single Fermi surface with spin-momentum locking.  (i.e. with a single left-moving mode and a single right-moving mode with opposite spin). This situation is similar to the helical channels of the two-dimensional topological insulator but it is important to note that here we need to explicitly break time-reversal symmetry in order to reach a helical state (here, in particular, a direct connection to the Kitaev model can be made in the limit $E_Z>>m\alpha^2,\Delta$ where the upper band of Fig. \ref{fig:6}b can be projected out of the problem).
\begin{figure}[!tt]
\centering
\includegraphics[width=1\columnwidth]{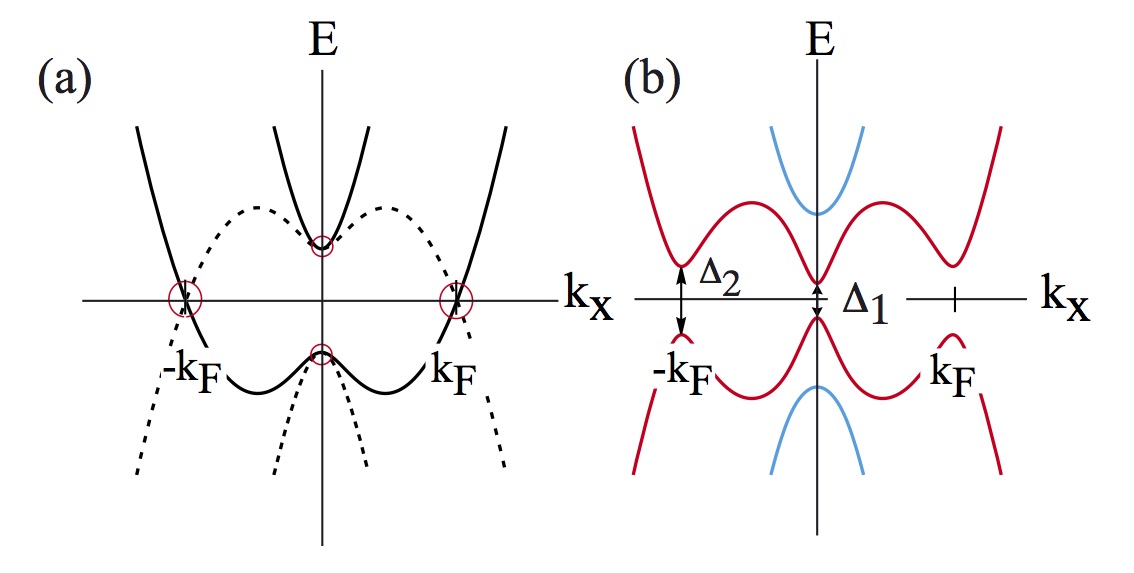}\\
\caption{Bogoliubov-de Gennes spectrum of a Rashba nanowire. a) $\Delta=0$: electron (solid lines) and hole (dashed lines) branches cross at finite energies (interband crossings) and at the Fermi energy (intraband crossings). b) A finite $\Delta\neq 0$ generates both intraband and interband pairings, Eq. (\ref{pairing-NW}), which lead to the opening of a gap $\Delta_2$ at the Fermi points $k_F=\sqrt{k^2_\mu+2k^2_{SO}+\sqrt{(k^2_\mu+2k^2_{SO})^2-k^4_\mu+k^4_Z}}$ (with $k_\mu=\sqrt{2m\mu}/\hbar$, $k_{SO}=m\alpha/\hbar^2$ and $k_Z=\sqrt{2mE_Z}/\hbar$), and modify the Zeeman gap at $k=0$. This new gap $\Delta_1$ is now determined by the competition between $E_Z$ and $\Delta$, and eventually closes and reopens for increasing $E_Z$, giving rise to a topological phase transition (see next figure).}
\label{fig:7}
\end{figure}
\begin{figure}[!tt]
\centering
\includegraphics[width=1\columnwidth]{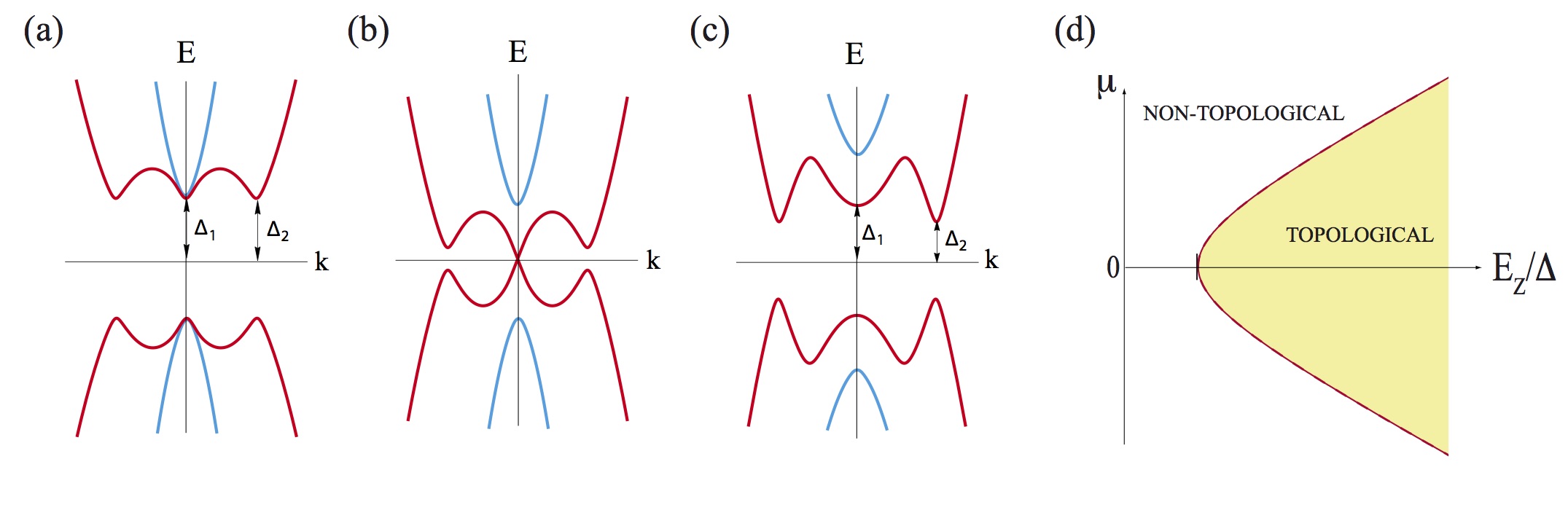}\\
\includegraphics[width=1\columnwidth]{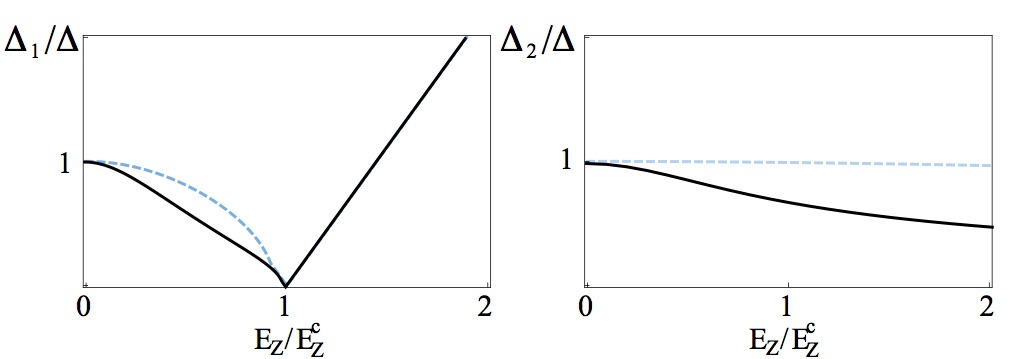}\\
\caption{Top: Bogoliubov-de Gennes spectrum of a Rashba nanowire for increasing Zeeman fields and phase diagram. a) $E_Z=0$: gaps at $k=0$ and $k=k_F$ have equal size, $\Delta_1=\Delta_2=\Delta$. The spectrum is degenerate at $k=0$ since interband pairing is zero. b) $E_Z=E^c_Z\equiv\sqrt{\Delta^2+\mu^2}$: the low momentum gap $\Delta_1$ closes signalling a topological transition. The gap $\Delta_2$ remains finite. c) For $E_Z>E^c_Z$, a Zeeman-dominated $\Delta_1$ reopens again and the system is a topological superconductor. The bands shown here correspond to $\mu=0$. For the more general case $\mu\neq 0$, $\Delta_1$ is at a small but finite $k$  in the trivial regime $E_Z<E^c_Z$, see footnote. d)  Phase diagram, the proximitized nanowire is a one-dimensional topological superconductor when $E_Z>\sqrt{\mu^2+\Delta^2}$. Bottom: gaps $\Delta_1$ and $\Delta_2$ as a function of Zeeman field for weak spin-orbit coupling (solid line, $\alpha=0.2eV\AA$) and strong spin-orbit coupling (dashed line, $\alpha=0.8eV\AA$). $\Delta_1$  closes at the critical field and reopens again. For weak spin-orbit coupling, $\Delta_2$  slowly decreases for increasing Zeeman while it remains roughly constant $\Delta_2\sim\Delta$ for the strong spin-orbit coupling case. The gaps depicted here correspond to a nonzero $\mu=2\Delta=0.5meV$, which explains why $\Delta_1$ weakly depends on spin-orbit in the trivial regime. Adapted from Ref. \cite{Cayao-Thesis}}
\label{fig:8}
\end{figure}

Intuitively, the emergence of topological superconductivity in proximitized nanowires can be understood by analyzing the various gaps that open up for $\Delta\neq 0$ as a result of the pairings in Eq. (\ref{pairing-NW}).
Let us denote the gaps near $k=0$ and near the Fermi momenta as $\Delta_1$ and $\Delta_2$, respectively. They can be obtained from the BdG spectrum of the system (Fig. \ref{fig:7})
\begin{equation}
E^2_\pm(k)=(\frac{\hbar^2k^2}{2m}-\mu)^2+\alpha^2k^2+E_Z^2+\Delta^2\pm2\sqrt{E_Z^2\Delta^2+(\frac{\hbar^2k^2}{2m}-\mu)^2(E_Z^2+\alpha^2k^2)}.
\end{equation}

\begin{figure}[!tt]
\centering
\includegraphics[width=0.8\columnwidth]{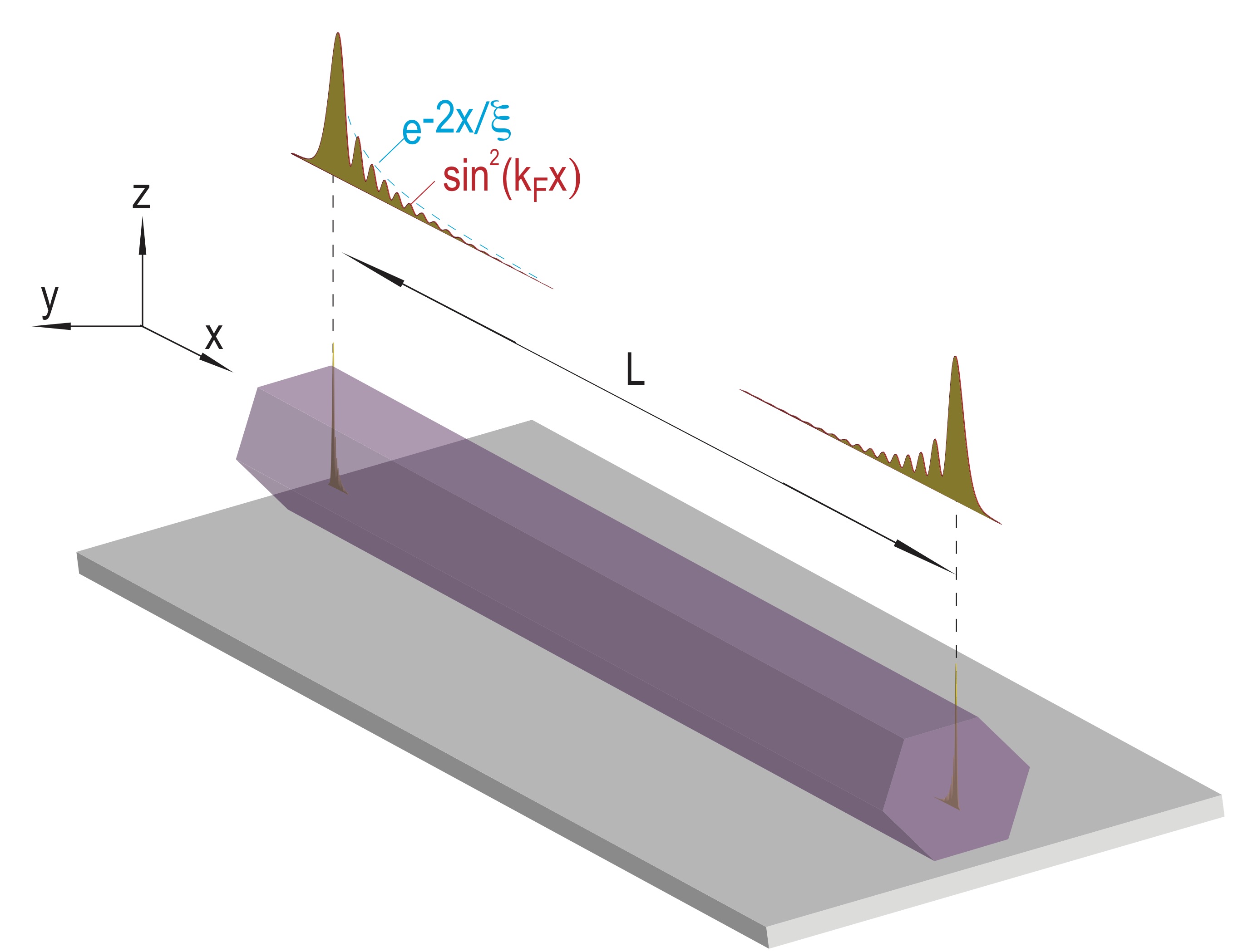}\\
\caption{A semiconducting nanowire with Rashba spin-orbit coupling (purple) in proximity with an $s$-wave superconductor (grey) becomes a one-dimensional topological superconductor with Majorana zero modes at its ends (yellow) when the applied Zeeman field $E_Z$ is larger than the critical field $E^c_Z\equiv\sqrt{\Delta^2+\mu^2}$. In realistic samples of finite length $L$, the Majoranas are not true zero modes but weakly overlapping modes of energy $\varepsilon\sim k_Fe^{-\frac{L}{\xi}}cos(k_F L)$, where the Majorana localization length $\xi$ depends on the strength of the spin-orbit interaction, see text.}
\label{fig:9}
\end{figure}

$\Delta_1$ is key to the emergence of topological superconductivity. Interestingly, there are two gapping mechanisms at low momentum, namely $E_Z$ and $\Delta$, that \emph{compete}. As a result of this competition, $\Delta_1$ can close, which signals a topological transition.  In particular, as $E_Z$ approaches the critical field
\begin{equation}
\label{criticalZeeman}
E^c_Z\equiv\sqrt{\Delta^2+\mu^2}, 
\end{equation}
the low momentum gap $\Delta_1$ vanishes linearly as $\Delta_1\approx |E_Z-E^c_Z|$ \cite{Lutchyn2010,Oreg2010}. For $E_Z>E^c_Z$, a Zeeman-dominated $\Delta_1$ reopens again and the system is a topological superconductor \footnote{Along this discussion, as in the original papers \cite{Lutchyn2010,Oreg2010}, we assume that the magnetic field only induces a Zeeman term such that orbital effects in the nanowire are neglected. Either a finite cross-section of the nanowire or deviations from perfect alignment of the magnetic field with the nanowire axis can induce orbital effects that may affect the above phase diagram. The role of such orbital effects in Majorana nanowires has been discussed in Refs. \cite{Lim2012,orbital1,orbital2,orbital3,orbital4}.}. Fig. \ref{fig:8} (top) illustrates one of this topological phase transitions as the Zeeman field increases \footnote{Note that, in general situations with $\mu\neq 0$, $\Delta_1$ is at a small but finite $k$ in the trivial regime $E_Z<E^c_Z$. However, as $E_Z$ approaches $E^c_Z$, $\Delta_1$ becomes centered at $k = 0$ and follows the linear dependence $\Delta_1\approx |E_Z-E^c_Z|$}.
Note that the topological phase occurs here for $E_Z>\sqrt{\Delta^2+\mu^2}$, as opposed to the topological insulator case where the topological superconducting phase appears when $E_Z<\sqrt{\Delta^2+\mu^2}$ (compare Fig. \ref{fig:8}d with Fig. \ref{fig:5}b). This difference can be traced back to the fact that we need to explicitly break time-reversal symmetry (through the Zeeman term) to reach a helical regime \footnote{Recently, another Zeeman-induced helical state akin to the Quantum Spin Hall state has been experimentally demonstrated in graphene's zero Landau level  \cite{Jarillo-QSHGraphene}. This helical state, that appears without the need of spin-orbit coupling,  is possible owing to the peculiar interaction-induced magnetic ordering of the zero Landau level which can be tuned from an insulating antiferromagnetic state to a ferromagnetic state with helical edges states upon the application of a strong in-plane Zeeman field. When proximitized with a superconductor, these helical edge states host a topological superconducting phase with Majorana states which are bounded by the degree of antiferromagnetic canting that gaps out the helical edges \cite{Graphene-Majoranas}.} as opposed to the two-dimensional topological insulator case. Since time reversal invariance is already broken in the topological phase, the system binds Majorana zero modes at boundaries with the topologically trivial vacuum gap, namely at the wire ends, when $E_Z>E^c_Z$, see Fig. \ref{fig:9}. The topological protection of the Majorana modes is guaranteed inasmuch as the gap to other finite energy quasiparticle excitations is well above the temperature. Right after the topological transition, this gap is given by $\Delta_1$ but for large enough Zeeman field it is replaced by $\Delta_2$ (in general its value is given by Min$(\Delta_1,\Delta_2)$). $\Delta_2$, in contrast to $\Delta_1$, never closes and, for strong spin-orbit coupling, remains roughly constant $\Delta_2\sim\Delta$. While its general form is too cumbersome to be written here, its $\mu=0$ value reads:
\begin{equation}
\Delta_2=\frac{2\Delta E_{SO}}{\sqrt{E_{SO}(2E_{SO}+\sqrt{E^2_Z+4E^2_{SO}})}}.
\end{equation}

For strong spin-orbit coupling, $\Delta_2$ remains roughly constant and roughly equal to the original gap $\Delta_2\sim\Delta$, see Fig. \ref{fig:8} (bottom).

The Majorana wave functions exponentially decay into the bulk of the superconductor as $\psi(x)\sim e^{-x/\xi_M}e^{\pm i k_Fx}$, with $k_F$ the Fermi wave vector associated with the zero mode and  $\xi_M$ their typical coherence length \footnote{The Majorana wave function presents an exponential decay provided that $E_Z>>E_Z^c$. Near the transition, both gaps $\Delta_1$ and $\Delta_2$ still contribute to the decay and the wave function presents a double-exponential decay, see Ref. \cite{Klinovaja-Loss}.}. This gives rise to a residual overlap between Majorana modes residing at opposite ends of the nanowire in realistic wires of finite length $L$. This overlap leads to the hybridization of Majoranas \cite{Lim2012,Prada2012,DasSarma2012,Rainis2013}  into Bogoliubov quasiparticles of energy 
\begin{equation}
\label{Majorana-overlap}
\delta\varepsilon\sim \hbar^2k_F \frac{e^{-2L/\xi_M}}{m\xi_M}cos(k_F L).
\end{equation}
This residual energy is expected to be exponentially small but its importance of course depends on the ratio $L/\xi_M$. The spatial extension of the Majoranas $\xi_M$ depends on various relevant energy scales of the problem and should be of the order of $\xi_M\sim \hbar v_F/\Delta$, with $v_F$ being the Fermi velocity. In the strong spin-orbit regime, where $E_{SO}>>E_Z$, the Fermi velocity is well approximated by $\alpha/\hbar$ and then $\xi_M\sim \alpha/\Delta$. For weak spin-orbit coupling (or large Zeeman), the correlation length acquires a prefactor that depends on the ratio between the Zeeman and the spin-orbit energies, namely $\xi_M\sim (E_Z/E_{SO}) \alpha/\Delta$ which can be written in terms of the spin-orbit length $l_{SO}=\hbar^2/(m\alpha)$ as $\xi_M\sim 2(E_Z/\Delta) l_{SO}$ \cite{Klinovaja-Loss,Mishmash}. Thus, for realistic values of the spin-orbit coupling, one expects a Majorana correlation length which should be of the order of the spin-orbit length and which should increase with Zeeman.

\begin{figure}[!tt]
\centering
\includegraphics[width=1\columnwidth]{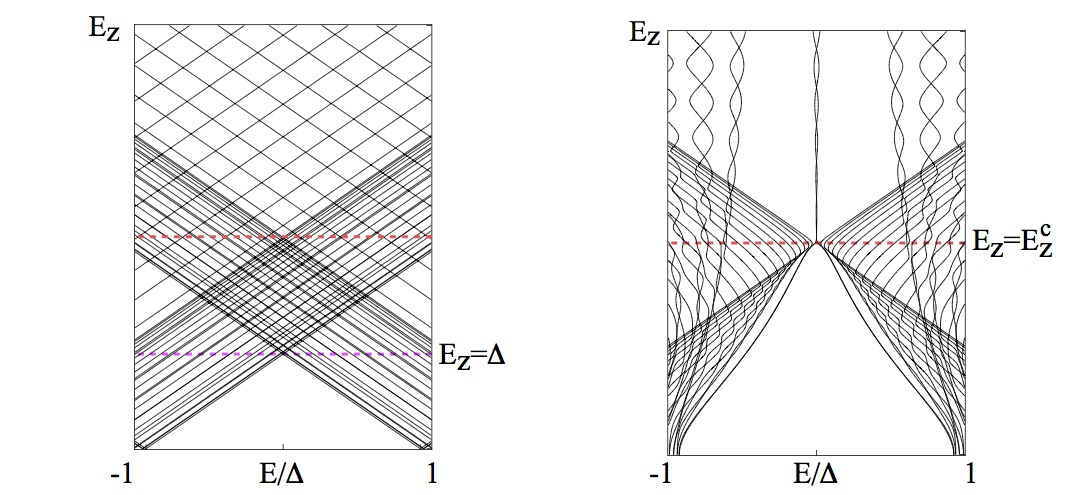}\\
\caption{Bogoliubov-de Gennes spectrum of a finite-length nanowire for increasing Zeeman fields. Without spin-orbit coupling (left panel), the superconducting gap closes at $E_Z=\Delta$ and never reopens again, since the Zeeman field induces depairing in an overall trivial $s$-wave superconductor. The sector $\Delta<E_Z<E_Z^c$ (region between purple and red dashed lines) contains BdG  levels with both spin components, while the spectrum is fully spin-polarized when $E_Z>E_Z^c$. With spin-orbit coupling (right panel), this spin-polarized spectrum becomes topological: the superconducting gap first closes at  $E^c_Z\equiv\sqrt{\Delta^2+\mu^2}$ but it reopens again at $E_Z>E^c_Z$ since the system is effectively a nontrivial $p$-wave superconductor. The nontrivial phase is characterized by near-zero Majorana modes which weakly overlap (small deviations from strict zero energy) owing to the finite length, see text. Adapted from Ref. \cite{Cayao-Thesis}}
\label{fig:10}
\end{figure}

One can understand the important role that spin-orbit coupling has in the mechanism leading to a topological transition by comparing $\alpha=0$ and $\alpha\neq 0$ BdG spectra as a function of Zeeman field (Fig. \ref{fig:10}). Without spin-orbit coupling (left panel),  the gap closes when $\Delta=E_Z$ (purple dashed line). This Zeeeman depairing effect is due to the Zeeman splitting of the BdG levels that cross zero energy when $E_Z=\Delta$. After the closing of the gap, the zero-energy crossings of Zeeman-split levels are not protected and the gap never reopens. Interestingly, while the regime $\Delta<E_Z<E_Z^c$ (region between purple and red dashed lines in the left panel of Fig. \ref{fig:10}) corresponds to BdG  levels with both spin components, one spin sector is completely removed when $E_Z>E_Z^c$ and the spectrum is fully spin polarised. This spin-polarised spectrum is the one that becomes topological when $\alpha\neq 0$. This is explicitly shown in the right panel, where it is clear that the finite spin orbit coupling shifts the gap closing point at $E_Z^c$  while inducing avoided crossings between levels at finite energy. This results in a reopening of the gap for $E_Z>E_Z^c=\sqrt{\Delta^2+\mu^2}$ which now contains a subgap level near zero energy. This BdG excitation that oscillates near zero energy \cite{Lim2012,Prada2012,DasSarma2012,Rainis2013} corresponds to the two overlapping Majoranas that we described previously  in Eq. (\ref{Majorana-overlap}) and is separated from the rest of excitations by a so-called topological gap (essentially the distance to the first finite energy excitation) which is governed by the amount of spin-orbit coupling. 

\section{Experimental detection protocols \label{detection}}
In the previous sections, we have discussed various ingenious proposals to implement topological superconductivity. These (and other) proposals provide with more than enough theoretical arguments to support the idea that the above platforms should support this exotic form of superconductivity. Thus, there is consensus that it should be possible to engineer devices with Majorana quasiparticles in the lab. An obvious question is, of course, how to make sure that we have created these exotic excitations and how to distinguish them from more mundane subgap excitations, such as trivial Andreev levels. In this section we shall discuss the most relevant protocols for Majorana detection: quantized conductance, zero-bias anomalies in tunneling spectroscopy experiments and the $4\pi$ Josephson effect.
\subsection{Normal metal-superconductor junctions: Quantized conductance and Zero bias anomalies in transport spectroscopy}
The emergence of Majorana modes as zero-energy subgap excitations  gives rise to resonant Andreev reflection: an incident electron is always Andreev reflected into a hole with unitary probability. As a result, conductance measurements in normal metal-superconductor (NS) junctions based on nanowires are expected to show perfect unitary quantized conductance $G=2e^2/h$. In what follows we elaborate on this idea.

At energies below the gap, an electron incident on a superconductor can be either reflected as an electron or as a hole. While the former process is a standard normal reflection, the latter (known as Andreev reflection) is subtler as it effectively transfers a Cooper pair into the superconductor (since the electron and the hole have opposite charges). This process gives rise to a finite conductance (the so-called Andreev conductance) across the normal-superconductor junction which can be fully described by a scattering matrix of the form
\begin{equation}
 \label{S-matrix1}
R= \left( \begin{array}{ccc}
r_{ee} & r_{eh}\\
r_{he} & r_{hh}  \end{array} \right),
\end{equation}
which has a block structure of $N \times N$ submatrices (with $N$ being the number of channels) describing normal (the matrices in the diagonal, $r_{ee}$ and $r_{hh}$) and 
Andreev (the matrices in the off-diagonal, $r_{eh}$ and $r_{he}$) reflection of  electrons and holes. Using this scattering matrix, the linear conductance can be written as 
\begin{equation}
\label{Andreev}
G=\frac{2e^2}{h}Tr[r_{eh}r^\dagger_{eh}]=\frac{2e^2}{h}\sum_{n=1}^N R_n,
\end{equation}
where the eigenvalues $R_n$ give the probability for Andreev reflection of the nth eigenmode at the Fermi level ($E=0$). It is important to stress that the factor of two is not due to spin degeneracy (which is included in the sum over channels), but due to the fact that Andreev reflection effectively transfers a charge $2e$ across the superconductor. For a single spinful channel, this formula describes all the possible values of Andreev conductance, from tunneling $G\sim 0$ to fully ballistic $G=4e^2/h$ depending on the transparency of the normal-superconductor interface.

Let us now discuss the case where we only have a single spinless channel, which is a relevant situation for nanowires when time-reversal symmetry is explicitly broken. In this case, particle-hole symmetry dictates that $R(E)=\tau^xR^*(-E)\tau^x$ which in turn implies $r_{ee}=r^*_{hh}$ and $r_{eh}=r^*_{eh}$ at zero energy. When combined with the unitarity of the reflection matrix  (since the transmission through the superconductor vanishes at energies below the gap), this implies that the determinant of the reflection matrix takes only two possible values $det R=\pm 1$: namely, the reflection matrix is either diagonal describing perfect normal reflection $|r_{ee}|=1$ and zero Andreev reflection $r_{eh}=0$, or off-diagonal describing perfect Andreev reflection $|r_{eh}|=1$ with zero normal reflection  $r_{ee}=0$. The first case corresponds to the trivial case and gives rise to zero conductance while the second case corresponds to the nontrivial case yielding perfect conductance $G=2e^2/h$. Note that symmetry imposes that one cannot smoothly evolve from one case to the other without closing the gap which makes $det R$ a topological index \cite{Akhmerov2011}. In the presence of interactions, these two limits are robust and represent renormalization group fixed points at low energies \cite{Fidkowski2012}.

The above result, where Andreev reflection is either perfect or zero, is a particular instance of the so-called B{\'e}ri degeneracy \cite{Beri}: in situations where Kramer's theorem does not hold, since time-reversal symmetry is broken, particle-hole symmetry dictates that $R_n$ in Eq. (\ref{Andreev}) is twofold degenerate with the exceptions $R_n=0$ and $R_n=1$. Using B{\'e}ri's degeneracy, one can understand the conductance quantization of $G=2e^2/h$ as resulting from a nondegenerate Andreev reflection eigenvalue, while other fully Andreev-reflected modes are necessarily twofold degenerate and thus contribute to a conductance $G=4e^2/h$. 

\begin{figure}[!tt]
\centering
\includegraphics[width=\columnwidth]{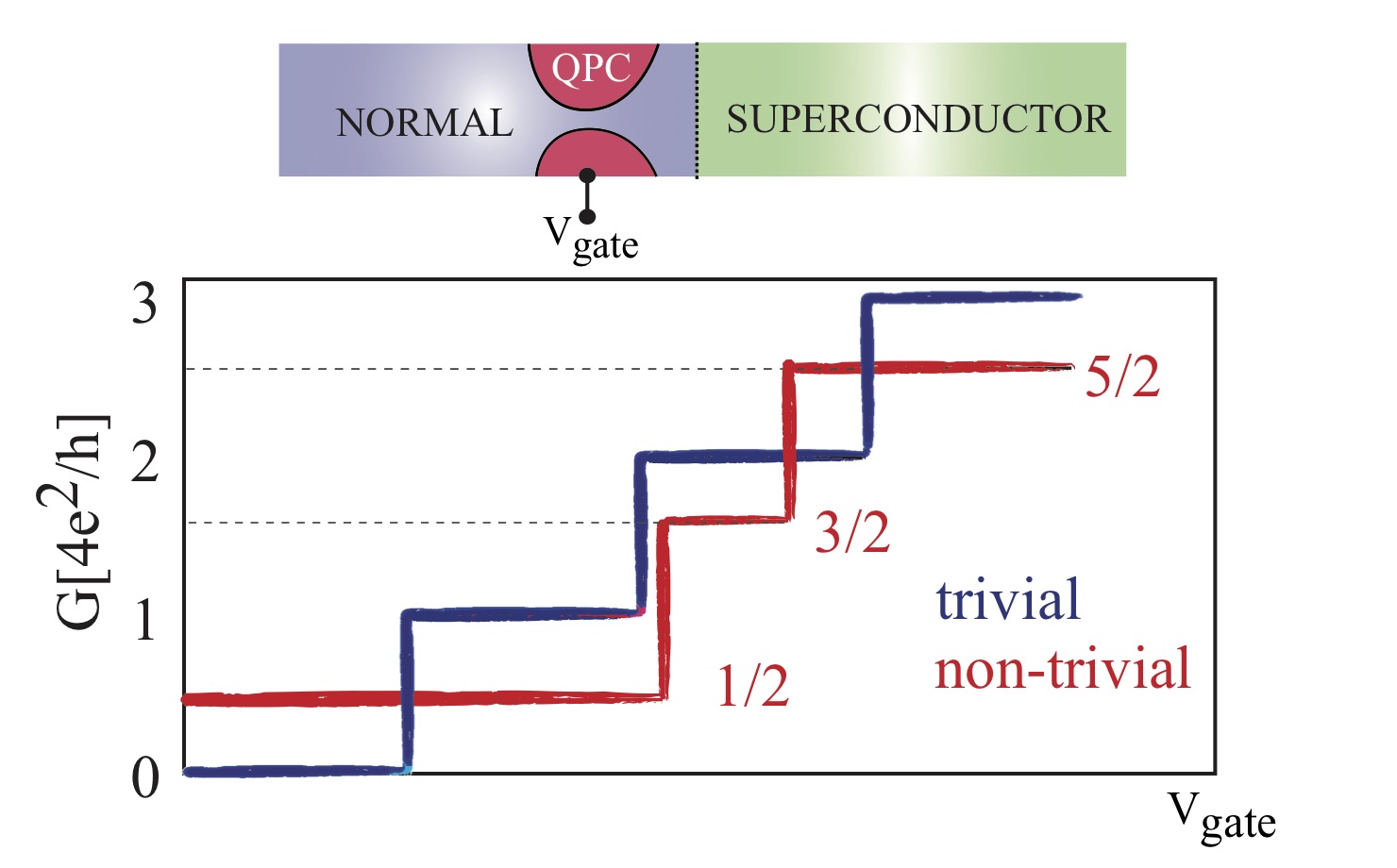}\\
\caption{Top: schematic of a normal-superconductor junction containing a quantum point contact (QPC). Bottom: sketch of the quantized conductance across the QPC for trivial (blue) and non-trivial (red) superconductors. The quantized steps directly probe topological superconductivity since the standard  quantization in integers of $4e^2/h$ that occurs when the superconductor is trivial shifts by $2e^2/h$ once the superconductor becomes topological, see Ref. \cite{Wimmer2011}. }
\label{fig:11}
\end{figure}

Interestingly, the conductance quantization in the topological regime extends from the tunneling to the fully transparent regime. This allows to write the quantized plateaus in a superconducting quantum point contact (QPC) geometry as \cite{Wimmer2011}:

\begin{equation}
\label{quantizedG}
G=\frac{4e^2}{h}\times\{ \begin{array}{ccc}
n \\
n+\frac{1}{2} \end{array}, n=0,1,2...
\end{equation}
where the upper (lower) row corresponds to the trivial (topological) regime. Thus, the conductance steps in normal-superconductor quantum point contacts (QPCs) are a direct probe of topological superconductivity since the quantized steps in conductance shift from integer to half-integer values upon transition from the topologically trivial to the nontrivial one (Fig. \ref{fig:11}). While these plateaus at half-integer steps are reminiscent of the quantum Hall plateaus in graphene, both originate from the existence of a zero mode, their topological protection against disorder is rather different (much weaker in the topological superconducting case). In fact, only the first $n=0$ plateau in Fig. \ref{fig:11} is protected against disorder (as opposed to the Quantum Hall case, where all plateaus are robust), see Ref. \cite{Wimmer2011}.

\begin{figure}[!tt]
\centering
\includegraphics[width=0.7\columnwidth]{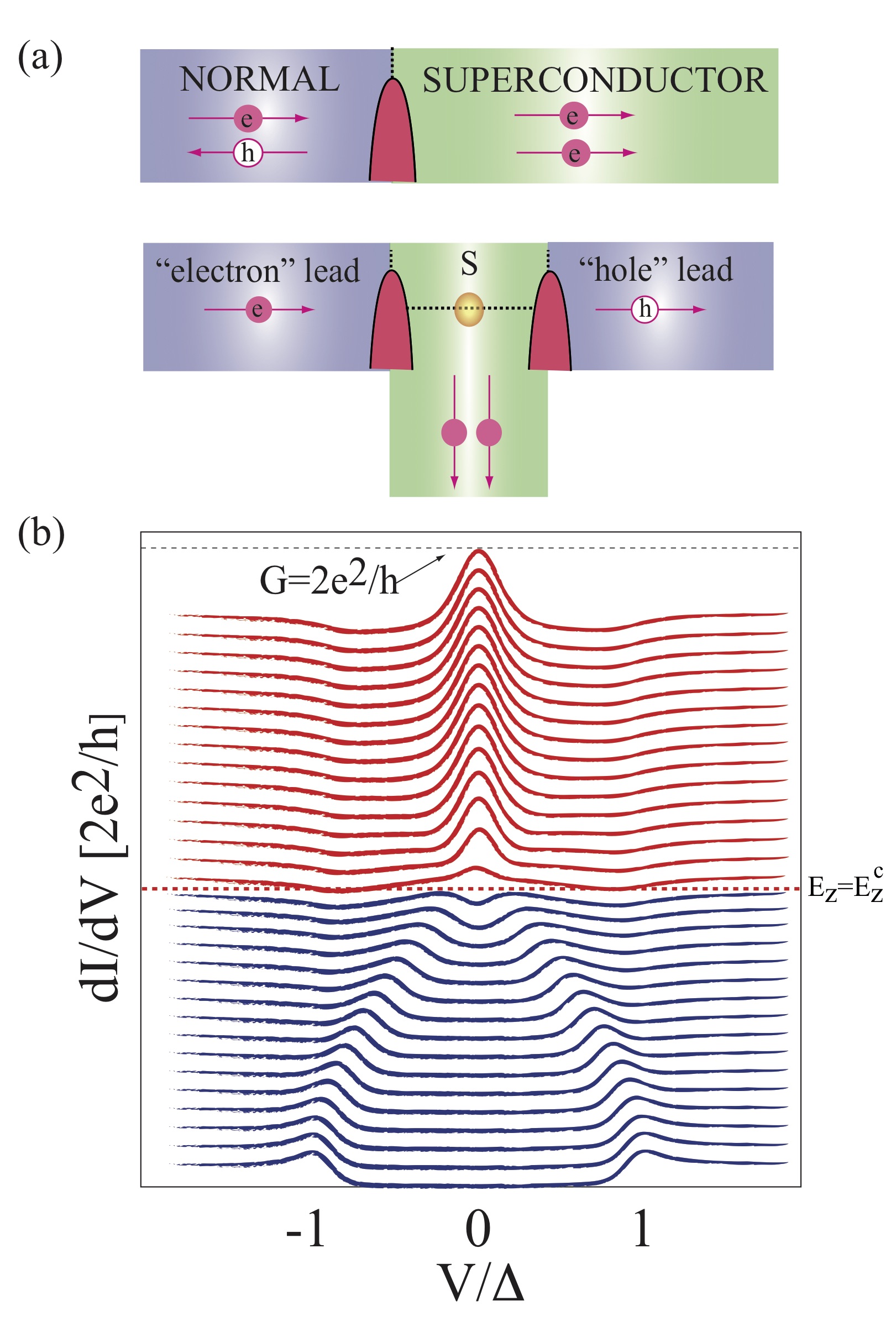}\\
\caption{Transport characteristics of Majorana normal-superconductor junctions. (a) Top: Schematic of Andreev reflection in a normal-superconductor junction: an incident electron is converted into a hole and a Cooper pair is created in the superconductor. Bottom: If the junction contains a Majorana bound state, Andreev reflection can be understood as a resonant process at zero energy through a "two lead" double barrier system where, conceptually, the original normal lead is separated into two "electron" and "hole" leads. 
(b) Sketch of  the expected behavior of differential conductance curves $dI/dV$ as a function of voltage for increasing magnetic fields. Blue and red denote trivial and non-trivial regions, respectively.The topological transition is seen as a closing of the superconducting gap at the critical Zeeman field and the emergent zero bias anomaly signals the appearance of Majorana zero modes in the system, see e. g. Ref. \cite{Stanescu2011}.}
\label{fig:12}
\end{figure}

At finite bias voltage $V$, the Andreev current can be written as 
\begin{equation}
\label{I-V}
I=e\int\frac{d\omega}{h}\frac{\Gamma_e\Gamma_h}{\omega^2+(\Gamma_e+\Gamma_h)^2/4}[f(\omega-eV)-f(\omega+eV)],
\end{equation}
where $f(\omega\pm eV)$ being Fermi functions and $\Gamma_{e/h}$ electron and hole tunneling rates. Electron-hole symmetry dictates that $\Gamma_e=\Gamma_h=\Gamma$ which results in perfect resonant Andreev reflection at $V=0$. This resonant process is completely analogous to resonant tunneling through a symmetric two barrier system (see Fig. \ref{fig:12}a), as implicit in the Lorentzian form in Eq. (\ref{I-V}),  which results in the expression \cite{Bolech2007,Law2009,Flensberg2010,Stanescu2011}:
\begin{equation}
\label{dI-dV}
\frac{dI}{dV}=\frac{2e^2}{h}\frac{\Gamma^2}{eV^2+\Gamma^2}.
\end{equation}

The above result implies that a quantized zero bias anomaly (ZBA) of conductance $G=2e^2/h$ should emerge in tunneling spectroscopy across a proximitized nanowire after the system undergoes a topological phase transition by e.g. increasing an external Zeeman field.  In particular, $dI/dV$ plots in the tunneling regime should be proportional to the local density of states at the end of the wire and, hence, map its BdG spectrum (like, for example, the one shown in Fig. \ref{fig:10}, right panel) as a function of Zeeman field. An example of the expected behaviour of $dI/dV$ for increasing Zeeman fields is shown in Fig. \ref{fig:12}b.

The above prediction is only valid at  very low temperatures $T<<\Gamma$. At finite temperature, the conductance can be written as:
\begin{equation}
\label{dI-dV-finiteT0}
\frac{dI}{dV}=\frac{2e^2}{h}\int d\omega \frac{\Gamma^2}{\omega^2+\Gamma^2}\frac{1}{4k_BTcosh^2((\omega-eV)/2k_BT)},
\end{equation}
which is just a thermally-broadened version of Eq. (\ref{dI-dV}).
Note that an increase of $\Gamma$ in Eq. (\ref{dI-dV}) does not change the unitary value $G=2e^2/h$ at $V=0$, it only increases the full width at half maximum (FWHM). This is no longer true at finite temperatures, where the maximum varies as $\frac{2e^2}{h}f(k_BT/\Gamma)$, where $f$ is a scaling function that depends only on the ratio of temperature to tunnel broadening.
\begin{figure}[!tt]
\centering
\includegraphics[width=\columnwidth]{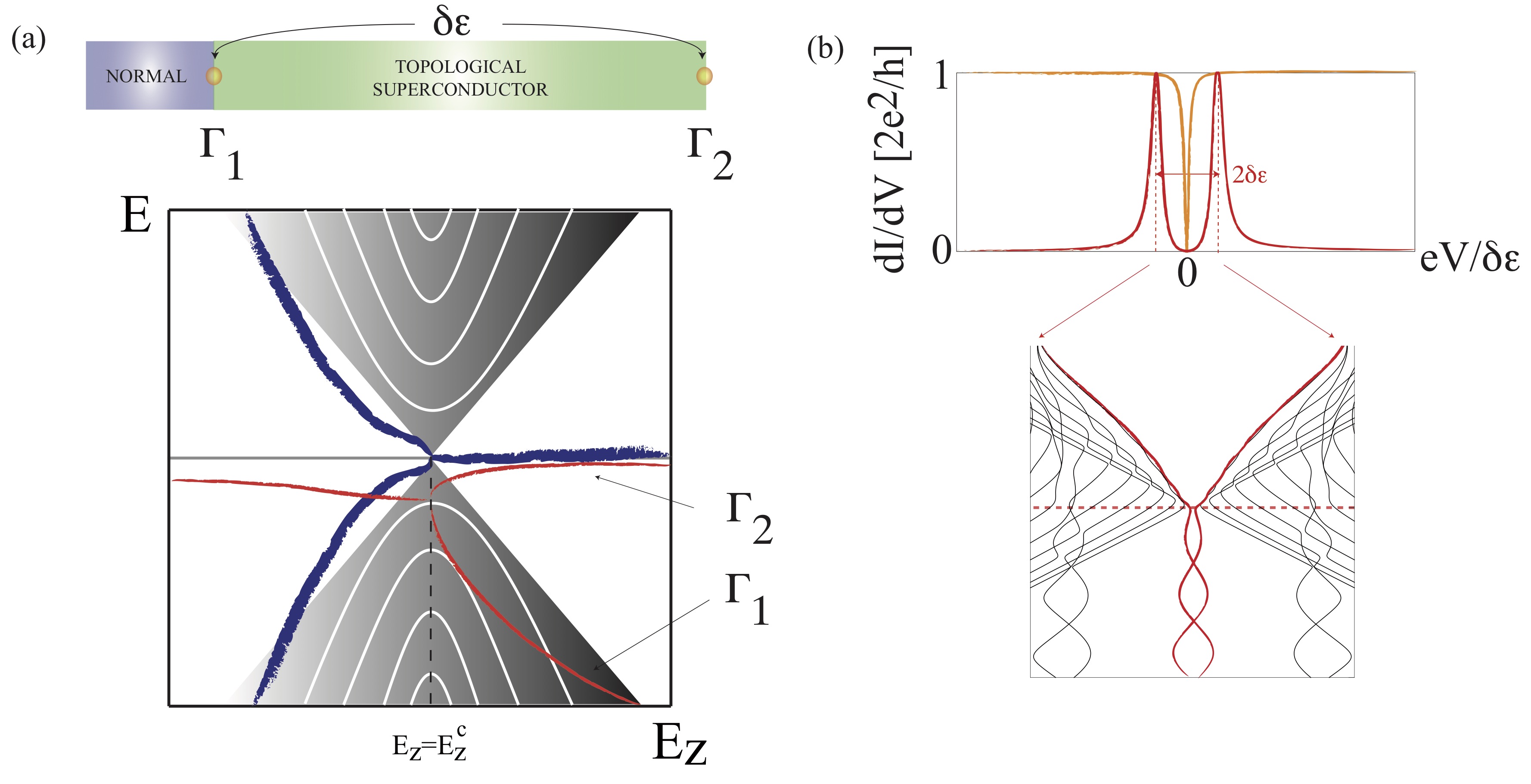}\\
\caption{(a) Top:  Schematic of a normal-topological superconductor junction of finite length containing two weakly overlapping Majoranas, overlap $\delta\varepsilon$, with different decay rates to the normal contact $\Gamma_1>>\Gamma_2$. Bottom: Pole bifurcation.  In a finite length nanowire coupled to the normal contact, the closing of the bulk gap characterizing a topological transition (grey areas) becomes a crossover which can be understood in terms of the Bogoliubov spectrum in the complex plane. The two lowest Bogoliubov excitations at finite energy (blue lines) coalesce and become zero modes (the real parts of higher excited states are sketched as white lines). At the same point, their imaginary parts (red lines) bifurcate such that the decay rates of the two emerging quasibound zero Majorana modes become distinct. In this language, the topological transition becomes a crossover as $\Gamma_2\rightarrow 0$ \cite{Pikulin:JL12,Pikulin:PRB13}. Interestingly, the pole bifurcation can occur at points in parameter space well separated from the critical field at which the topological transition is expected in a bulk system \cite{SanJose2016}. (b) Sketch of  the expected behavior of differential conductance curves $dI/dV$ of a finite length nanowire for $\delta\varepsilon/\Gamma_1<<1$ (transparent regime, orange) and $\delta\varepsilon/\Gamma_1>>1$ (tunneling regime, red). The tunneling regime maps the local density of states of the wire, which allows to extract the BdG spectrum with overlapping Majoranas (bottom) from $dI/dV$ measurements.}
\label{fig:13}
\end{figure}

The maximum zero bias conductance $G=2e^2/h$ becomes $G=0$ for finite length wires, as dictated by B{\'e}ri's degeneracy. This zero conductance can be understood from a slightly different perspective by analyzing the complex poles of the scattering matrix $S(\epsilon)$, where all the information about Andreev reflection is encoded. These poles can be written as  $\epsilon_n=E_n-i\Gamma_n$ and represent the real energy $E_n$ and decay rate $\Gamma_n$ into the normal contact of quasibound Bogoliubov excitations in the wire. Particle-hole symmetry requires that these complex poles come in pairs symmetrically arranged around the imaginary axis ($\epsilon_n$ and $-\epsilon_{n}^*$). In this language, topology is characterized by the number of poles, $N_Y$, with zero real part (at a topological phase transition $N_Y$ changes by $\pm 1$). However, in a finite size system, like a finite length nanowire, topological transitions become crossovers where $N_Y\rightarrow N_Y\pm 2$. This pole transition \cite{Pikulin:JL12,Pikulin:PRB13} occurs through the fusion of the two poles $\epsilon_n$ and $-\epsilon_{n}^*$ with finite real part, which meet at the imaginary axis, and then split vertically (namely a bifurcation \footnote{This mechanism stabilizes parity crossings in an isolated system into extended zero modes when the system is coupled to a reservoir \cite{Mi2014,Tarasinski2015}. If the reservoir is helical \cite{Cayao2015}, this mechanism can be used to obtain true bound Majoranas without the need of a topological transition \cite{SanJose2016}. These bifurcations, also known as exceptional points, are extensively studied in the context of non-hermitian photonics see e. g. Refs.\cite{Malzard2015,Zhen2015}} at which the real parts become pinned to E = 0 and the decay rates of the two quasibound states become distinct $\Gamma_1\neq\Gamma_2$, see Fig. \ref{fig:13}a). A topological crossover may occur \emph{after} a pole transition, when one of the two bifurcated poles moves along the imaginary axis into the origin $\Gamma_2\rightarrow 0$ (at which point it becomes completely decoupled from N). However, this can only happen for infinitely long wires. For finite length wires, both poles have finite imaginary part and Andreev reflection in a NS junction can be understood as a destructive interference between both zero modes at $V=0$ and hence zero conductance \cite{Pikulin:PRB13,Ioselevich:NJOP13}. For transparent contacts with 
$\Gamma_1>>\delta\varepsilon$, with $\delta\varepsilon$ being the finite Majorana overlap, the conductance shows a $G=2e^2/h$ plateau (perfect Andreev reflection) with a superimposed dip of width 
$\sim\delta\varepsilon^2/\Gamma_1$ at  $V=0$, owing to the presence of the far end Majorana, see Fig. \ref{fig:13}b (orange curve). In the opposite tunnel limit, $\Gamma_1<<\delta\varepsilon$, the conductance consists of two well-separated Lorentzian peaks, Fig. \ref{fig:13}b (red curve). This is the most favorable regime to extract overlapping Majoranas from $dI/dV$. In the limit $\Gamma_2\rightarrow 0$, this Fano shape  can be written as \cite{Flensberg2010,Ioselevich:NJOP13}: 

\begin{equation}
\label{dI-dV-finiteL}
\frac{dI}{dV}=\frac{2e^2}{h}\frac{(2eV\Gamma_1)^2}{[(eV)^2-\delta\varepsilon^2]^2+(2eV\Gamma_1)^2}.
\end{equation}

\subsection{Superconductor-normal-superconductor junctions: the $4\pi$ Josephson effect \label{4pi}}
A Josephson junction, a device in which two superconductors are coupled through a weak link, supports an electrical current flow in the absence of any applied voltage. This so-called Josephson current is proportional to the sine of the superconducting phase difference and takes values between $\pm I_c$, the critical current of the junction
\begin{equation}
\label{Jospehson-current}
I(\phi)=I_c sin\phi .
\end{equation}

The presence of Majorana bound states in junctions made of topological superconductors affects the current-phase relation of the Josephson effect in a very peculiar way. In order to illustrate the idea, let us consider again the Kitaev model in 
Eq. (\ref{kiatev0}) with $\mu=0$ and $t=\Delta$ (topological phase) and write the tunneling Hamiltonian describing a junction of two such Kitaev wires coupled through a weak link $H=H_L+H_R+H_J$ with ($\alpha\in L,R$)

\begin{equation}
\label{kiatev_Josephson}
H_\alpha=-\frac{t}{2}\,\sum_{j=1}^{N-1}\Big[\,\big(c^{\dagger}_{\alpha,  j}c_{\alpha, j+1}+c^{\dagger}_{\alpha,j+1}c_{\alpha,j}\big)\,+\,\,(e^{i\phi_\alpha}c_{\alpha,j}c_{\alpha,j+1}\,+e^{-i\phi_\alpha}\,c^{\dagger}_{\alpha,j+1}c^{\dagger}_{\alpha,j})\Big]\,,
\end{equation}
with $\phi_\alpha$ being the superconducting phase of each segment. The term that couples the left and right segments is a standard electron tunneling Hamiltonian of the form:
\begin{equation}
\label{kitaev-tunneling}
H_J=-t_J\big(c^{\dagger}_{L N}c_{R1}+c^{\dagger}_{R 1}c_{LN}\big).
\end{equation}
If we now write the fermion operators $c_{R1}$ and $c_{LN}$ in terms of Majorana operators $c_{LN}=\frac{e^{-i\phi_L/2}}{2}(\gamma^A_L+i\gamma^B_L)$ and $c_{R1}=\frac{e^{-i\phi_R/2}}{2}(\gamma^B_R+i\gamma^A_R)$, as we did in section \ref{Kitaev}, it is straightforward to get the following low-energy projection
\begin{eqnarray}
c_{LN}\rightarrow\frac{1}{2}e^{-i\phi_L/2}\gamma^A_L\nonumber\\
c_{R1}\rightarrow\frac{i}{2}e^{-i\phi_R/2}\gamma^A_R,
\end{eqnarray}
since the B Majoranas hybridize with the operators in neighboring sites and form finite-energy fermions. This yields an effective Josephson coupling term of the form (in what follows we remove the superscripts)
\begin{equation}
\label{kiatev_Josephson2}
H^{eff}_J=-i\frac{t_J}{2}cos\big(\frac{\phi_R-\phi_L}{2}\big)\gamma_L\gamma_R.
\end{equation}
If we now remember the relation between the fermion parity operator and the overlap between two Majorana operators, c. f. Eq. (\ref{parityoperator}), $P\equiv 1-2\hat{n}=1-2f^\dagger f=-i\gamma_L\gamma_R$, we can write the Josephson coupling in a very appealing form ($\phi\equiv\phi_R-\phi_L$):
\begin{equation}
\label{kiatev_Josephson3}
H^{eff}_J=-\frac{t_J}{2} cos\big(\frac{\phi}{2}\big)(2\hat{n}-1),
\end{equation}
which \emph{relates explicitly the Josephson coupling with the fermionic occupation number operator}. Remarkably, this Hamiltonian is $4\pi$-periodic for fixed fermionic parity (namely, changing the phase $\phi$ by $2\pi$ does not conserve the energy!), as illustrated in Fig. \ref{fig:16} (red and blue lines denote states with opposite fermionic parity \footnote{Note that the ground state is, of course, $2\pi$-periodic, but it is inaccessible when $\phi\rightarrow\phi+2\pi$ if fermionic parity is conserved.}). 

Using this result, the Josephson current reads:
\begin{equation}
\label{kiatev_Josephson4}
I_J=\frac{2e}{\hbar}\frac{d\langle H^{eff}_J\rangle}{d\phi}=\frac{et_J }{2\hbar}sin\big(\frac{\phi}{2}\big)(2\langle n \rangle-1)=\pm\frac{et_J }{2\hbar}sin\big(\frac{\phi}{2}\big),
\end{equation}
where $\phi\equiv\phi_R-\phi_L$ and the sign reflects the fermionic parity. Eq. (\ref{kiatev_Josephson4}) reflects a so-called fractional Josephson effect \cite{Kwon,Kitaev2,Fu-Kane2} which gives rise to $4\pi$-periodic Josephson currents (as opposed to the standard Josephson current in Eq. (\ref{Jospehson-current}) which is $2\pi$-periodic). 

\begin{figure}[!tt]
\centering
\includegraphics[width=\columnwidth]{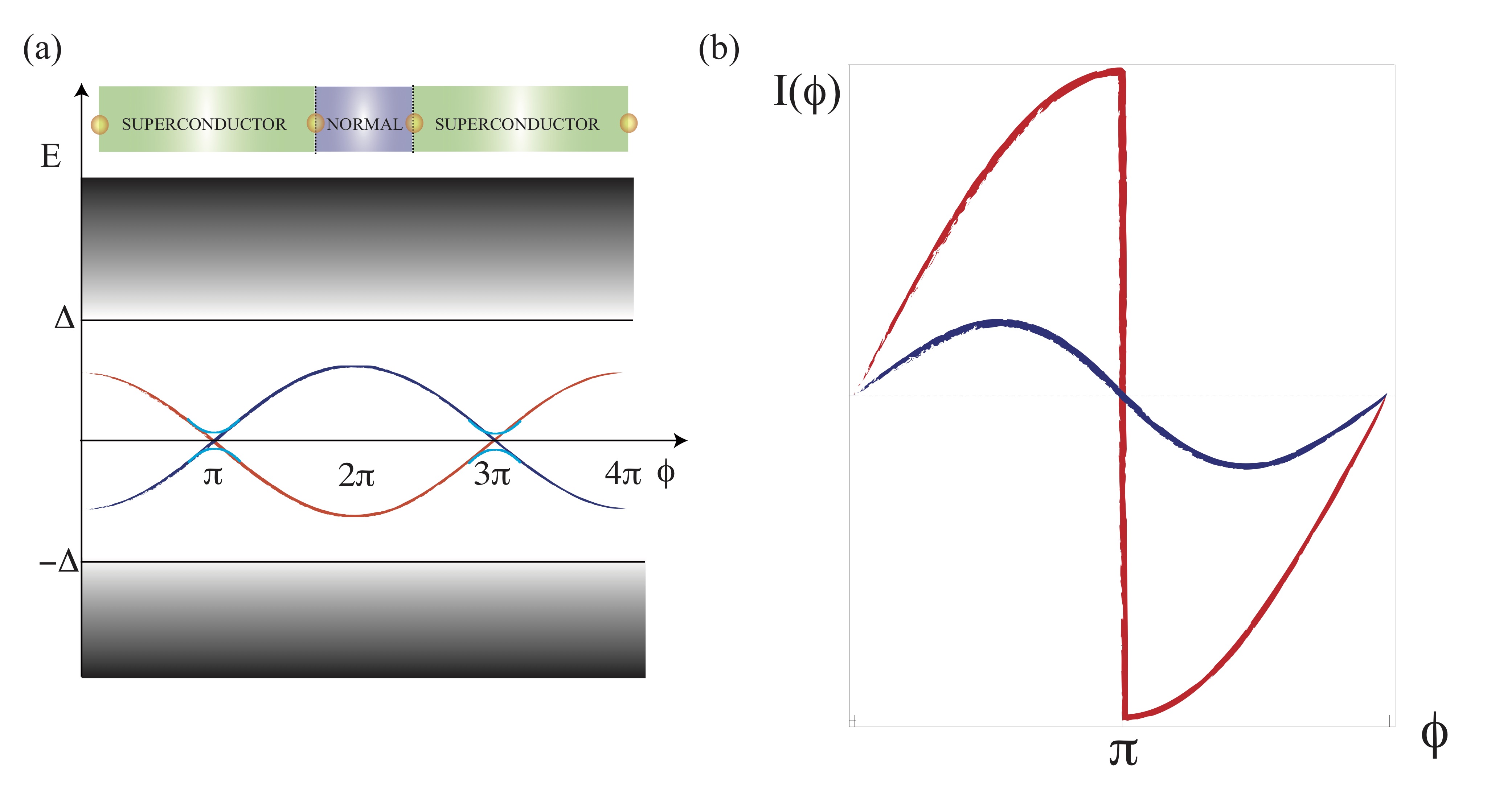}\\
\caption{ (a) Low-energy spectrum of a SNS Josephson junction containing Majoranas as a function of superconducting phase difference. Red and blue denote different fermionic parities which give rise to perfect crossings at $\phi=\pi$ and $\phi=3\pi$. Any finite length of the two topological superconducting regions gives rise to two additional Majoranas at the nanowire ends (upper sketch) results in residual splittings (dashed lines).  (b) The equilibrium supercurrents in realistic (finite length, quasiparticle poissoning, etc) Majorana  junctions are always $2\pi$ periodic. Nevertheless, one could in principle distinguish between trivial and nontrivial situations as the $I(\phi)$ curves in the tunneling regime change from sine-like $I(\phi)\sim sin\phi$ (trivial) to sawtooth-like $I(\phi)\sim sin(\phi/2)$ (nontrivial). Furthermore, the scaling of the critical current with the normal transmission of the junction changes from $T_N$ to $\sqrt{T_N}$ as the superconductors become topological. This allows for a mapping between $I_c$ and the topological phase diagram of the junction \cite{SanJose2014}.}
\label{fig:14}
\end{figure}
Intuitively, the fractional Josephson current can be understood as a change in the unit of transferred charge from $2e$ in a conventional Josephson junction to $e$ in a junction with Majoranas, which doubles the fundamental periodicity from $\sim sin\phi$ to $\sim sin(\phi/2)$ \cite{Kwon,Kitaev2,Fu-Kane2}. This change in the unit of transferred charge is also reflected in the fact that the critical current becomes proportional to $t_J$ instead of $t^2_J$. Using the Kitaev model above, the weak link can be parametrized as $t_J=\tau t= \tau \Delta$, with $\tau\leq 1$ being a parameter that controls the junction's transparency $T_N=\tau^2$. This means that the critical current for a Majorana junction is $I_c\equiv \frac{e\sqrt{T_N}\Delta}{2\hbar}$ instead of the standard $I_c\equiv \frac{e T_N\Delta}{\hbar}$ (namely, square root of transmission across the weak link instead of transmission itself). In tunnel junctions with $T_N<<1$, the topological contribution to the critical current is therefore larger than the one from trivial modes. This effect allows for a mapping between $I_c$ and the topological phase diagram of the junction \cite{SanJose2014}. Apart from this, the critical current contains further information about nontrivial topology and Majoranas. This includes signatures of the gap inversion at the topological transition and an oscillatory pattern that originates from Majorana overlaps for finite length nanowires \cite{Cayao17}.

A direct measurement of $4\pi$ periodicity in nanowire Josephson junctions would constitute a strong signature of Majoranas in the junctions. Unfortunately, the $4\pi$ effect is very fragile and hard to observe experimentally. First, any finite length of the two topological superconducting regions gives rise to two additional Majoranas at the nanowire ends, allowing for the hybridization of two states of the same fermionic parity \cite{SanJose2012,Pikulin2012,Virtanen2013}. This results in residual splittings at  $\phi=\pi$ (Fig. \ref{fig:14}a dashed blue lines) which, despite being exponentially small, destroy the fractional effect as the system remains in the ground state for all phases. A second, and potentially more severe, problem is random quasiparticle poissoning (stochastic quasiparticle tunneling events into the junction) which spoils fermionic parity conservation. Owing to these realistic effects, the equilibrium supercurrent is always  $2\pi$ periodic, Fig. \ref{fig:14}b.

\begin{figure}[!tt]
\centering
\includegraphics[width=\columnwidth]{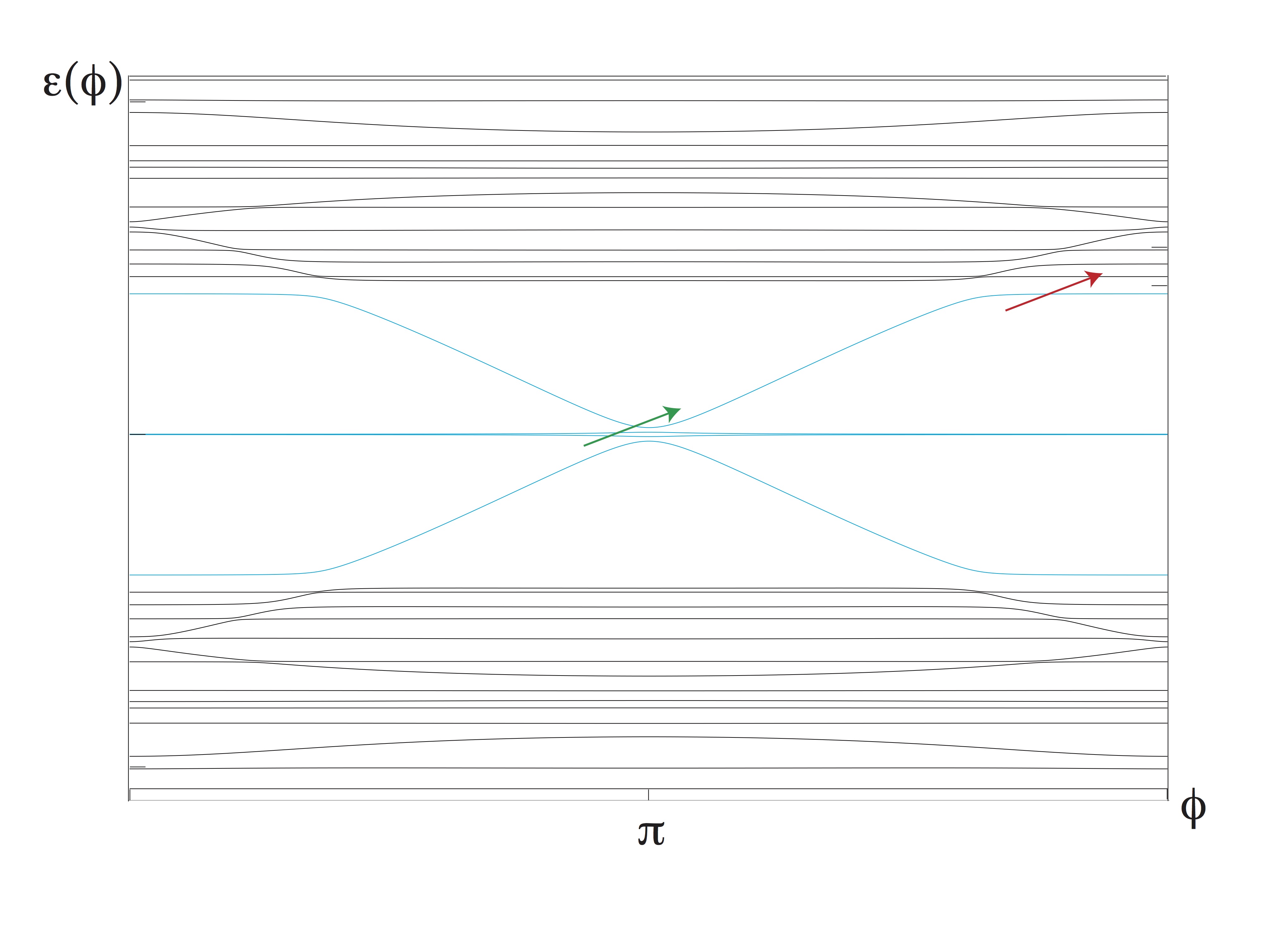}
\caption{Andreev level spectrum of a realistic SNS junction based on semiconducting nanowires \cite{SanJose2012,SanJose2013}. Above the low-energy Majorana sector sketched in Fig. \ref{fig:14}a there is a dense quasiparticle spectrum at higher energies. A voltage bias results in non-adiabatic Landau-Zener transitions (green and red arrows) between different states.}
\label{fig:15}
\end{figure}

\begin{figure}[!tt]
\centering
\includegraphics[width=\columnwidth]{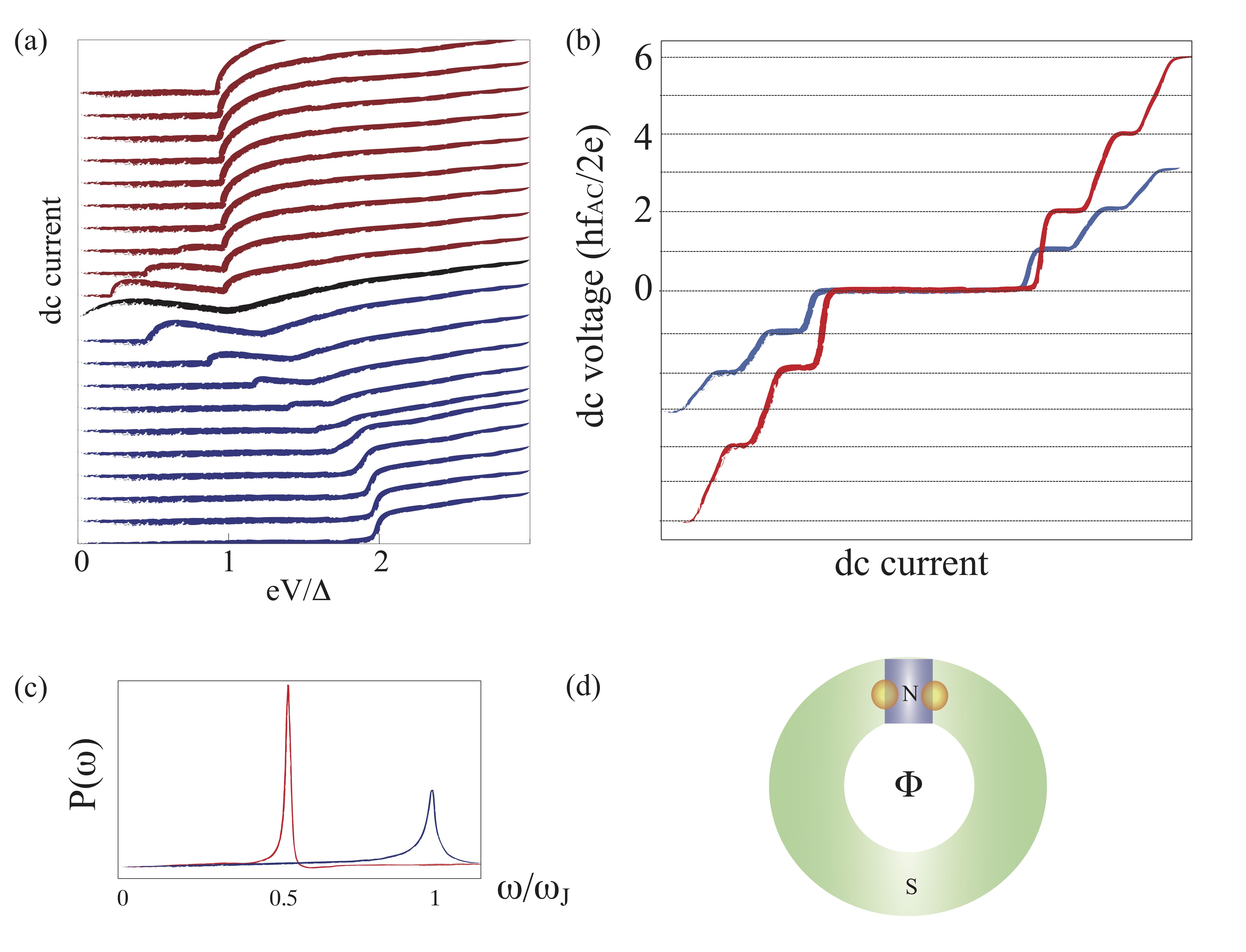}
\caption{ Various theoretical predictions for the fractional Josephson effect.  (a) The Andreev current through a SNS junction in the tunneling regime for increasing Zeeman fields (curves offset for clarity) directly measures the topological transition (closing of the gap, black curve). The conventional step at $eV=2\Delta$ in the trivial regime (blue curves) halves to $eV=\Delta$ in the topological regime (red curves) owing to the presence of midgap zero energy Majorana modes.  (b) Shapiro steps in junctions driven by an ac current: as the junction transitions from trivial (blue) to topological (red), the average voltage develops Shapiro steps only at even multiples of the driving frequency $\hbar\omega_{AC}/2e=h f_{AC}/2e$ owing to the change of periodicity. (c)The peak at $\omega=\omega_J$ in the spectrum of ac Josephson radiation in the trivial case (blue) moves to $\omega=\omega_J/2$ in the topological case (red) and becomes narrower. (d) A SNS junction incorporated into a ring geometry is free of external Majoranas. The periodicity of the current as a function of the external flux threading the ring $\Phi$ reflects wether the junction is trivial ($\Phi_0=h/2e$ periodicity) or topological  ($2\Phi_0=h/e$ periodicity).}
\label{fig:16}
\end{figure}

The fractional Josephson effect may be recovered by biasing the junction with a voltage $V$ which makes the phase difference  time dependent $\phi(t)=\frac{2eVt}{\hbar}=\omega_Jt$, where $\omega_J$ is the Josephson frequency. This time-dependence allows to sweep the phase fast enough through the anticrossing such that Landau-Zener processes induce nonadiabatic transitions (green arrow in Fig. \ref{fig:15}) and thus restore $4\pi$-periodicity \cite{SanJose2012,Pikulin2012,Virtanen2013}. However, similar Landau-Zener processes can induce transitions to higher excited states (red arrow Fig. \ref{fig:15}) which results in an unavoidable $2\pi$ periodicity in the steady state.  Owing to the above realistic considerations one has to devise alternative experiments, beyond direct measurements of the current-phase relationship (see Fig. \ref{fig:14}b), to probe this remarkable fractional Josephson effect.

One option is to study the dc response of a voltage biased SNS junction containing Majoranas. In particular, the dissipative multiple Andreev reflection current $I_{dc}$ exhibits unique features that can be used to detect the topological transition (gap inversion) and the formation of Majorana bound states. Notably, the current in the tunneling limit, where only single Andreev reflections contribute, can be directly used to detect Majoranas through a halving of the standard steps at $eV=\pm 2\Delta$ of conventional junctions that become $eV=\pm\Delta$ in the topological regime (Fig. \ref{fig:16}a) \cite{Houzet1,SanJose2013,Zazunov2016}. This halving results from the alignment of the BCS singularity of the lead with a Majorana zero mode right in the middle of the superconducting gap (instead of the standard alignment at $eV=\pm 2\Delta$ of two BCS singularities). Interestingly, the maximum conductance jump at $eV=\pm\Delta$ is universal and has a value $\frac{dI}{dV}=(4-\pi)\frac{2e^2}{h}$ \cite{Peng2015}. An obvious advantage over the quantized conductance for normal leads, Eq. (\ref{quantizedG}), is that thermal broadening is exponentially suppressed by the superconducting gap. However, when the junction transparency deviates from the tunnel regime there is a finite conductance for $e|V|<\Delta$, arising from multiple Andreev reflections \cite{SanJose2013}, which is no longer quantized \cite{Setiawan2017}.

Concerning dynamics, one option is to drive the junction by a current with both dc and ac components $I(t)=I_{DC}+I_{AC}cos(\omega_{AC}t)$. When this high frequency current is applied to the junction, the ac Josephson current phase-locks to the external drive giving rise to regions of constant voltage in the I-V characteristics. In a conventional junction, the average voltage develops plateaus (known as Shapiro steps) at integer multiples of $\frac{\hbar\omega_{AC}}{2e}$, namely at integer multiples of the Josephson frequency $\omega_J=\frac{2eV}{\hbar}=n\omega_{AC}$. In the topological case, the $4\pi$-periodic current-phase relation has a component at half the Josephson frequency $\omega=\omega_J/2$, which results in a doubling of the Shapiro steps \cite{Dominguez2012,Dominguez2017,Sau2017,Pico2017} as compared to a conventional junction with $\omega=\omega_J$. Thus only even multiples of  $\hbar\omega_{AC}/2e$ are expected (or, equivalently, missing odd steps as the junction transitions from trivial to topological, see Fig. \ref{fig:16}b). While several mechanisms (higher harmonics, non-linearities, capacitance effects, etc) can lead to subharmonic Shapiro steps, only a component at half the Josephson frequency can be responsible for the doubling of the Shapiro steps.

Another option is to directly detect the noise spectrum \cite{Blanter-Buttiker,Marcos2011} of the emitted high-frequency radiation (Fig. \ref{fig:16}c), which is expected to show a narrow peak at $\omega=\omega_J/2$ instead of the standard peak at $\omega=\omega_J$ \cite{SanJose2012,Pikulin2012,Zazunov2016,Houzet2}. This could be accomplished by e.g. direct on-chip detection of the frequency-resolved spectral density of the Majorana junction by a second Josephson junction which is used as a detector \cite{Deblock2003,Geresdi2017}. Using this setup, frequency-sensitive measurements can be achieved by analysing the photon-assisted tunneling current of the detector junction which is determined by the spectral density of the coupled microwave radiation \cite{Aguado2000}.

Finally, we mention the possibility of incorporating the SNS Josephson junction into a ring geometry. If the ring is made of a conventional superconductor, the current flowing through the junction should exhibit $h/2e$ periodicity as a function of the external flux $\Phi$ (since the superconducting phase across the junction and the external flux threading the ring are related by the condition $\phi=-2e\Phi/\hbar=-2\pi\Phi/\Phi_0$, where $\Phi_0=h/2e$ is the flux quantum associated with the transfer of Cooper pairs). In a topological superconductor ring the current is $2\Phi_0=h/e$ periodic \footnote{Note that while this is the same $\phi_0$-periodicity as the persistent current through a \emph{normal} ring, this is still a supercurrent that does not decay with the length of the ring $L$, as opposed to a persistent current that typically decays as $\sim1/L$.}, which corresponds to $4\pi$ periodicity in $\phi$  \cite{Sau2012,Pientka2013,Dmytruk2016}. Although the ring geometry has obvious advantages over a standard SNS geometry (since it is free from additional Majoranas, see Fig. \ref{fig:16}d), it still is prone to quasiparticle poissoning and reverts to $h/2e$ periodicity when fermion parity is not conserved. Remarkably, if the Majoranas overlap through both the weak link and the interior of the superconducting ring, a situation that one would encounter in mesoscopic rings with $L\sim\xi_M$, the $\phi_0=h/e$ periodicity is recovered even in the absence of fermion parity conservation \cite{Sau2012,Pientka2013}.

The ring geometry has another potential advantage since one can include a small ac component to the flux $\Phi(t)=\Phi_{DC}+\Phi_{AC}cos(\omega t)$ and study the finite-frequency dynamics of the junction, without driving it out of equilibrium \footnote{Note that this is impossible under a voltage bias, which leads to non-trivial dynamics (a phase linearly increasing with time $\phi(t)=\phi_0+\frac{2eV}{\hbar}t$ which generates Landau-Zener transitions), as opposed to small (linear response) periodic driving of the flux which corresponds to phase oscillations of controlled amplitude $\phi_{AC}$ around a fixed value $\phi_{0}$, $\phi(t)=\phi_0+\phi_{AC}cos(\omega t)$. For a discussion about the two different biasing schemes see e.g. Ref. \cite{Virtanen2013}.}. The key is to consider the linear-response magnetic susceptibility  $\chi(\omega)= \partial I / \partial \Phi_{AC}=\chi^{I}(\omega)+i\chi^{II}(\omega)$, whose finite-frequency properties allow to extract information about the dynamics of the Andreev states in the junction. At low frequencies the response is purely inductive, and the dc susceptibility defines the Josephson inductance $L(\phi)$:
\begin{equation}
\frac{1}{L(\phi)}=\chi(\omega\rightarrow 0)=-\frac{2\pi}{\Phi_0}\frac{\partial I (\phi)}{\partial \phi},
\end{equation}
where $I (\phi)=\sum_n  f_n I_n(\phi)=-\frac{2e}{\hbar}\sum_n  f_n\frac{\partial\varepsilon_n(\phi)}{\partial\phi}$ is the Josephson current of the junction written in terms of individual contributions $I_n(\phi)$ of each Andreev level $\varepsilon_n(\phi)$ ($f_n$ are occupation factors). At finite frequencies, two physical processes, relaxation of Andreev level populations and transitions between different levels, contribute to the dynamical response.
Both physical contributions can be clearly identified by writing $\chi(\omega)$ in the basis of Andreev levels using the Kubo formula \cite{Dassoneville2013,Ferrier2013} \footnote{These ideas date back to the physics of persistent currents in mesoscopic normal rings, see e. g. Ref. \cite{Buttiker,Trivedi88}.}. At low frequencies the largest contribution comes from diagonal terms of the form
\begin{equation}
\label{diagonal}
\chi_D(\omega)=\frac{-i\omega\tau_{in}}{1-i\omega\tau_{in}}\sum_n I^2_n(\phi)\frac{\partial f_n}{\partial\varepsilon_n},
\end{equation}
which describes thermal relaxation of the Andreev level populations with a characteristic inelastic time $\tau_{in}$ \cite{Dassoneville2013,Ferrier2013}. At high enough frequencies, transitions between Andreev levels become important and the non-diagonal terms of the susceptibility govern microwave absorption \cite{Zazunov2016,Dmytruk2016,Tewari2012,Kos2013,Vayrynen2015,Pientka2016}. 
The physical meaning of these quantities can be easily understood by recalling the relation between the susceptibility and the impedance of the junction $-e\chi(\omega)=i\omega Y(\omega)$. For example, the dissipative part of the susceptibility is related to the ac conductance $G(\omega)=ReY(\omega)$ and the noise spectrum $S_I(\omega)$ via the fluctuation-dissipation theorem:
\begin{equation}
\label{FDT}
S_I(\omega)=4k_BTG(\omega)=4k_BT\frac{\chi^{II}(\omega)}{\omega}=4k_BT\sum_n I^2_n(\phi)\frac{\partial f_n}{\partial\varepsilon_n}[\frac{\tau_{in}}{1+(\omega\tau_{in})^2}].
\end{equation}

While the above ideas have been explored in the context of mesoscopic superconductivity \footnote{For example, Eq. (\ref{FDT}) has been used in the past in the context of ballistic superconducting quantum point contacts \cite{Martin-Rodero96,Averin96}. Its relevance in the context of quasiparticle poissoning in $4\pi$ topological Josephson junctions based on quantum spin hall insulators is briefly discussed in Ref. \cite{Fu-Kane2}.}, their potential for studying dynamics of topological Josephson junctions remains relatively unexplored (exceptions are Refs. \cite{Zazunov2016,Tewari2012,Vayrynen2015,Pientka2016} that study microwave absorption and Refs. \cite{Dmytruk2016,Murani2016,Dmytruk2017} that focus on the different contributions to $\chi(\omega)$). 

\subsection{Coulomb blockade effect in Majorana Islands \label{CB}}
For short enough proximitized nanowires in a "floating" geometry (nanowire isolated from ground by a capacitor), the interplay between superconductivity and Coulomb blockade allows to study in detail deviations from perfect ground state degeneracy in the form of splittings of near-zero energy overlapping Majoranas. The idea is very simple: following basic ideas from Coulomb blockade theory, the electrostatic energy of a normal nanowire segment (isolated from normal-metal leads by tunnel barriers and with effective capacitance $C$) varies with gate voltage as $E_C(N -N_g)^2$, where $N$ is the number of excess electrons in the island, $E_C=e^2/2C$ is the charging energy and $N_g = CV_g/e$ is the gate-induced charge in dimensionless units. At low temperatures $T<<E_C$ and for low bias voltages, the system is in the Coulomb blockade regime since charge fluctuations are strongly suppressed and transport through the island is blocked by the large charging energy. Current flow is only possible at special degeneracy points which correspond to gate voltages where the energy of having $N$ and $N+1$ electrons become degenerate. In the presence of superconductivity the main principle of Coulomb blockade still applies with the sole difference that the energy for having $N$ electrons depends on fermionic parity: if $N$ is even all quasiparticles pair up into Cooper pairs at zero energy while an odd ground state needs to accommodate an extra quasiparticle at energy cost $E_0$. This parity-dependent energy can be concisely written as:
\begin{equation}
E_N=E_C(N -N_g)^2+p_NE_0,
\end{equation}
with $p_N=0$ for even $N$ and $p_N=1$ for odd $N$. This superconducting even-odd Coulomb blockade effect has been demonstrated in small mesoscopic superconducting islands since the early 90s \cite{Eiles93,Tuominen92,Lafarge93}. The ratio $E_0/E_C$ determines the different transport regimes. For $E_0>E_C$, the ground state always has even parity for any $V_g$ and transport occurs via tunnelling of Cooper pairs at degeneracies of the even-N parabolas, namely when $E_N= E_{N + 2}$ at points where $N_g$ is an odd integer, resulting in a $2e$-periodic Coulomb-blockade effect. Conversely, if $E_0<E_C$ the crossings around integer values of $N_g$ now support an odd ground state, which results in the splitting of the $2e$-periodic Coulomb diamonds, namely alternation of even and odd diamonds. In this regime, the Coulomb-peak spacing is proportional to $E_C+2E_0$ for even diamonds and $E_C-2E_0$ for odd diamonds \cite{Tuominen92,Lafarge93}.

\begin{figure}[!tt]
\centering
\includegraphics[width=\columnwidth]{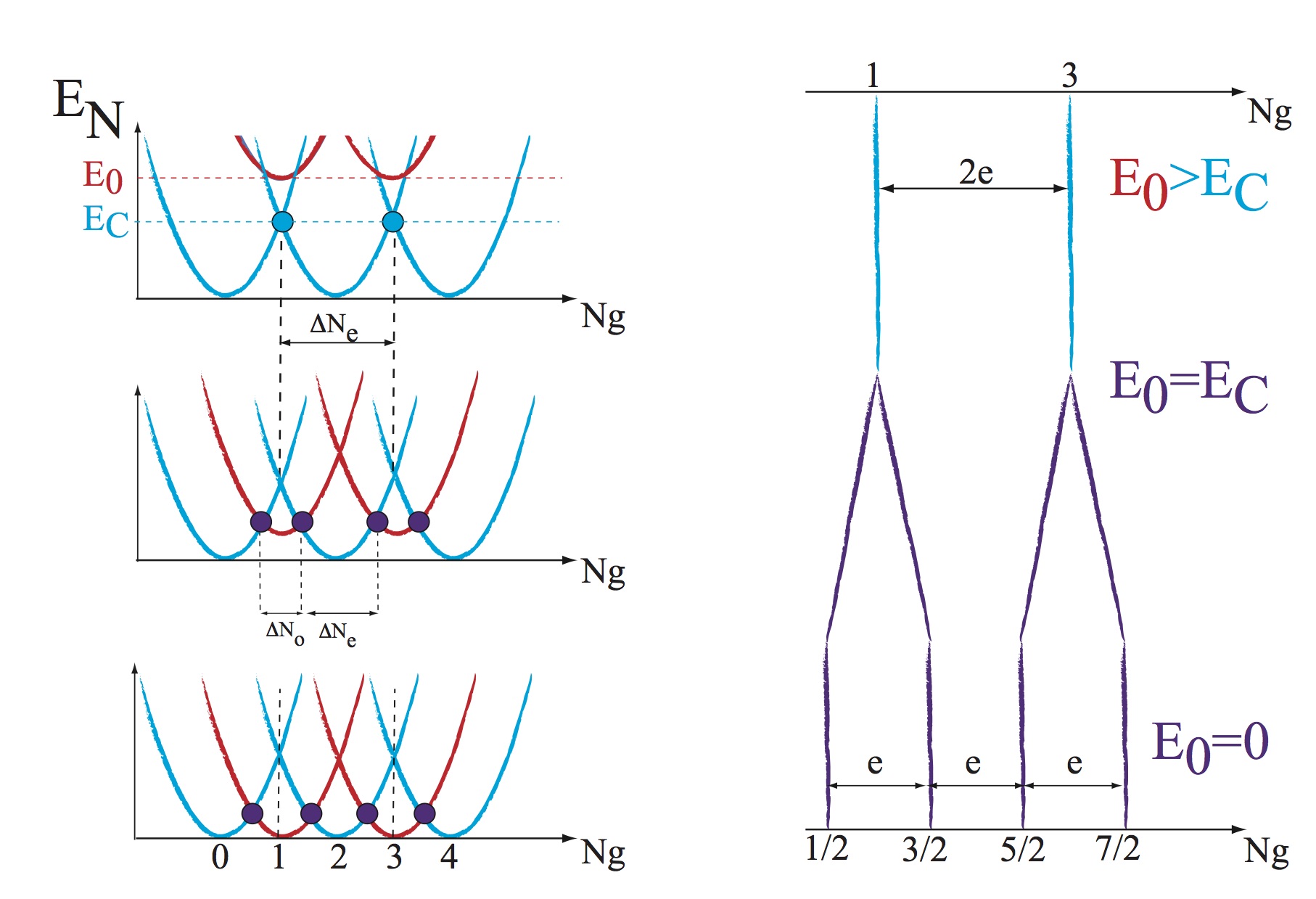}\\
\caption{Coulomb Blockade spectroscopy in Majorana islands. When the energy to accommodate an extra quasiparticle is larger than the charging energy $E_0>E_C$ (left panel), Coulomb peaks (right panel) appear at degeneracies between charge states differing by $2e$. As $E_0$ gets reduced below $E_C$, odd ground states are possible around the crossings at integer values of $N_g$. This splits the $2e$-periodic Coulomb peaks. In a Majorana island, $E_0$ can be reduced by an external magnetic field $E_0=\Delta(B)$ down to $E_0=0$ such that the peak spacing becomes $1e$-periodic at points where $N_g$ is a half-integer, signaling the presence of zero energy Majorana quasiparticles.}
\label{fig:17}
\end{figure}

Despite being a very well known phenomenon, this even-odd effect acquires a new twist in the context of Majorana islands: at zero magnetic field, the minimal energy cost for having an extra quasiparticle is just the superconducting gap $E_0=\Delta$.  Interestingly, the application of a magnetic field allows to reduce the gap such that the ratio $\Delta(B)/E_C$ can be \emph{tuned down to zero} when we reach the topological transition point and the gap closes. After the transition, a quasiparticle excitation at zero energy (the Majorana) is possible and we expect that the peak spacing should become $1e$-periodic, at points where $N_g$ is a half-integer, like in a normal island \cite{Fu2010}, see Fig. \ref{fig:17}. Deviations from true zero energy Majoranas are expected since the island has a finite length: as we already discussed, overlapping Majoranas give rise to a finite energy splitting that distinguishes different fermionic parities of the corresponding Bogoliubov excitation. Thus, by measuring the peak spacing between alternate even and odd diamonds as a function of magnetic field one can directly map the expected oscillatory behaviour $E_0(B)$ of the Majorana splitting. Furthermore, these oscillating energy splittings should be exponentially suppressed with wire length, which allows to measure $\xi_M$, see Eq. (\ref{Majorana-overlap}). 

The expected Coulomb blockade peaks have a maximum conductance of $G=e^2/h$, as opposed to the universal conductance value of $G=2e^2/h$ that we found for noninteracting wires. The full crossover from the noninteracting value $G=2e^2/h$  to the Coulomb blockade regime $G=e^2/h$ has been described in Ref. \cite{hutzen2012}. The peak conductance in the Coulomb blockade regime occurs when two degenerate island charge states are accessible and the model effectively maps to a spinless resonant tunneling problem.  Since this level represents the delocalized Bogoliubov excitation formed by the two Majoranas that are far apart, this resonant process is dubbed in some articles "electron teleportation"  \cite{Fu2010,Semenoff2007}, in a somewhat abuse of terminology. Here it is important to point out that, in order to observe nonlocal transport, Coulomb charging is an essential ingredient. Without charging effects, nonlocal transport is completely suppressed in favor of local Andreev reflection for true zero modes, such that currents through
the left and right leads are completely uncorrelated \cite{Bolech2007}. In order to have nonlocal current correlations it is essential that the Majorana splitting exceeds the tunneling broadening as discussed in Refs. \cite{Nilsson2008,Tewari2008,Liu2013}.
It is also worth mentioning that entanglement can be generated quite efficiently in these Majorana islands as a consequence of elastic cotunneling \cite{Plugge2015}, i. e. when the system is gated in a Coulomb blockade valley region. If the superconducting island contains more than two Majoranas, nonperturbative corrections to the above mechanism generates a so-called topological Kondo effect, as discussed in Refs.   \cite{Beri-Cooper2012,Altland-Egger2013,Beri2013,Altland-Egger2014,Zazunov2014}.

Interestingly, the transmission phase shift of the nonlocal resonant process described above is predicted to depend on fermionic parity (with a phase shift change by $\pi$ signalling a change in fermionic parity). These phase shift changes form the basis of various ingenious schemes for parity readout. The main idea behind these schemes is to use two-path electron interferometry, with one of the paths involving a Majorana island and the other path serving as a reference, and measure conductance. If transport through the interferometer is phase coherent, the conductance picks up a flux-dependent contribution $f(\Phi)$ which depends periodically, with $h/e$ periodicity, on the external magnetic flux $\Phi$ enclosed by the two interfering paths and that also depends on the parity of the island $G(\Phi)\sim i\gamma_1\gamma_2 f(\Phi)$  \cite{Vijay2016}. Similar schemes use the persistent current around a closed loop \cite{Vijay2016,Jacquod2013}, in a way reminiscent of the ring geometries discussed in the preceding section. Beyond parity readout, the above schemes hold promise for topological quantum computation since the unitary transformation that implements braiding operations can be realized by performing a sequence
of projective measurements of Majorana bilinear operators, namely \emph{without performing physical braiding of Majoranas} \cite{Bonderson2008}. This has obvious advantages over previous schemes which rely on detecting non-Abelian statistics and performing braiding by moving Majoranas around each other in complicated geometries, such as T-junctions \cite{Alicea-braiding,Aasen2016}. Using these ideas various recent papers have already demonstrated the possibility of performing measurement-based  topological quantum computation \cite{Karzig2016,Plugge2017}. 

\begin{figure}[!tt]
\centering
\includegraphics[width=1\columnwidth]{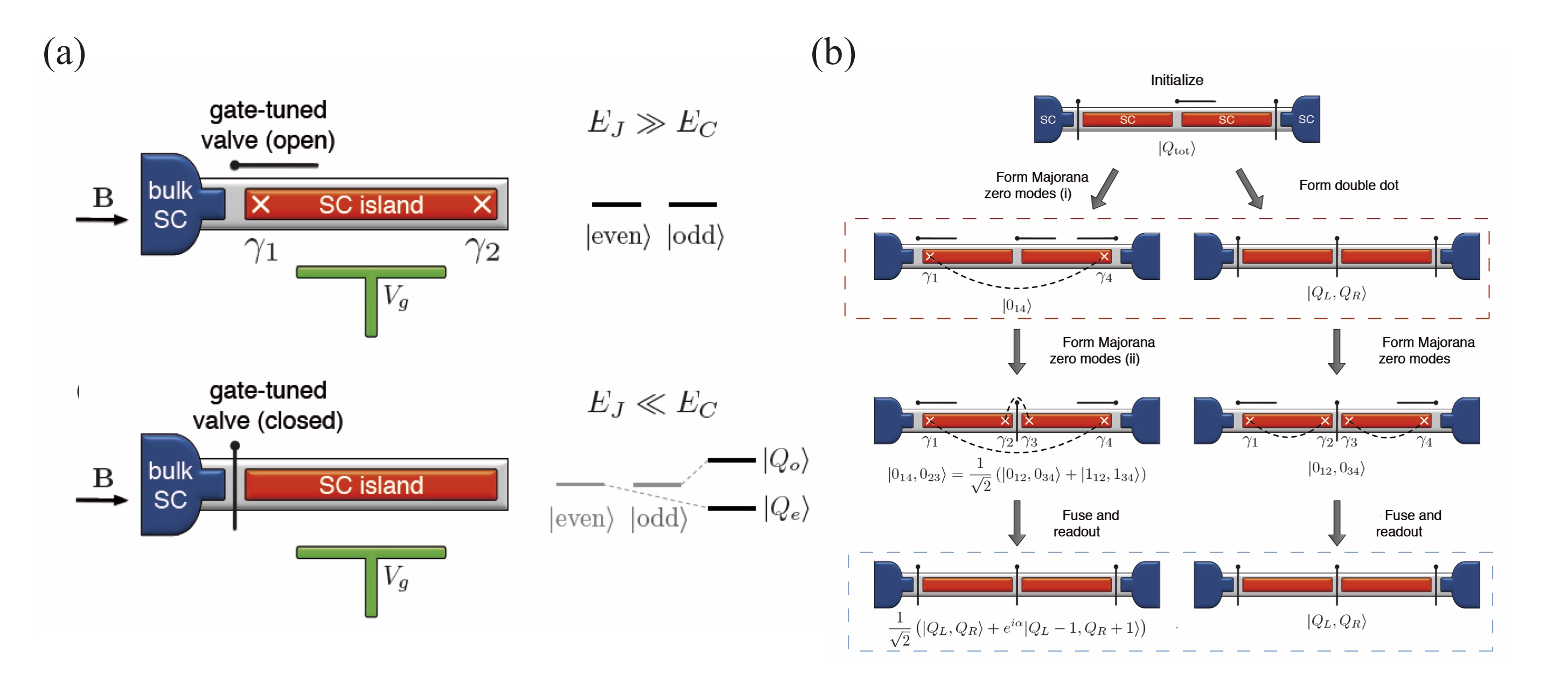}
\caption{Checking fusion rules with Majorana islands. (a) Superconducting island coupled to a bulk superconductor. A gate-tunable coupling (valve) controls the ratio of Josephson energy $E_J$ to the island charging energy $E_C$. This allows to tune the system from the regime $E_J\gg E_C$  (degenerate ground state for even and odd fermion parities, top) to the opposite case $E_J\ll E_C$ (non-degenerate states with island charges $Q_e$ and $Q_o$, bottom). (b) Protocol to detect the nontrivial Majorana fusion rules via the parity-to-charge scheme. Sequences of open and closed valves that differ only by one intermediate state (red dashed box) result in different final charge states (blue dashed box). Adapted from Ref. \cite{Aasen2016}.}
\label{fig:18}
\end{figure}

Another route to demonstrate the ground state degeneracy in the topological phase is to perform fusion experiments. As we mentioned before, this is an interesting option that also avoids performing complicated Majorana exchanges (and thus easier than braiding). In order to check the nontrivial fusion rules of Majoranas, Aasen {\it et al} proposed in Ref. \cite{Aasen2016} to use Majorana islands coupled to a bulk superconductor. Using gate-tunable couplings between the island and the superconductor, it is possible to modify the ratio of Josephson energy $E_J$ to the island charging energy $E_C$, since charging effects can be quenched by making the contact to the bulk superconductor more transparent. As a result, the ground state degeneracy for even and odd fermion parities in the Majorana island is preserved when the system is tuned in the regime $E_J\gg E_C$, whereas the opposite limit $E_J\ll E_C$ restores charging effects and converts these parity eigenstates into non-degenerate states with island charges $Q_e$ and $Q_o$, see figure \ref{fig:18}a. This parity-to-charge conversion can be used in systems of two coupled islands to perform specific protocols that allow one to check Majorana fusion rules by tuning on and off the couplings (which, as suggested by the authors, can be visualised as "opening"  and "closing" valves). In particular, on and off sequences that differ only by one intermediate state result in different final charge states, see figure \ref{fig:18}b, owing to nontrivial fusion rules of Majoranas.

\section{Experimental progress in induced one-dimensional topological superconductivity \label{progress}}

\begin{figure}[!tt]
\centering
\includegraphics[width=1\columnwidth]{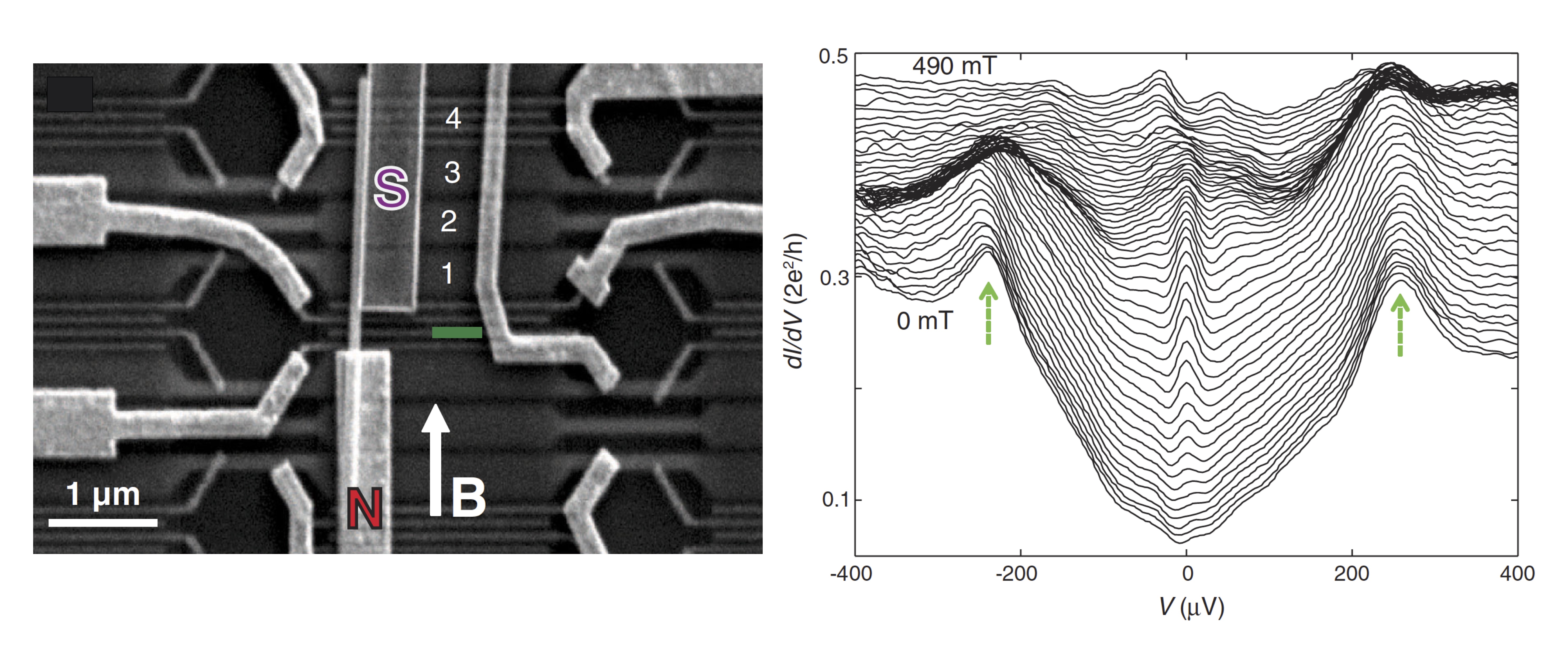}
\caption{Delft experiment. Left: Scanning electron microscope image of the device with normal (N)
and superconducting (S) contacts covering an InSb nanowire. The underlying gates, numbered 1 to 4, are used to deplete the wire. The tunnel barrier between normal and superconducting contacts is marked by a green line. Right: $dI/dV$ versus $V$
taken at different B fields along the nanowire axis (from 0 to 490 mT in 10 mT steps; traces are offset for clarity, except for the lowest trace at B = 0). For B fields between 100 and 400 mT, a clear ZBA at zero voltage emerges in the middle of the superconducting gap. Green arrows indicate the induced gap peaks. Adapted from Ref. \cite{Mourik2012}.}
\label{fig:19}
\end{figure}
\subsection{Semiconducting Nanowires I: the first generation}
The first evidence of emerging ZBAs in tunneling spectroscopy through proximitized nanowires was reported in 2012 by the Kouwenhoven group in Delft \cite{Mourik2012}. An example of this experiment is shown in Fig. \ref{fig:19}. The top panel shows a scanning electron microscope image of the device fabricated with an InSb nanowire contacted with a normal (N) gold electrode and a superconducting (S) electrode. An external magnetic field is applied parallel to the nanowire while different voltages (applied in the underlying gates, numbered 1 to 4 in the figure) vary the electron density and define a tunnel barrier between normal and superconducting contacts (green line). Since InSb is a semiconductor with strong spin-orbit coupling and the superconductor (niobium titanium nitride) has a high critical field (large magnetic fields do not destroy superconductivity), this nanowire device is an experimental implementation of the above theoretical proposals in subsection \ref{Rashba}. Typical estimated parameters for this device are a Rashba coupling $\alpha\approx 0.2 eV\AA$, a large $g$-factor $g\approx 50$ (which gives $E_Z/B\approx 1.5meV/T$) and an induced superconducting gap $\Delta\approx 250\mu eV$. According to the Lutchyn and Oreg models \cite{Lutchyn2010,Oreg2010}, we expect that the proximitized nanowire will enter a topological phase for magnetic fields $B\sim 0.15T$ where the Zeeman energy $E_Z$ exceeds the induced gap $\Delta$ (c. f. Eq. (\ref{criticalZeeman}) assuming $\mu=0$).

The bottom panel shows conductance data where different curves represent $dI/dV$ for increasing B fields from 0  (bottom trace) to 490 mT (top trace), in 10mT steps. At $B=0$, bottom curve, the two peaks at $V\approx \pm 250\mu eV$ (green arrows)correspond to the BCS peaks in the density of states of the nanowire, demonstrating proximity-induced superconductivity. For B fields between 100 and 400 mT along the nanowire axis, a clear ZBA at zero voltage emerges in the middle of the superconducting gap. This emergent ZBA is consistent with the existence of zero-energy Majorana bound states in the nanowire. Subsequent experiments \cite{Deng2012,Das2012,Finck2013,Churchill2013} showed similar phenomenology, at least superficially. 

Despite the initial excitement, the experiments on signatures of Majoranas in nanowires were hotly discussed and their interpretation was in many cases challenged. Importantly, many features of the Delft experiment (and the others) did not agree with the predicted behaviour (compare Fig. \ref{fig:19} bottom with the theoretical expectation in Fig. \ref{fig:12}b). The main discrepancies with the predictions based on the Lutchyn and Oreg models \cite{Lutchyn2010,Oreg2010} are as follows: First, the zero bias peak in all the experiments is much lower than the predicted quantized conductance at $G_0=2e^2/h$. In the Delft experiment, for example, the peak has a very small height of $G\sim 0.05G_0$. Second, as we have discussed before, the topological transition is characterised by a gap inversion (closing and reopening of the excitation gap). Such gap inversion is never seen in the data. Third, a Rashba coupling $\alpha\approx 0.2 eV\AA$ yields a spin-orbit length $l_{SO}\approx 200nm$ which is not negligible when compared with the finite-length section of the nanowire proximitized by the superconductor.  This should result in a finite overlap between Majoranas which gives rise to finite energy oscillations around zero energy (see Eq. \ref{Majorana-overlap}). However, no oscillations are seen in the experiment.
The first two deviations from the ideal expectations can be easily understood by taking into account multiple subband occupation in the nanowire combined with the possibility of smooth (finite width and height) tunneling barriers between the normal contact and the nanowire, as demonstated by Prada {\it et al} in Ref. \cite{Prada2012}:  if multiple subbands are filled, their coupling to a smooth barrier is not constant since the lowermost (topmost) band will have the strongest (weakest) tunnel coupling owing to its large (small) momentum. Since the Majoranas originate from the topmost occupied subband (which is the first one that becomes topological  for fixed $\mu$), they are probed deep in the tunneling regime and their signal is masked by the (stronger) contribution from lower bands which give a large background conductance \footnote{Multiple subbands in a quasi-one dimensional nanowire may simultaneously fulfill the topological condition. In such case, two different regimes are relevant: 1) the strength of spin orbit interaction $E_{SO}$ is small as compared to the typical subband spacing (equivalently, the nanowire width is much smaller than the spin orbit length $W<<l_{SO}$). In this case, interband coupling is very weak and the subbands are almost decoupled, a regime formally known as topological BDI class. As a consequence, Majoranas in different subbands are uncoupled, and may coexist in the system. 2) For large $E_{SO}$ as compared to the typical subband spacing $W>>l_{SO}$, interband coupling becomes strong and Majoranas are coupled. This regime is formally known as topological D class. In this regime, systems with an even number of subbands are trivial (all Majoranas are paired) while systems with an odd number of subbands are non-trivial (some Majoranas remain unpaired). For a full discussion see e. g. Ref. \cite{Stanescu-review}.}. Similar considerations where discussed in Refs. \cite{Rainis2013,Pientka2012}. 

In order to understand the absence of Majorana oscillations in the data, one could assume stronger SO couplings. For example, assuming that the SO energy is several hundreds of $\mu eV$, roughly one order of magnitude larger than the quoted value in Ref. \cite{Mourik2012}, one can simulate Majorana peaks with almost no oscillatory behaviour \cite{Prada2012,Rainis2013}. However, the exact experimental value of $E_{SO}$ is still unknown.
Moreover, for strong SO coupling the Majorana localisation length is no longer governed by the $l_{SO}$ length but rather by the ratio $\xi_M\sim \alpha/\Delta$, which implies that wires of length $L\gtrsim 1\mu m$ are needed (see the discussion after Eq. \ref{Majorana-overlap}). Electronic interactions, both intrinsic and extrinsic, also suppress Majorana oscillations. In the first case, the Coulomb repulsion inside the nanowire tends to counteract changes in electron density induced by increasing the Zeeman field. In the limit of very strong Coulomb repulsion the system is completely incompressible and prefers to remain with a constant electron density  such that $\mu$ adjusts itself upon increasing $B$. This 
gives rise to an almost complete suppression of Majorana oscillations \cite{DasSarma2012}. Recently, it has also been pointed out that extrinsic interactions with bound charges in the dielectric surroundings may completely suppress Majorana hybridisation around parity crossings and give rise to zero-energy pinning \cite{Dominguez2017b}. In particular, the zero-energy pinning mechanism is the result of the electrostatic energy cost from the interaction of the finite charge added to the nanowire at parity crossings
\begin{equation}
Q_n=e\int_0^L dx\left\{f(\epsilon_n)|u_n(x)|^2+[1-f(\epsilon_n)]|v_n(x)|^2\right\},
\end{equation}
where $f(\epsilon_n)$ is its occupation probability,
and image charges in the dielectric environment. This finite charge at parity crossings is due to deviations from the strict Majorana condition (perfect particle-hole symmetry at zero energy) that give rise to a finite energy charged Bogoliubov quasiparticle \footnote{These parity jumps are accompanied by charge-density jumps $\delta N$ in the nanowire that could be measured, by e.g. a single-electron transistor, to extract information about Majorana overlaps \cite{Lin2012,Ben-Shach2015}.}. 

Another important problem in the data is the large residual quasiparticle density of states, very different from the ideal BCS density of states, which results in a very soft gap (finite conductance below the superconducting gap) and very broad coherence peaks. This residual quasiparticle of density of states implies that alternative mechanisms giving rise to ZBAs should be seriously considered. 

One of such alternative mechanisms is disorder which creates subgap states at $E_Z=0$ that cluster around zero energy as Zeeman increases, thus mimicking Majorana behaviour \cite{Pientka2012,Liu2012,Pikulin2012b,Bagrets2012}. One important argument against the disorder interpretation is that disorder is random such that disorder-induced ZBAs are expected to be fragile against changes in model parameters. This is in contrast with the experimental observations, where ZBAs are robust against both gate voltage and magnetic field.
\begin{figure}[!tt]
\centering
\includegraphics[width=\columnwidth]{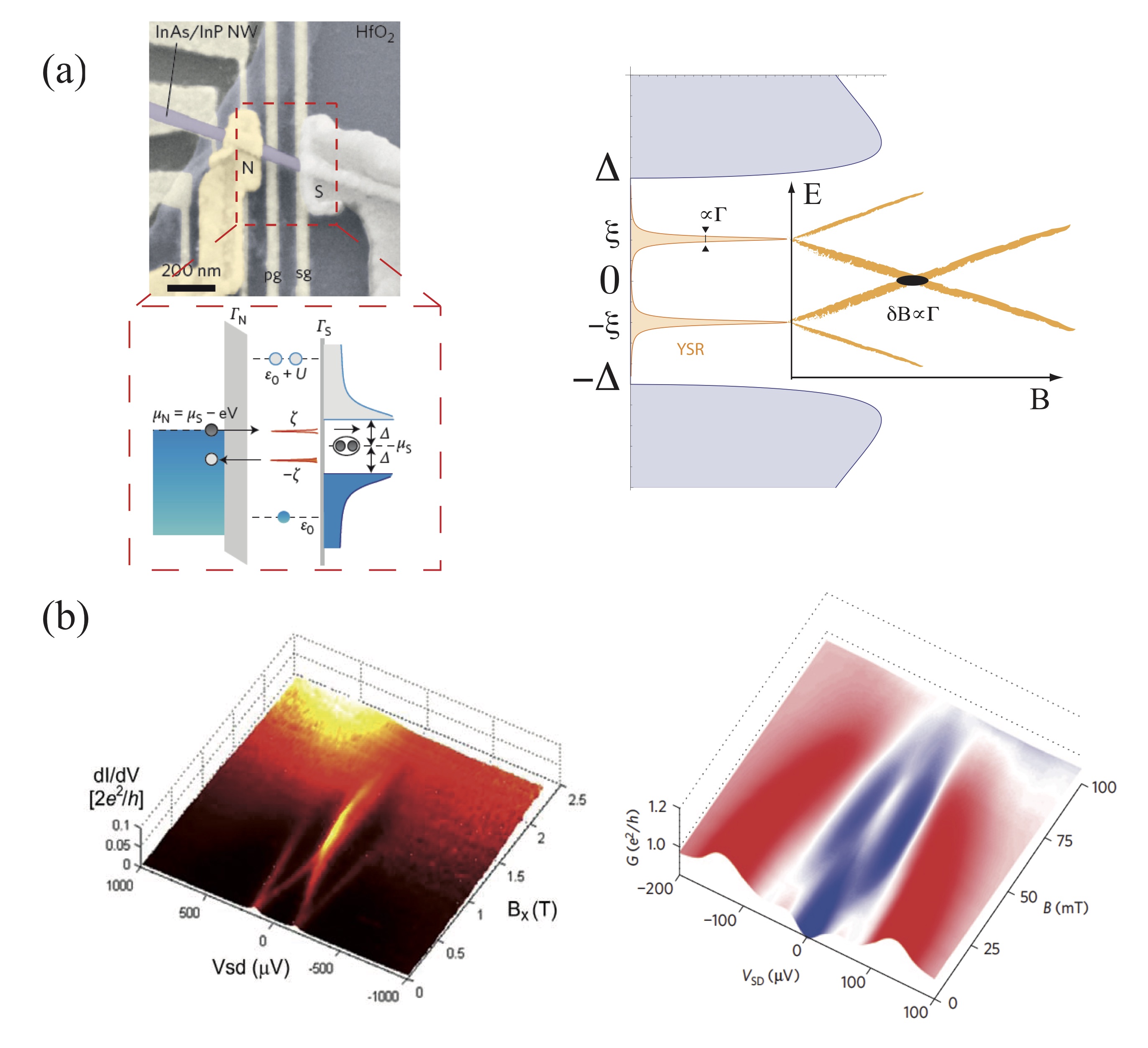}\\
\caption{(a) Left Panel, top: Scanning electron micrograph of an InAs/InP core/shell nanowire contacted with normal and superconducting leads. A quantum dot is (QD) naturally formed in the nanowire section ($\sim 200nm$) between the S and N contacts. Left Panel, bottom: schematics of the resulting N-QD-S device with asymmetric tunnel couplings to the normal metal and superconductor leads, $\Gamma_N$ and $\Gamma_S$, respectively. $\Delta$ is the superconducting gap, $U$ is the charging energy, $\mu_{N(S)}$ is the chemical potential of the N(S) lead and $\varepsilon_0$ is the QD energy level relative to $\mu_S$. The subgap peaks located at $\pm \xi$ represent Andreev levels, which in this case are Yu-Shiba-Rusinov subgap states since they carry spin. In tunnel spectroscopy $dI/dV$ measurements the alignment of $\mu_N$ to a subgap level yields a peak in the differential conductance. This process involves an Andreev reflection at the QD-S interface, which transports a Cooper pair to the S lead and reflects a hole to the N contact. Adapted from Ref. \cite{Lee2014}. Right panel: Schematic density of states in the quantum dot, showing the induced superconducting gap $\Delta$ with a Bogoliubov pair of subgap Yu-Shiba-Rusinov bound states $\xi$. Right: since the Yu-Shiba-Rusinov states are spinful, they split under an applied magnetic field and may cross may cross zero energy signalling a singlet-doublet transition. This results in a zero bias peak that extends over a magnetic field range $\delta B$ of the order of to the  broadening $\Gamma$ of the levels. (b) $dI/dV$ versus magnetic field plots from the experiments in Ref. \cite{Lee2014} where the Zeeman splitting of Yu-Shiba-Rusinov excitations is invoked to explain the data versus data from Ref. \cite{Das2012}, where a Majorana explanation is invoked. Both mechanisms are indistinguishable in practice.}
\label{fig:20}
\end{figure}
A second alternative is the Kondo effect. This effect can be very relevant if the normal-superconducting junction contains a quantum dot that forms, even unintentionally, when gating the nanowire. In this case, the superconducting proximity effect competes with Coulomb blockade, which comes from the electrostatic repulsion among the electrons of the QD, and ultimately with the Kondo effect. As a result of this competition, the system undergoes a quantum phase transition when the superconducting gap $\Delta$ is of the order of the Kondo temperature $T_K$ \cite{Lee2016}. The underlying physics behind such transition ultimately relies on the physics of the Anderson model where the standard metallic host is replaced by a superconducting one, namely the physics of a (quantum) magnetic impurity in a superconductor. A characteristic feature of this hybrid system is the emergence of subgap bound states, the so-called Yu-Shiba-Rusinov (YSR) states, which cross zero energy across the quantum phase transition, signaling a switching of the fermion parity and spin (doublet or singlet) of the ground state. Interestingly, in parameter regions where superconductivity is expected to suppress the Kondo effect, namely $\Delta (B=0)>T_K$, an external magnetic field may decrease the gap (owing to depairing) leading to a magnetic-field induced Kondo enhancement $\Delta (B\neq 0)\lesssim T_K$ which results in ZBAs \cite{Lee2012}. Furthermore, the presence of an extra normal contact, usually considered as a non-perturbing tunnelling weak probe, leads to nontrivial Kondo enhancement \cite{Zitko2015}. Nevertheless, by studying the Zeeman-splitting of the Kondo peak (with should be very large for large $g$-factors) one should be able to distinguish Kondo peaks from Majorana peaks.

Another important aspect related to YSR states is that their Zeeman splitting may in some cases mimic Majorana states. Again, it all boils down to the physics of parity crossings. In this case, YSR excitations with \emph{doublet} character \footnote{These occur in regions of the phase diagram where the ground state is a singlet since quasiparticle YSR excitations have always the opposite fermionic parity.} become Zeeman-split upon applying a magnetic field and may cross zero energy (Fig. \ref{fig:20}a). This parity crossing results in a ZBA anomaly that remains pinned to zero energy for a Zeeman field range of the order of the broadening of the YSR, $\delta B\sim\Gamma$. This, together with a strong anticrossing from the gap closing for increasing B-fields, results in ZBAs for a finite range of Zeeman, since the anticrossing with the continuum prevents the YSR excitations moving away from zero energy, thus mimicking Majoranas \cite{Lee2014}. We believe that this mechanism could falsify many of the Majorana claims in the literature, particularly in systems where quantum dot formation is rather clear \cite{Das2012,Finck2013}. Fig. \ref{fig:20}b shows $dI/dV$ versus magnetic field plots from the experiments in Ref. \cite{Lee2014}, where the Zeeman splitting of YSR excitations is invoked to explain the experiments, versus data from Ref. \cite{Das2012}, where a Majorana explanation is invoked. Both sets of data are remarkably similar and, in many aspects, both mechanisms are indistinguishable in practice.
\begin{figure}[!tt]
\centering
\includegraphics[width=\columnwidth]{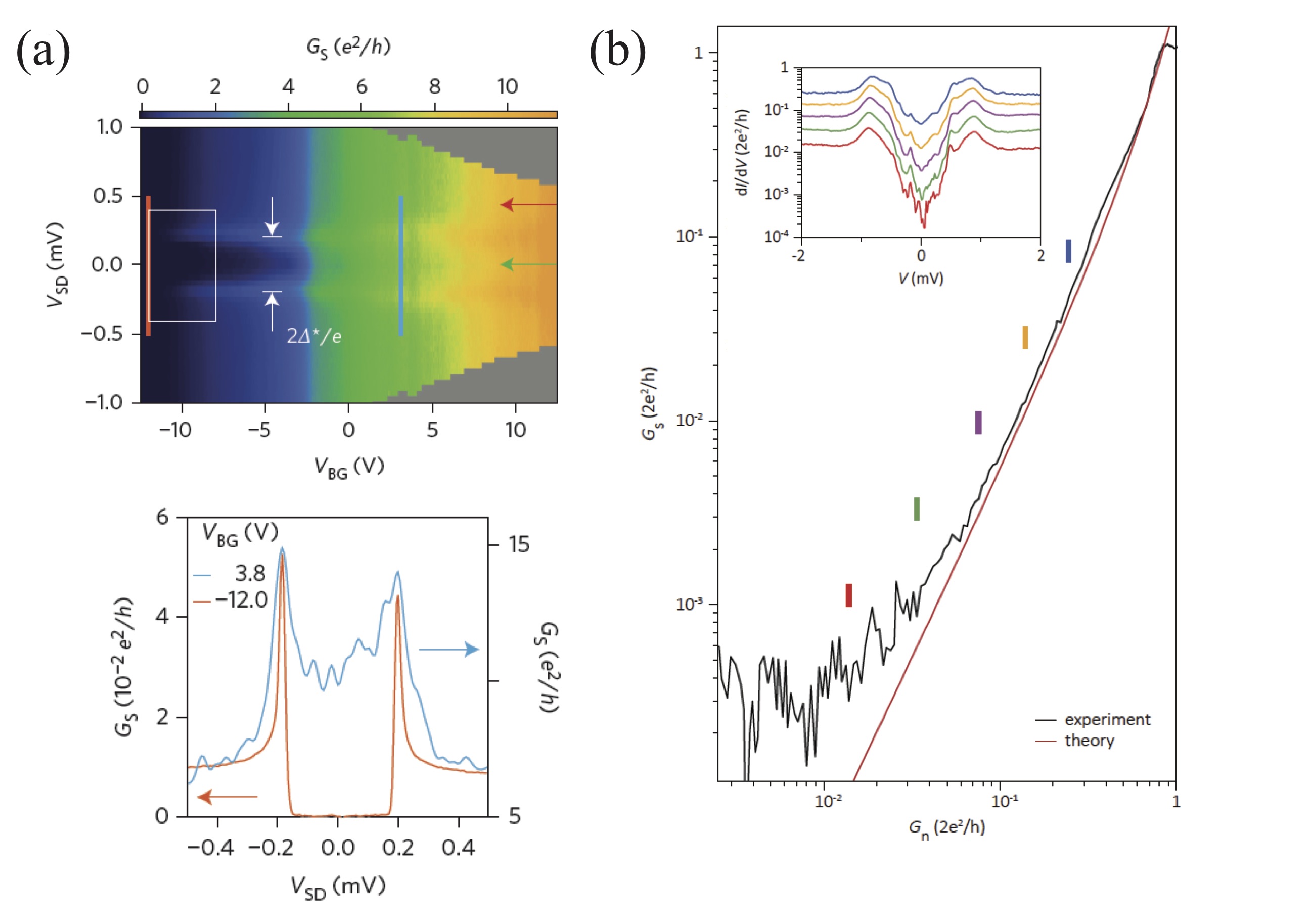}\\
\caption{Hard gaps. (a) Differential conductance $G_S$ of an epitaxial nanowire device as a function of backgate voltage $V_{BG}$ and sourceÐdrain voltage $V_{SD}$. Increasing $V_{BG}$, the conductance increases from the tunneling to the Andreev regime (orange and blue plots in the bottom). Adapted from Ref. \cite{Chang2015}. (b) Subgap conductance $G_s$ as a function of the normal (above-gap) conductance $G_n$. Red curve is the theory prediction for a single channel NS contact, Eq. (\ref{NS-Andreev}). Inset shows different $dI/dV$ taken at different values of $G_n$. Adapted from Ref. \cite{Zhang2016}.}
\label{fig:21}
\end{figure}
\subsection{Semiconducting Nanowires II: the second generation}
During the last couple of years, the soft gap problem has essentially been solved: various fabrication breakthroughs, such as the epitaxial growth of crystalline superconductor shells directly on the surface of the nanowires \cite{Krogstrup2015} or the careful engineering of high-quality semiconductor-superconductor interfaces, have allowed for dramatically cleaner devices, with longer mean free paths and a robust induced superconductivity. Importantly, barrier-free electrical contact to the superconductor and hard gaps have been reported in both InAs-Al \cite{Chang2015}  and InSb-NbTiN \cite{Zhang2016,Gul2017} interfaces. Both kind of devices exhibit very good induced hard gaps and almost perfect Andreev conductance as the transmission is gated from the tunneling regime to the transparent one. Fig. \ref{fig:21}a shows data from Copenhagen and  Fig. \ref{fig:21}b shows data from Delft. In both cases, the agreement with the single channel formula for the enhanced subgap conductance in an NS contact owing to Andreev reflection \cite{Beenakker92}
\begin{equation}
\label{NS-Andreev}
G_s=\frac{4e^2}{h}\frac{G_n^2}{(\frac{4e^2}{h}-G_n)^2},
\end{equation}
where $G_n=\frac{2e^2}{h} T_N$, is almost perfect (see Fig. \ref{fig:21}b).

\begin{figure}[!tt]
\centering
\includegraphics[width=\columnwidth]{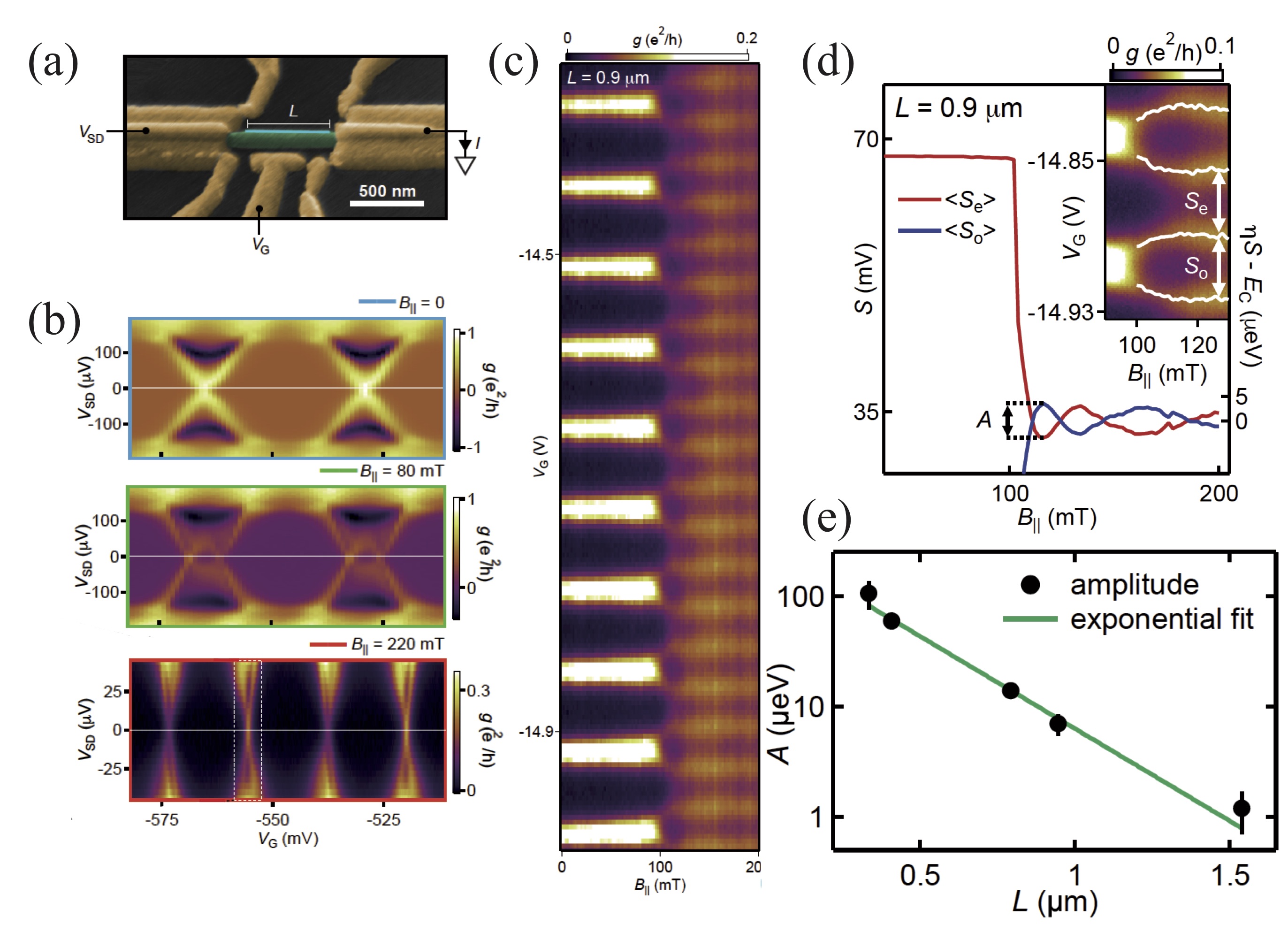}\\
\caption{Copenhagen experiment with Majorana islands. (a) Electron micrograph (false color) of a Majorana box device (yellow denotes gold contacts and green is the InAs nanowire). (b) Differential
conductance, as a function of gate voltage, $V_G$, and source-drain bias, $V_{SD}$, for increasing parallel magnetic fields, $B_{|| }$. For $B_{|| }= 0$ the Coulomb diamonds are evenly spaced. For increasing magnetic fields,  odd diamonds appear until the diamonds become evenly spaced for the largest $B_{|| }$ with a period in gate voltage that has halved. This effect can be clearly seen in (c) where the series of $2e$-periodic Coulomb blockade peaks below  $B_{|| }\sim 100 mT$ become $1e$-periodic for larger magnetic fields. (d) The average peak spacing for even and odd Coulomb valleys reveals oscillations around the $1e$-periodic value as a function of applied magnetic field. (e) The oscillation amplitudes for five devices of different length allow to extract a typical Majorana length of $\xi=260 nm$ through an exponential fit,  in good agreement with the estimations. Adapted from Ref. \cite{Albrecht2016}.}
\label{fig:22}
\end{figure}

The first materials combination, from the Marcus lab in Copenhagen, indeed showed improved Majorana signatures, extracted through Coulomb blockade spectroscopy \cite{Albrecht2016} and by using an additional quantum dot to perform detailed transport spectroscopy \cite{Deng2016}. Fig. \ref{fig:22} shows an example of the first kind of experiments in Majorana islands (small floating superconductors fabricated with epitaxial Al-InsAs nanowires, green region in Fig. \ref{fig:22}a). Fig.\ref{fig:22}b shows the conductance as a function of gate voltage, $V_G$, and source drain bias, $V_{SD}$. At zero magnetic field (top panel), the conductance show a series of evenly spaced Coulomb diamonds as expected for $\Delta>E_C$. As the magnetic field increases ($B_{||}=80mT$ middle panel), the even diamonds shrink and a second set of diamonds appears inside the previous crossings, yielding the even-odd spacing of Coulomb blockade zero-bias conductance peaks, as we discussed in subsection \ref{CB}. For large magnetic fields ($B_{||}=220mT$ bottom panel), Coulomb diamonds are again periodic, but with $1e$ periodicity. It is important to point out that this regime with $1e$ Coulomb Blockade periodicity appears at magnetic fields which are not high enough to destroy superconductivity, which supports the interpretation of an emergent quasiparticle excitation at zero energy. The full sequence of Coulomb Blockade peaks as a function of $B_{||}$ in Fig.\ref{fig:22}c clearly shows the evolution from $2e$ to $e$ periodicity. An important test to prove the nontrivial origin of the zero quasiparticle excitation is to extract the expected Majorana overlap (since the Majorana island has a relatively short length in the range  $L\sim 1\mu m$). In order to extract the expected overlap, the authors 
average (over an ensemble of adjacent peaks) even and odd Coulomb peak spacings.  This average reveals oscillations around the $1e$-periodic value as a function of applied magnetic field (Fig.\ref{fig:22}d) which is qualitatively consisting with an oscillating energy $E_0$ around zero due to hybridized Majoranas. The data also shows very good agreement with the predicted exponential decay that characterizes topological protection. This is demonstrated in Fig.\ref{fig:22}e, where the oscillation amplitudes for five different devices, along with a fit to an exponential function, 
$A = A_0e^{-L/\xi}$ allow to extract a typical Majorana length of $\xi=260 nm$ in good agreement with the estimations. Arguably, these experiments reported in Ref.  \cite{Albrecht2016} are much more convincing than the previous generation of experiments since they allow to go well beyond the ZBA paradigm.

\begin{figure}[!tt]
\centering
\includegraphics[width=\columnwidth]{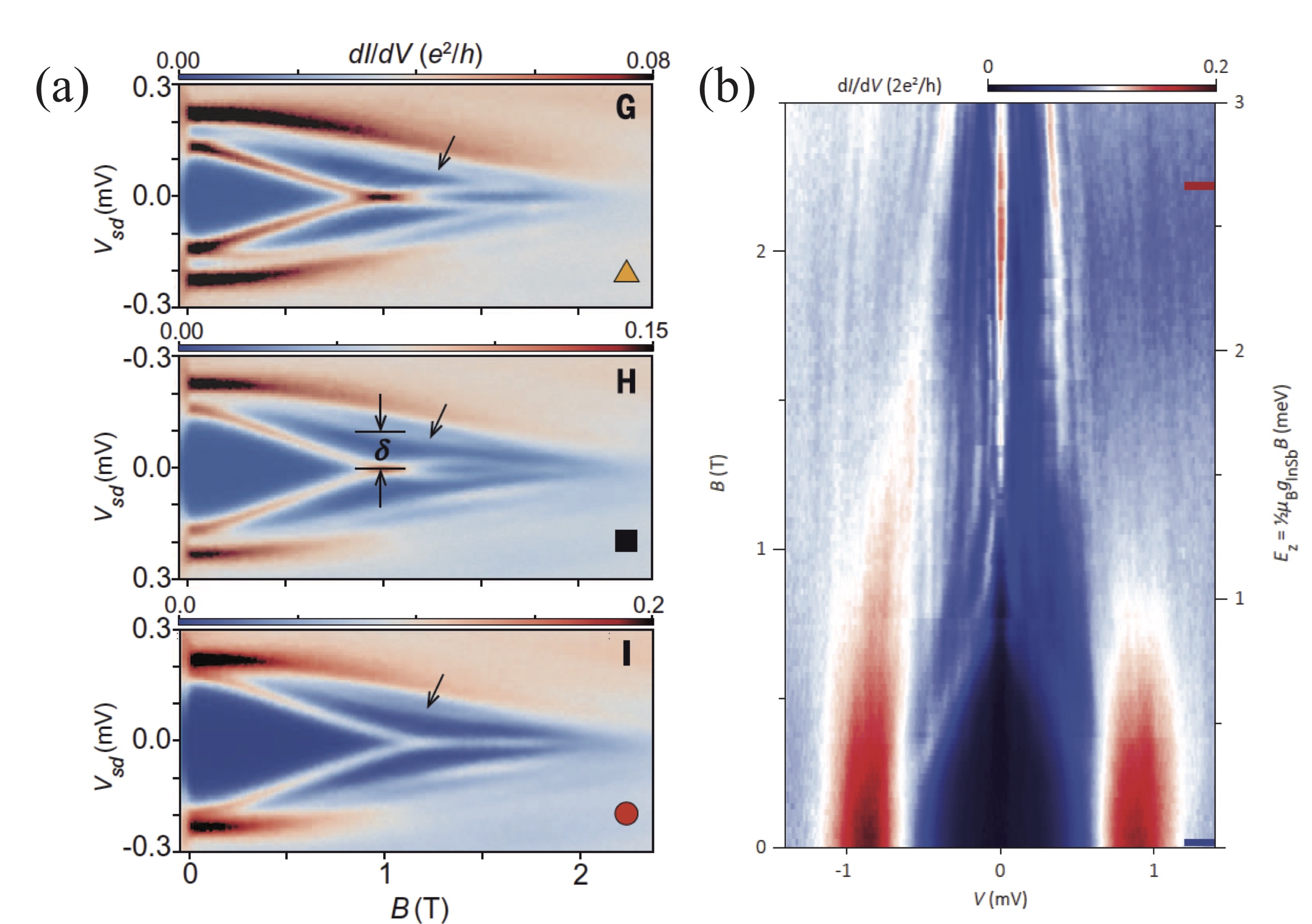}\\
\caption{Robust zero bias anomalies. a) Quantum dots coupled to nanowires allow to study in detail the emergence of Majoranas (bottom) from coalescing YSR states (top), like the ones in Fig. \ref{fig:20}, as a function of Zeeman field. Adapted from Ref. \cite{Deng2016}. (b) Robust zero bias anomaly in an InSb nanowire proximitized with NbTiN. Adapted from Ref. \cite{Zhang2016}.}
\label{fig:23}
\end{figure}

An alternative point of view, also from Copenhagen, is to perform detailed transport spectroscopy of Majoranas with the aid of a quantum dot at the end of the nanowire \cite{Deng2016}. This is a very interesting alternative since this setup allows to study in detail the emergence of Majoranas from coalescing YSR states (see Fig. \ref{fig:23}a) as a function of either the Zeeman field or the gate voltage applied to the wire (which changes the wire's chemical protential). Recently, the interpretation of these experiments in terms of Majoranas has been challenged by the Das Sarma's group by pointing out that Andreev/YSR levels sticking to zero energy are essentially indistinguishable from Majoranas \cite{Liu-DasSarma2017}. This idea is essentially a follow-up of the discussion in Fig. \ref{fig:20}, where we pointed out the similarity between parity crossings induced by a Zeeman field and emergent Majoranas, adapted to proximitized wires.  Interestingly, it has been proposed that this geometry with a quantum dot at the end of the nanowire may be used as a powerful spectrometer that fully quantifies the degree of Majorana non-locality through a local transport measurement \cite{Prada2017,Clark2017}, by analysing in detail the typical spectral patterns showing anticrossings between quantum dots states and Majorana peaks \cite{Baranger2011,Vernek2013,Ruiz-Tijerina2015}. Interestingly, since the quantum dot sub-gap states are spin-polarised, the full spin structure of the Majorana wave function \cite{Sticlet2012} can also be probed using this geometry \cite{Prada2017}. We believe that systematic experiments trying to quantify the degree of Majorana non-locality through anticrossings between quantum dot levels and zero-energy states would be able to settle 
the nagging question of trivial Andreev levels sticking to zero energy versus Majoranas by fully studying the crossover from fully localised trivial Andreev levels to fully non-local Majoranas for increasing Zeeman fields  \cite{Deng2017}.

\begin{figure}[!tt]
\centering
\includegraphics[width=0.75\columnwidth]{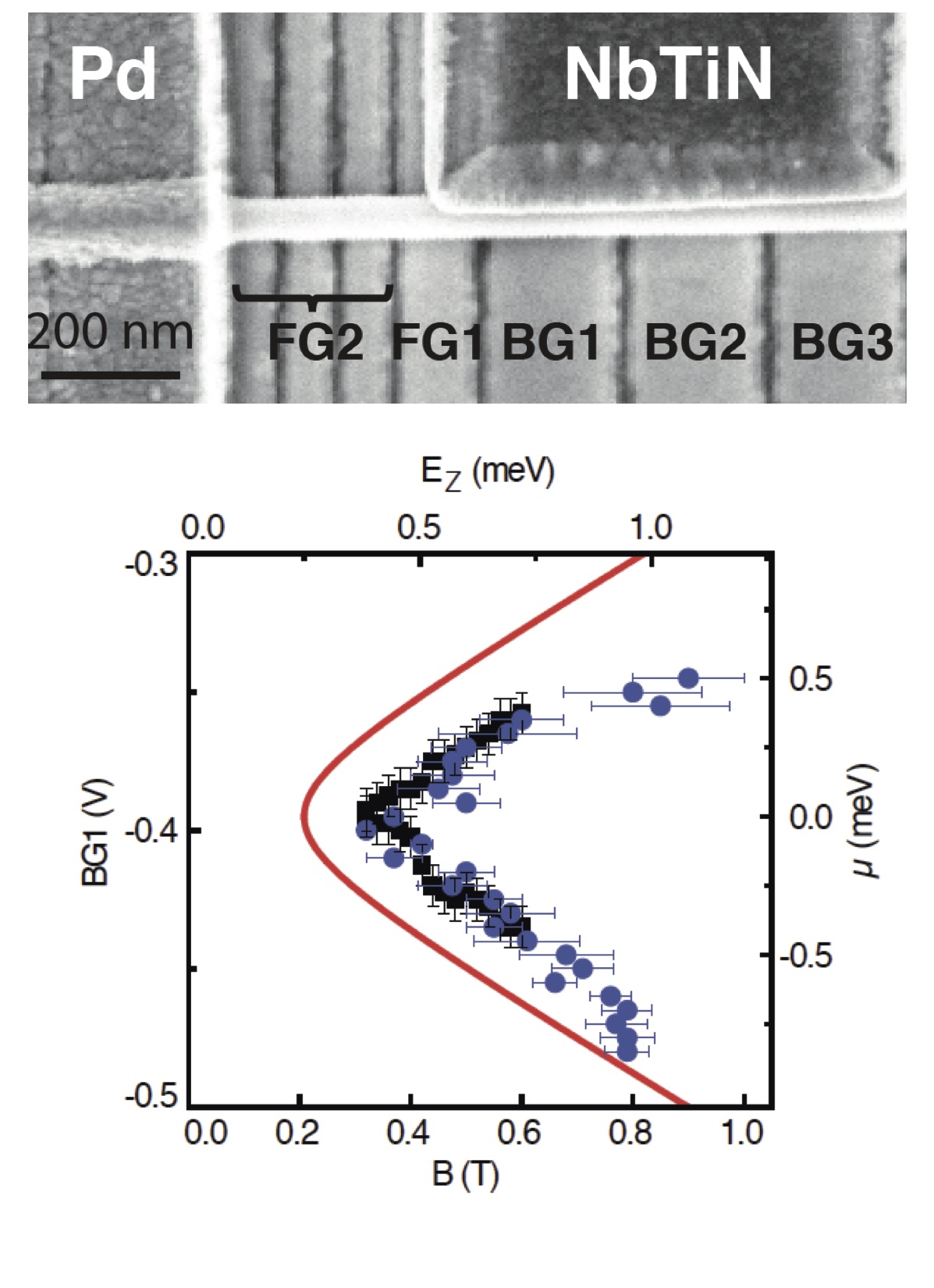}\\
\caption{Top: Scanning electron micrograph of the device consisting of an InSb nanowire half-covered by a superconductor NbTiN, and normal metal Pd contact. Bottom: By plotting the zero-bias peak onset points as a function of gate voltage and magnetic field the topological phase diagram can be mapped out (compare with Fig. \ref{fig:8}d). The top axis $E_Z$ is calculated from magnetic field using $g = 40$. The right axis is calculated from $B_{G1}$ according to $10 meV/V$, and set to be zero at the parabolic vertex, around $B_{G1}\sim -0.4V$. Adapted from Ref. \cite{Chen2016}.}
\label{fig:24}
\end{figure}

The second materials combination (InSb-NbTiN), from the Kouwenhoven lab, is also very promising. Without $B$ field, the experiments from Delft \cite{Zhang2016} demonstrate ballistic transport in the form of a quantized conductance in the normal state that is strongly enhanced (almost doubled) owing to almost perfect Andreev reflection. In the tunneling regime, an induced hard-gap with a strongly reduced subgap density of states is demonstrated at $B=0$. At finite $B$ fields, very robust Majorana peaks, rigid over a large region in the parameter space of gate voltage and magnetic field, are observed with conductance values reaching almost $G\sim 0.5e^2/h$ above the background conductance.Very recently, similar InSb-NbTiN devices were used in the Frolov lab in Pittsbugh to map out the full topological phase diagram (see Fig. \ref{fig:8}d) through the dependence of the zero-bias peak on the chemical potential and magnetic field \cite{Chen2016}. An example of this mapping is shown in Fig. \ref{fig:24}.

Further progress with the InSb nanowires includes the fabrication of sharp and narrow tunnel barriers in InSb nanowires with built-in $Ga_xIn_{1-x}Sb$ \cite{Car2017}. Such sharp barriers would help to overcome the smooth tunnel barrier \cite{Prada2012}, and the inevitable weak ZBAs in multiband nanowires, that we discussed before. Moreover, conductance quantization data with some features compatible with a helical gap were recently observed in quantum point contacts fabricated with InSb nanowires \cite{Kammhuber2017}.  An Andreev molecule in an InSb nanowire double quantum dot (which is the minimal building block to artificially engineer a Kitaev one-dimensional topological p-wave superconductor with a chain of quantum dots \cite{Sau-NatComm2012,Fulga2013}) has also been reported in Ref. \cite{Su2016}.

Another interesting route is to perform high frequency studies in nanowire junctions. Microwave spectroscopy measurements that directly reveal the presence of spinful Andreev bound states in InAs nanowire-based Josephson junctions have been reported in Ref. \cite{Woerkom2016}. A second experiment demonstrating an on-chip microwave coupling circuit to measure the microwave radiation spectrum of an InSb nanowire junction was presented in Ref. \cite{Woerkom2017}. This setup allows to measure the Josephson radiation emitted by an InSb-NbTiN junction using photon assisted quasiparticle tunneling in an AC-coupled superconducting tunnel junction. These high-frequency studies hold promise as a new tool for studying Majorana bound states in nanowires.

Finally, it is also interesting to mention the recent experiments from the Giazotto's lab in Pisa \cite{Strambini2017}, where an anomalous enhancement of the critical supercurrent through a Josephson junction formed by an InAs nanowire and Ti/Al superconducting leads was clearly observed for increasing magnetic fields. Such enhancement is compatible with a magnetic field-induced topological transition, as described in Ref. \cite{SanJose2014}.

\subsection{Proximitized edges of two-dimensional topological Insulators}
During the last few years, inducing superconductivity through the proximity effect into the surface or edge states of a topological insulator has been revealed as a powerful tool for studying topological superconductivity. The first experiments demonstrating the
Josephson effect in topological insulators were reported almost in parallel with the early ZBA experiments in nanowires. These include junctions made of  $Bi_2Se_3$ \cite{Sacepe2011,Williams2012} and $Bi_2Te_3$ \cite{Veldhorst2012,Qu2012}. However, these bismuth-based topological insulators are known to exhibit a residual bulk conductance that is a serious drawback to investigate induced superconductivity in topological surface states. Strained HgTe has, on the other hand, an effectively insulating bulk which makes it a more suitable candidate to induce topological superconductivity. Proximity induced superconductivity in strained bulk $HgTe$ was first demonstrated in 2013 in the group of Laurens Molenkamp in W\"urzburg, \cite{Oostinga2013} in a Nb-HgTe-Nb Josephson junction. 

Subsequent measurements of the current-phase relationship using a scanning SQUID technique revealed  clear deviations from the standard sinusoidal shape of a tunneling junction, with current-phase behaviour close to a skewed sawtooth curve (see discussion in subsection \ref{4pi}). These results suggested that the supercurrent is carried by a ballistic mode, consistent with a helical Andreev bound state. Further evidence from nontrivial Andreev states comes from Shapiro steps in Josephson junctions irradiated with microwaves \cite{Wiedenmann2016}. At high frequencies, the Shapiro steps appear at standard voltages $V_n= n hf/2e$. At lower frequencies the Shapiro response of the junction changes and shows only partial agreement with the theory that we described in section \ref{4pi}: only the \emph{first step} disappears while the other odd steps remain visible. This is probably related to the large number of channels supported by the junction, where there is only one gapless Andreev level contributing to a $4\pi$ supercurrents, while many other trivial modes in the junction contribute with standard $2\pi$ supercurrents \footnote{A similar observation was reported in 2012 by Rokhinson {\it et al} who studied Shapiro steps in irradiated Nb/InSb/Nb Josephson junctions fabricated lithographically from a shallow InSb quantum well \cite{Rokhinson2012}. Quite likely, the disappearance of only the first odd step can be also attributed to many channels contributing to the critical current.}. Theoretically, this can be understood from a resistively shunted junction (RSJ) model \cite{Dominguez2017} which shows that lower critical currents (less channels) are required in order to have more higher index odd plateaus disappearing. Recently, capacitive effects have been included within a resistively capacitively shunted junction (RCSJ) model for the junction dynamics in Ref. \cite{Pico2017}. The inclusion of these capacitance effects results in a strong reduction of the first step when the junction is underdamped, while the higher odd steps are less affected, in agreement with the experiments.

\begin{figure}[!tt]
\centering
\includegraphics[width=\columnwidth]{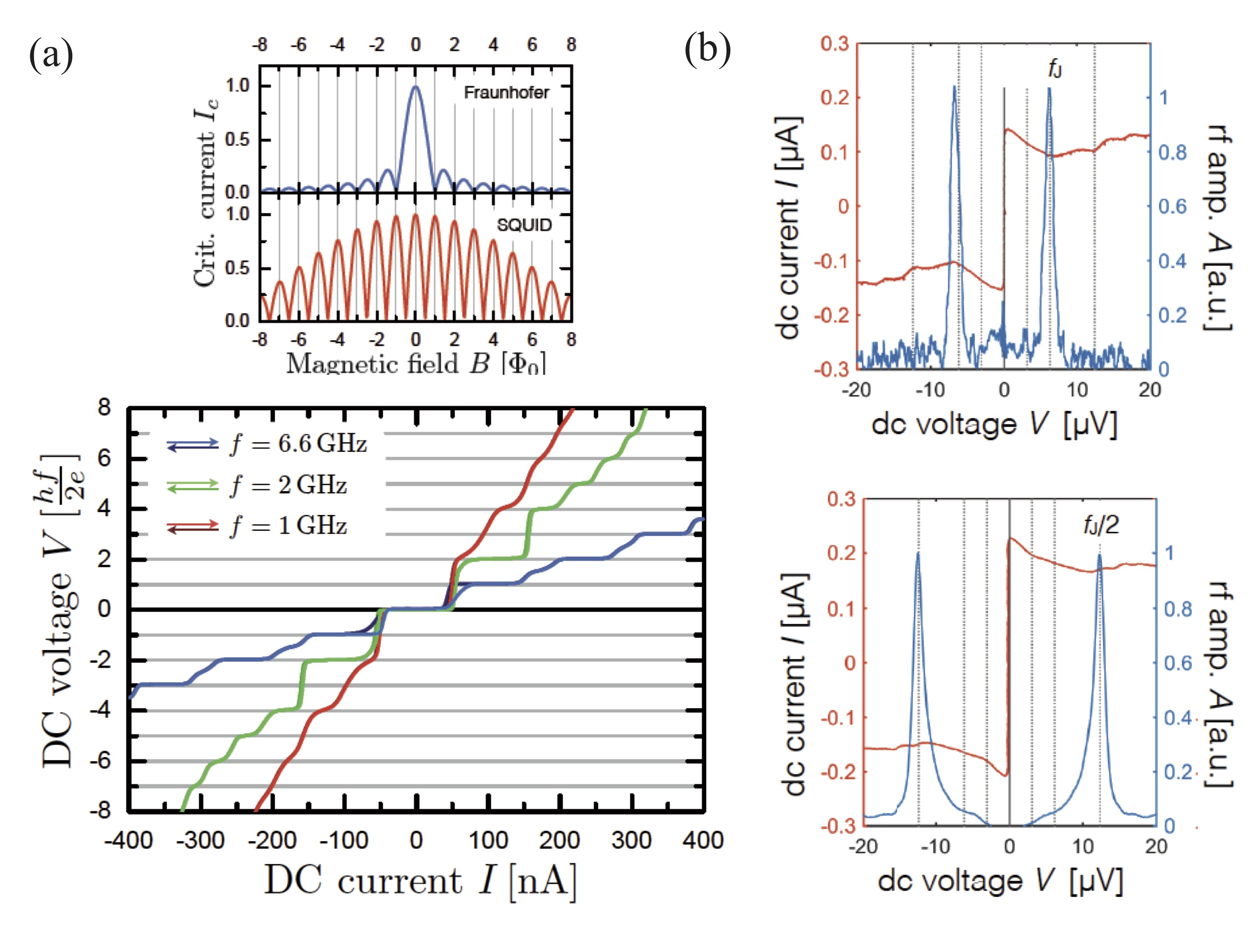}\\
\caption{$4\pi$ effect experiments from W\"urzburg. (a) Top: Critical current $I_c$ as a function of the magnetic field B of a Josephson junction fabricated with a HgTe quantum well, a 2D topological insulator that exhibits the quantum spin Hall (QSH) effect. For a uniform planar current (trivial regime), a Fraunhofer pattern (blue line) is obtained. For current flowing on the edges (QSH regime), a SQUID pattern (red line) is recovered. Bottom: The I-V curves in the presence of radio-frequency excitation for three different frequencies exhibit a different sequence
of Shapiro steps. For the largest frequency (f = 6.6 GHz) all integer Shapiro steps are visible while for the lowest one (f = 1 GHz) odd steps ($n = \pm 1, \pm 3, \pm 5$) are clearly missing. (b) I-V curves (red lines) and emission spectrum (blue lines) for a nontopological weak link (top) and a
quantum spin Hall weak link (bottom). The trivial case shows an emission peak at the expected Josephson frequency $f_J$ while the quantum spin Hall weak link has an emission peak at half the Josephson frequency $f_J/2$. Adapted from Ref. \cite{Bocquillon2017} and  Ref. \cite{Deacon2017}.}
\label{fig:25}
\end{figure}

The above results suggest that quantum spin Hall insulators, which ideally do not present the many-channel problem, should be better suited for exploring topological superconductivity. An important prerequisite to demonstrate that the supercurrent is only carried by the helical edge modes is to perform Fraunhofer spectroscopy which provides information on the spatial dependence of the critical current density: when a planar junction is dominated by bulk modes, the uniform critical current flow results in a standard Fraunhofer pattern while a current flow only through the edges of the sample results in a dc SQUID response \cite{Hart2014,Pribiag2015}. After confirming this predicted Fraunhofer behaviour  Fig. \ref{fig:25}a (top), Bocquillon {\it et al} reported very convincing experimental evidence of $4\pi$-periodic supercurrents in Josephson junctions fabricated with $HgTe$ quantum wells, a two-dimensional topological insulator that exhibits the quantum spin Hall effect, proximitized with Nb  \cite{Bocquillon2017}. Results from this experiments are shown in Fig. \ref{fig:25}a (bottom), where a clear Shapiro response with only even steps at low frequencies (up to $n = 9$ missing odd steps are reported) is demonstrated.

Further evidence of  $4\pi$-periodic supercurrents in this kind of junctions was recently reported  by the same group where the emitted Josephson radiation was directly measured \cite{Deacon2017}, see Fig. \ref{fig:25}b. The trivial case, where the junction is fabricated with a thin quantum well where the band structure is not inverted, is plotted in the top panel and shows an emission peak is at the expected Josephson frequency. In the topologically inverted structure (bottom panel), the radiation is emitted at half the Josephson frequency, corresponding to a doubling of the period from $2\pi$ to $4\pi$ (see subsection \ref{4pi} and Fig. \ref{fig:16}c for a more detailed explanation).

While these experiments have provided convincing evidence of a supercurrent with $4\pi$ periodicity, some questions remain hitherto unanswered. The main objection to the interpretation is that there is no explicit breaking of time reversal symmetry, which, within the Fu and Kane model \cite{Fu-Kane2}, would imply that no $4\pi$ signal should exist (the zero modes are not detached from the continuum and Landau-Zener processes like the red arrows in Fig. \ref{fig:15} are unavoidable). The experiments
are nevertheless quite convincing that $4\pi$ supercurrents are present. The obvious answer is that  probably time reversal symmetry is spontaneously broken. This is also in agreement with the conductance quantization of the quantum spin Hall edge modes, which is not perfect. A possible explanation has been put forward in Ref. \cite{Peng2016}, where the role of density modulations and their induced puddles of electrons have been considered. These puddles act as small quantum dots and, therefore, when they host an odd number of electrons, charging effects turn them into magnetic impurities which are exchange coupled to the helical edge channels. Remarkably, the presence of such magnetic impurities turns the Josephson current $8\pi$-periodic, instead of the dissipative $2\pi$ periodicity in pristine junctions. Similar results have been reported in  Ref. \cite{Liu2017}. Very recently, it has also been proposed that realistic smooth edge potential, as opposed  to sharp edge potentials used in standard calculations, gives rise to edge reconstruction and, consequently, spontaneous time-reversal symmetry breaking \cite{Wang2017}.

\subsection{Atomic chains}
Chains of magnetic atoms on the surface of a superconductor provide yet another approach to engineer topological superconductivity. This platform for topological superconductivity largely relies on the physics of magnetic adatoms in superconductors and, more specifically, on the formation of subgap YSR states that appear inside the superconducting gap $\Delta$. In the classical limit, the YSR bound state energy is given by the expression
\begin{equation}
\varepsilon_0=\pm\Delta\frac{1-\alpha^2}{1+\alpha^2},
\end{equation} 
where $\alpha=\pi\rho_0JS$ with $\rho_0$, $J$ and $S$ being the normal density of states, the exchange interaction and the spin of the impurity, respectively. These bound states are spin
polarized along the direction of the impurity spin $S$ and their wave function is localised around the impurity, decaying as $1/r$ for distances smaller than the coherence length and exponentially at distances longer than $\xi$, namely 
 \begin{equation}
 \Psi_{YSR}(r)\sim \frac{e^{-r/\xi}}{r},
 \end{equation} 
 with an energy-dependent coherence length defined as
 \begin{equation}
\xi=\frac{\hbar v_F}{\sqrt{\Delta^2-\varepsilon_0^2}}.
\end{equation} 
Similar to the quantum dot case that we described previously, when the energy of the YSR state crosses zero (which happens here at $\alpha=1$) the system undergoes a quantum phase transition where the fermionic parity of the ground state changes from even to odd. A chain of magnetic impurities leads to the formation of a band of such YSR states inside the gap of the superconductor. For very dilute atomic chains, the physics can be described starting from weakly overlapping Shiba states associated with individual magnetic impurities. In this case,  it can be shown that the system supports a topological phase for very deep Shiba states ($\alpha\sim 1$), assuming that 
superconducting pairing in such spin-polarized chain can be induced via proximity to an s-wave superconductor. This can be accomplished if some noncollinear spin configuration is present. The combined effect of
the superconductivity and the noncollinear spin configuration maps the Shiba chain onto an effective Kitaev chain model that can support a topological phase with Majorana bound states \cite{Choi2011,Nadj-Perge2013,Pientka2013b,Braunecker2013,Klinovaja2013,Vazifeh2013}. Physically, noncollinear spin ordering can be induced, for example, if the superconducting electrons mediate Ruderman-Kittel-Kasuya-Yoshida (RKKY) interactions between the localized magnetic moments of the chain. This RKKY interaction induces a spiral magnetic order with a pitch that depends on dimensionality (the simplest case being $2k_F$ in one dimension). In the more realistic case of a two-dimensional substrate, the formation of a helical magnetic chain results from the interplay of various effects such as RKKY interaction and spin-orbit coupling (resulting in Dzyaloshinskii-Moriya interactions \cite{Kim2014}). More simply, a ferromagnetic chain can also host a topological superconducting phase if the superconducting substrate has a strong spin-orbit coupling \cite{Li2014,Brydon2015}. If, on the other hand, the magnetic atoms in the chain are densely packed, it is more appropriate to think in terms of bands with large bandwidth \cite{Hui2015}. This limit is very similar to the nanowire proposal and also hosts topological phases with Majorana bound states at the ends of the chain. 

\begin{figure}[!tt]
\centering
\includegraphics[width=\columnwidth]{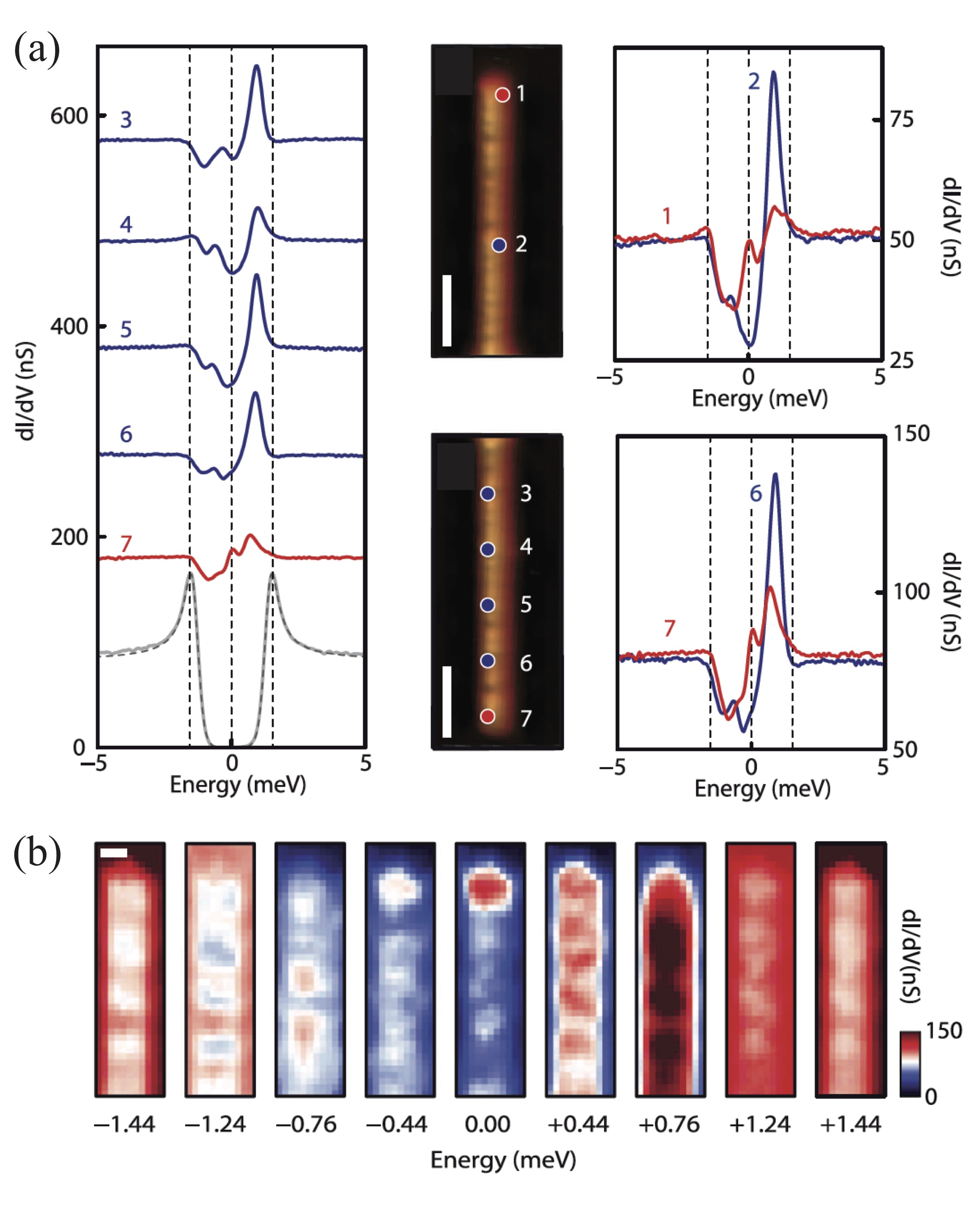}\\
\caption{STM experiments of Fe chains on top of Pb (Princeton). (a) STM spectra measured through $dI/dV$ (left and right panels) on different atomic chain locations (the center panels show the topography of the upper and lower end of the chain). The red spectrum shows a zero bias anomaly for both ends of the chain (positions 1 and 7). (b) Spatial and energy-resolved conductance maps of one of the atomic chains close to its end. Scale bar 10 $\AA$. Adapted from Ref. \cite{Nadj-Perge2014}.}
\label{fig:26}
\end{figure}

These theoretical ideas (which are largely independent of the specific mechanism responsible for the helix formation) have recently been implemented in various experiments. The Yazdani lab in Princeton reported in 2014 signatures of Majorana states in chains of iron (Fe) atoms on the superconducting Pb(110) surface \cite{Nadj-Perge2014}.  Scanning tunnelling microscopy (STM) spectroscopy of these chains shows that the onset of superconductivity in the Pb substrate is always accompanied by the emergence of zero bias peaks in tunnelling spectroscopy data. Since STM measurements are spatially resolved, the Princeton experiment allows to directly visualise the Majorana end modes associated with these zero-bias peaks in tunnelling spectroscopy,  see Fig. \ref{fig:26}. Contrary to naive expectations, these Majoranas are strongly localised at the end of the chain which generated an interesting scientific debate \footnote{The estimated exchange field and bandwidth presented in the theory accompanying the experiments in Ref. \cite{Nadj-Perge2014} are much larger than the energy gap of the induced superconductivity on the Fe chain (which is estimated to be very small $\sim 0.2 meV$). According to the simplest estimates, the superconductivity coherence length is expected to be much longer that the atomic chain itself, because of the small energy gap (more specifically, and since the induced gap is only a small percentage of the bulk gap of the Pb substrate, the localization length of
the end states is expected to be much longer than the coherence length of Pb). This expectation seems at odds with the experimental observation of very short localization lengths (the zero bias peak signatures disappear at very short distances $\sim 10\AA$, which is orders of magnitude smaller than the coherence length of Pb). One possible mechanism that might explain the discrepancy is the strong Fermi velocity renormalization due to interaction effects \cite{Peng2015}. This renormalization results in localization lengths much smaller than the coherence length of the host superconductor.}. Subsequent STM experiments with superconducting tips have reported similar zero-energy states with longer localization lengths $\lesssim 25nm$ (Basel group \cite{Pawlak2016}) and with a rich subgap structure, including zero-energy states in some cases (Berlin group \cite{Ruby2015}). Higher resolution low-temperature STM experiments, also from Yazdani's lab \cite{Feldman2017}, clearly demonstrate that the zero-bias peak has no detectable splitting from hybridization with other subgap states. Recent data from the same group shows that spin-resolved STM allows to distinguish between trivial YSR states and Majorana end modes \cite{Yazdani-Mallorca,Yazdani-Science2017}.

The concept of Shiba chains can be also extended to two dimensions where Shiba lattices \cite{Shibalattice1,Shibalattice2,Shibalattice3} have been proposed as a new playground for topological superconductivity with chiral Majoranas and nonzero Chern numbers $C$. Interestingly, and similar to the integer quantum Hall effect, these systems with nonzero Chern number display a universal thermal Hall conductivity $\kappa_{xy}=C\frac{\pi^2k_B^2}{3h}T$. Recently, it has been predicted that chiral Majoranas in such Shiba lattices also possess a quantized charge conductance that is proportional to the topological Chern number and carry a supercurrent whose chirality reflects the sign of $C$ \cite{Shibalattice4}.
\section{Outlook}
There has already been remarkable experimental progress since the early theoretical proposals that demonstrated the possibility of artificially engineering topological superconductivity.
During the last few years, the second generation of experiments from Copenhagen  \cite{Albrecht2016,Deng2016}, Delft \cite{Zhang2016}  and Pittsburgh \cite{Chen2016} have reported remarkably clear signatures of Majorana zero modes in semiconducting nanowires using various techniques (Coulomb blockade, Andreev transport, quantum dot spectroscopy, etc). Moreover, the progress in other experimental platforms is also outstanding. This includes various high-frequency experiments from W\"urzburg \cite{Bocquillon2017,Deacon2017} that have reported very convincing evidence of $4\pi$-periodic supercurrents in Josephson junctions based on quantum spin Hall insulators. Furthermore, impressive STM experiments on ferromagnetic atomic chains deposited on top of superconductors \cite{Nadj-Perge2014,Pawlak2016,Feldman2017,Yazdani-Mallorca} reveal clear features of zero-energy Majorana states localized at the ends of the chain. 

Although not discussed in this review, we also mention in passing the experimental progress in other hybrid systems, such as the recent report of chiral Majorana edge states in superconductor-quantum anomalous Hall insulator structures \cite{chiralZhang2016} (implemented using a magnetic topological insulator thin film in contact with a superconducting electrode) or recent spin-polarized STM measurements of vortices in superconductor-topological insulator heterostructures  \cite{Hua2016} which reveal a spin selective Andreev reflection \cite{Law2014} consistent with the existence of Majorana zero modes inside the vortex core. Furthermore, chiral Majorana modes surrounding topological superconducting domains made of a single atomic layer of Pb covering magnetic islands of Co/Si(111) have been reported in the experiments of Ref. \cite{Cren2016}. Tremendous progress in graphene-based hybrid systems, which includes the observation of supercurrents in the quantum Hall regime \cite{Amet2016} and the recent experimental observation of Andreev levels in superconductor-graphene-superconductor junctions \cite{Jarillo2017}, also deserves to be mentioned here. Also worth mentioning are the recent experiments in nanowires lithographically patterned on hybrid two-dimensional electron gases based on InAs/Al quantum well heterostructures \cite{2DEG1,2DEG2}, where further evidence of Majoranas has been reported \cite{2DEG3,2DEG4}. Interestingly, the high degree of tunability in these samples allows for very systematic studies, such as the scaling of ZBAs predicted in Eq. (\ref{dI-dV-finiteT0}), which is yet another piece of evidence that supports the Majorana interpretation.

Taken together, all these experiments indicate that inducing topological superconductivity by means of the proximity effect has been achieved, and that emergent Majorana quasiparticles have been created and measured in various solid state platforms.  Also, alternative platforms for engineering topological superconductivity and intrinsic topological superconductors are, undoubtedly,  still full of surprises. The field is ripe and these state-of-the-art experiments lay the foundation for tackling next challenges. These include demonstrating spatial non-locality of Majoranas or implementing the various schemes for fusion rule experiments and braiding. The holy grail in the field is, of course, to demonstrate quantum computation with exponentially-protected topological qubits. 
On a more mundane level, a melting pot of fundamental physical phenomena is lurking in these hybrid devices. As we have seen along the review, interesting competitions between superconductivity, charging effects, Kondo physics, spin-orbit physics, disorder, and the like, give rise to a great variety of unusual properties that, in some cases, mimic Majorana physics and in most cases complicate the simpler models \cite{Fu-Kane1,Lutchyn2010,Oreg2010,Fu-Kane2} where induced topological superconductivity was predicted. It is clear that demonstrating non-Abelian statistics would take the field to a completely different level since a Majorana-based decoherence-free qubit would constitute a disruptive alternative to the already existing solid state qubits. While this happens, more down-to-earth studies aiming at fully characterizing and understanding the complex physics occurring in such hybrid platforms are crucial. These include a clear demonstration of helical states in nanowires, a direct microwave spectroscopy of subgap states in topological junctions, a demonstration of full quantum coherent control of these low-energy quasiparticle excitations, etc. 

It goes without saying that there is still a long road until the Ithaca of topological quantum information processing with Majoranas becomes a reality, but, quoting Constantine P. Cavafy, "when you depart for Ithaca, wish for the road to be long, full of adventure, full of knowledge.".
\acknowledgments
I am indebted to many colleagues and collaborators from whom I have learnt a great deal about Majoranas and related topics. In particular, I would like to thank Jorge Cayao, Silvano De Franceschi, Mingtang Deng, Karsten Flensberg, Sergey Frolov, Leo Kouwenhoven, Eduardo Lee, Alfredo Levy-Yeyati, Rosa L\'opez, Charles Marcus, Vincent Mourik, Pascal Simon and Rok \v{Z}itko. Special thanks go to Elsa Prada and Pablo San Jose, my close collaborators on the subject, for endless discussions about Majoranas and lots of fun working together. I would also like to thank Jorge Cayao for proofreading the manuscript. Financial support from the Spanish Ministry of Economy, Industry and Competitiveness (MINEICO) through Grant No. FIS2015-64654-P is acknowledged.

\end{document}